\documentclass[structabstract]{aa} 
\usepackage{amsmath}
\usepackage{amssymb}
\usepackage[varg]{txfonts}
\usepackage{natbib}
\usepackage{acro}[=v2] 
\usepackage{graphicx}
\usepackage{dcolumn}
\usepackage{bm}
\usepackage{hyperref}
\usepackage{float}
\usepackage{enumitem}
\usepackage[skip=10pt plus1pt, indent=10pt]{parskip}
\usepackage{lipsum}
\usepackage{xcolor}
\newcommand{\syidx}[1]{%
	\acs{#1}%
}

\DeclareAcronym{SYMed}{
	short=\rho,
	long=Energy density,
	class=symbol
}
\DeclareAcronym{SYMedbg}{
	short=\bar{\rho},
	long=Energy density of background universe,
	class=symbol
}
\DeclareAcronym{SYMedavg}{
	short=\langle\rho\rangle,
	long=Energy density averaged over oscillation period,
	class=symbol
}
\DeclareAcronym{SYMpr}{
	short=p,
	long=Pressure,
	class=symbol
}
\DeclareAcronym{SYMprbg}{
	short=\bar{p},
	long=Pressure of background universe,
	class=symbol
}
\DeclareAcronym{SYMpravg}{
	short=\langle p\rangle,
	long=Pressure averaged over oscillation period,
	class=symbol
}
\DeclareAcronym{SYMdenspar}{
	short=\Omega,
	long=Density parameter for component $i$,
	class=symbol
}
\DeclareAcronym{SYMDE}{
	short=\Lambda,
	long=Dark Energy,
	class=symbol
}
\DeclareAcronym{SYMH}{
	short=H,
	long=Hubble parameter in cosmic time,
	class=symbol
}
\DeclareAcronym{SYMHc}{
	short=\mathcal{H},
	long=Hubble parameter in conformal time,
	class=symbol
}
\DeclareAcronym{SYMG}{
	short=G,
	long=Gravitational constant,
	class=symbol
}
\DeclareAcronym{SYMc}{
	short=c,
	long=Speed of light,
	class=symbol
}
\DeclareAcronym{SYMa}{
	short=a,
	long=Scalefactor a(t),
	class=symbol
}
\DeclareAcronym{SYMg}{
	short=g,
	long=FLRW metric,
	class=symbol
}
\DeclareAcronym{SYMgbg}{
	short=\bar{g},
	long=FLRW metric of background,
	class=symbol
}
\DeclareAcronym{SYMLdens}{
	short=\mathcal{L},
	long=Lagrangian density,
	class=symbol
}
\DeclareAcronym{SYMsfbec}{
	short=\psi,
	long=wave function of BEC,
	class=symbol
}
\DeclareAcronym{SYMsf}{
	short=\varphi,
	long=Scalar Field,
	class=symbol
}
\DeclareAcronym{SYMsfperturb}{
	short=\phi,
	long=Perturbations of Scalar Field,
	class=symbol
}
\DeclareAcronym{SYMvpot}{
	short=V,
	long=Potential of the Scalar Field Dark Matter,
	class=symbol
}
\DeclareAcronym{SYMlambdasi}{
	short=\lambda,
	long=Strength of self-interaction,
	class=symbol
}
\DeclareAcronym{SYMdenscontrast}{
	short=\delta,
	long=The density contrast describes the relative deviation of density from the average density of the universe,
	class=symbol
}
\DeclareAcronym{SYMcurvpar}{
	short=\kappa,
	long={Curverture parameter: -1..open universe, +1..closed universe, 0..flat universe},
	class=symbol
}
\DeclareAcronym{SYMeos}{
	short=w,
	long={The equation of state (EOS) relates pressure to energy density},
	class=symbol
}
\DeclareAcronym{SYMderivD}{
	short=D,
	long={The covariant derivative with respect to $\mu$},
	class=symbol
}
\DeclareAcronym{SYMkron}{
	short=\delta_{ij},
	long={Kronecker delta},
	class=symbol
}
\DeclareAcronym{SYMmpsynh}{
	short=h,
	long={Metric perturbation in synchronous gauge},
	class=symbol
}
\DeclareAcronym{SYMmpsyneta}{
	short=\eta,
	long={Metric perturbation in synchronous gauge},
	class=symbol
}
\DeclareAcronym{SYMmpnewpot}{
	short=\Phi,
	long={Metric perturbation in newtonian gauge -- potential},
	class=symbol
}
\DeclareAcronym{SYMmpnewlapse}{
	short=\Psi,
	long={Metric perturbation in newtonian gauge -- laps function},
	class=symbol
}
\DeclareAcronym{SYMct}{
	short=\tau,
	long={Conformal time},
	class=symbol
}
\DeclareAcronym{SYMt}{
	short=t,
	long={Cosmological time},
	class=symbol
}
\DeclareAcronym{SYMemt}{
	short=T,
	long={Energy momentum tensor},
	class=symbol
}
\DeclareAcronym{SYMemtbg}{
	short=\bar{T},
	long={Energy momentum tensor for background},
	class=symbol
}
\DeclareAcronym{SYMemtvelocity}{
	short=\theta,
	long={Velocity divergence of the energy-momentum tensor},
	class=symbol
}
\DeclareAcronym{SYMemtstress}{
	short=\sigma,
	long={Anisotropic stress of the energy-momentum tensor},
	class=symbol
}
\DeclareAcronym{SYMradius}{
	short=R,
	long={Radius of the Universe},
	class=symbol
}
\newcommand{\glo}[1]{%
	{\acs{#1}}%
}
\newcommand{\emidx}[2][]
{%
	\ifx&#1&%
	\q{#2}%
	\index{#2}%
	\else%
	#1#2#1%
	\index{#2}%
	\fi%
}

\newcommand{\TBDgl}{%
	\colorbox{yellow!30}{%
		\parbox{\dimexpr\linewidth-2\fboxsep\relax}{%
			Glossary to be done
		}%
	}
}
\DeclareAcronym{staticuniverse}
{
	short=static universe,
	long=\textit{Is a universe, that does not expand},
	extra=\\\TBDgl,
	class=glossary
}
\DeclareAcronym{flrwuniverse}
{
	short=\ac{FLRW} universe,
	long=\textit{Is a universe that is described by the \glo{flrwmetric}},
	extra=\\\TBDgl,
	class=glossary
}

\DeclareAcronym{flrwmetric}
{
	short=\ac{FLRW} metric,
	long=\textit{This metric is the most general type of a metric describing a homogeneous and isotropic spacetime},
	extra=\\\TBDgl,
	class=glossary
}

\DeclareAcronym{cosmoprinciple}
{
	short=cosmological principle,
	long=\textit{The cosmological principle refers to a universe, which is homogeneous (all locations in space are equivalent) and isotropic (all directions in space are equivalent)},
	extra=\\\TBDgl,
	class=glossary
}

\DeclareAcronym{LCDMmodel}
{
	short=$\Lambda$CDM model,
	long=\textit{Is a cosmological model, belonging to the family of the \glo{flrwuniverse} and has \ac{CDM} and \ac{DE}},
	extra=\\\TBDgl,
	class=glossary
}

\DeclareAcronym{LSFDMmodel}
{
	short=$\Lambda$SFDM model,
	long=\textit{Is a cosmological model based on the \glo{LCDMmodel}, where the CDM component is \acidx{SFDM} instead of the standard \acidx{WIMP} particles},
	extra=\\\TBDgl,
	class=glossary
}

\DeclareAcronym{friedmannequations}
{
	short=Friedmann equations,
	long=\textit{Friedmann derived a set of equations, that describe the evolution of the universe under the assumption of the \glo{cosmoprinciple}},
	extra=\\\TBDgl,
	class=glossary
}

\DeclareAcronym{friedmannequation}
{
	short=Friedmann equation,
	long=\textit{The term Friedmann equation refers to the first of the two Friedmann equations},
	class=glossary
}

\DeclareAcronym{decelerationequation}
{
	short=deceleration equation,
	long=\textit{The deceleration equation refers to the second of the two Friedmann equations},
	class=glossary
}

\DeclareAcronym{scalarfield}
{
	short=scalar field,
	long=\textit{The scalar field describes bosonic matter, that behaves like a \ac{BEC}},
	extra=\\\TBDgl,
	class=glossary
}

\DeclareAcronym{inflaton}
{
	short=Inflaton,
	long=\textit{The Inflaton is a \glo{scalarfield} that caused the inflationary epoch of the universe},
	extra=\\\TBDgl,
	class=glossary
}
\DeclareAcronym{inflation}
{
	short=inflation,
	long=\textit{The inflation is a very short period in time in the very early universe, where the universe expanded by many orders of magnitude},
	extra=\\\TBDgl,
	class=glossary
}
\DeclareAcronym{inflationaryphase}
{
	short=inflationary phase,
	long=\textit{The inflationary phase denotes the era of inflation in the evolution of the universe. see \glo{inflation}},
	extra=\\\TBDgl,
	class=glossary
}
\DeclareAcronym{cosmologicalconstant}
{
	short=cosmological constant,
	long=\textit{The cosmological constant is denoted by the symbol $\Lambda$ and is responsible for the accelerated expansion in the current epoch of the universe},
	extra=\\\TBDgl,
	class=glossary
}

\DeclareAcronym{criticaldensity}
{
	short=critical density,
	long=\textit{A universe with its density as big as the critical density will expand forever, but expansion stops at infinity. This universe is also called to be flat},
	extra=\\\TBDgl,
	class=glossary
}

\DeclareAcronym{densityparameter}
{
	short=density parameter,
	long=\textit{The density parameter...},
	extra=\\\TBDgl,
	class=glossary
}

\DeclareAcronym{lambdadominateduniverse}
{
	short=$\Lambda$~dominated universe,
	long=\textit{This term denotes our universe in the current epoch, where the \glo{darkenergy} dominates over all other constituents of the universe},
	extra=\\\TBDgl,
	class=glossary
}

\DeclareAcronym{radiationdominateduniverse}
{
	short=radiation dominated universe,
	long=\textit{In the early universe, its energy content was dominated by radiation (photons and relativistic neutrinos). The expansion of the universe diluted radiation more intensively than matter. Therefore, this era came to an end and the universe was then dominated by matter},
	extra=\\\TBDgl,
	class=glossary
}

\DeclareAcronym{matterdominateduniverse}
{
	short=matter dominated universe,
	long=\textit{After radiation has been diluted, matter (baryonic matter and Dark Matter) became the dominating energy in the universe. This era is lasting until today},
	extra=\\\TBDgl,
	class=glossary
}

\DeclareAcronym{darkenergy}
{
	short=Dark Energy,
	long=\textit{Dark Energy is a type of energy, which was discovered in the late 1990's and which is thought to be the reason for the accelerated expansion of the universe we see today},
	extra=\\\TBDgl,
	class=glossary
}

\DeclareAcronym{densitycontrast}
{
	short=density contrast,
	long=\textit{The density contrast is a quantity, that measures the relative deviation of the density at a point in space from the average density of the universe at a specific point in time. Mostly, it is denoted by the symbol $\delta$},
	extra=\\\TBDgl,
	class=glossary
}

\DeclareAcronym{flatnessproblem}
{
	short=flatness problem,
	long=\textit{The flatness problem is a problem in cosmology},
	extra=\\\TBDgl,
	class=glossary
}

\DeclareAcronym{horizonproblem}
{
	short=horizon problem,
	long=\textit{The horizon problem is a problem in cosmology},
	extra=\\\TBDgl,
	class=glossary
}

\DeclareAcronym{primordialpowerspectrum}
{
	short=primordial power spectrum,
	long=\textit{The primordial power spectrum is a description of the density perturbations at the end of \glo{inflation}},
	extra=\\\TBDgl,
	class=glossary
}

\DeclareAcronym{powerspectrum}
{
	short=power spectrum,
	long=\textit{The power spectrum is ...},
	extra=\\\TBDgl,
	class=glossary
}

\DeclareAcronym{harrisonzeldovichspectrum}
{
	short=Harrison Zel’dovich spectrum,
	long=\textit{The Harrison Zel’dovich, developed by \textsc{Harrison} and \textsc{Zel’dovich} is a theoretical description of a power spectrum with a power law},
	extra=\\\TBDgl,
	class=glossary
}

\DeclareAcronym{linearstructureformation}
{
	short=linear structure formation,
	long=\textit{Linear structure formation denotes a regime of structure formation, where the \glo{densitycontrast} is smaller than unity, such that all equations used to model the corresponding processes can be of linear order},
	extra=\\\TBDgl,
	class=glossary
}

\DeclareAcronym{cosmicmicrowavebackground}
{
	short=cosmic microwave background,
	long=\textit{The cosmic microwave background is the relic radiation of the Big Bang},
	extra=\\\TBDgl,
	class=glossary
}

\DeclareAcronym{virialradius}
{
	short=virial radius,
	long=\textit{The virial radius defines the size of an astrophysical object, that is in dynamical equilibrium. On the basis of theoretical calculations, the \glo{densitycontrast} is defined to be $200$},
	extra=\\\TBDgl,
	class=glossary
}

\DeclareAcronym{virialmass}
{
	short=virial mass,
	long=\textit{The virial mass defines the size of an astrophysical object, that is in dynamical equilibrium. On the basis of theoretical calculations, the \glo{densitycontrast} is defined to be $200$},
	extra=\\\TBDgl,
	class=glossary
}

\DeclareAcronym{hubbleflow}
{
	short=Hubble flow,
	long=\textit{The notion "Hubble flow" describes the situation, where matter "follows" the expansion of the universe},
	extra=\\\TBDgl,
	class=glossary
}

\DeclareAcronym{hubbleparameter}
{
	short=Hubble parameter,
	long=\textit{The notion "Hubble parameter" describes ...},
	extra=\\\TBDgl,
	class=glossary
}

\DeclareAcronym{hubblesphere}
{
	short=Hubble sphere,
	long=\textit{The notion "Hubble sphere" describes ...},
	extra=\\\TBDgl,
	class=glossary
}

\DeclareAcronym{hubbleconstant}
{
	short=Hubble constant,
	long=\textit{When \textsc{Hubble} discovered the \glo{hubbleslaw}, he found that the relation between the distance of a galaxy and the velocity the galaxy moves away from us, is defined by a constant. This constant was then called the Hubble constant. Meanwhile it is known that this quantity is not a constant, but varies with time. Therefore, it is now called generally \glo{hubbleparameter}.},
	extra=\\\TBDgl,
	class=glossary
}

\DeclareAcronym{hubbleslaw}
{
	short=Hubble's law,
	long=\textit{When \textsc{Hubble}, for the first time, observed the expansion of the universe, he found that galaxies move faster away from us the farther they are from us. This relation is called Hubble's law},
	extra=\\\TBDgl,
	class=glossary
}

\DeclareAcronym{conformaltime}
{
	short=conformal time,
	long=\textit{The notion "Conformal time" describes ...},
	extra=\\\TBDgl,
	class=glossary
}

\DeclareAcronym{MILNEuniverse}
{
	short=Milne universe,
	long=\textit{The Milne universe is a cosmological model developed by \textsc{Edward Arthur Milne} in 1933, where it was the first time to formulate the \glo{cosmoprinciple}},
	extra=\\\TBDgl,
	class=glossary
}

\DeclareAcronym{cosmoterm}
{
	short=cosmological term,
	long=\textit{\textsc{Einstein} added this term to his equations of \acidx{GR} to guarantee a \glo{staticuniverse}, which includes the \glo{cosmoconstant}},
	extra=\\\TBDgl,
	class=glossary
}

\DeclareAcronym{cosmoconstant}
{
	short=cosmological constant,
	long=\textit{\textsc{Einstein} added the \glo{cosmoterm} to his equations of \acidx{GR} to guarantee a \glo{staticuniverse}, he used this constant to generate an repulsive force to avoid the contraction of the universe caused by gravity. He used the symbol $\Lambda$ for this constant, which is nowadays used to denote DE},
	extra=\\\TBDgl,
	class=glossary
}

\DeclareAcronym{bigbang}
{
	short=Big Bang,
	long=\textit{The Big Bang was introduced by \textsc{Lema\^{i}tre} as he realized the expansion of the universe and concluded that the universe formed in a single event out of a "primeval atom"},
	extra=\\\TBDgl,
	class=glossary
}

\DeclareAcronym{hotbigbangmodel}
{
	short=hot Bing Bang model,
	long=\textit{This model states that the universe came into existence in an \glo{bigbang} event creating a extremely hot and dense environment. The subsequent evolution is determined by the laws of thermodynamics, where due to the expansion the universe cooled and diluted.},
	extra=\\\TBDgl,
	class=glossary
}

\DeclareAcronym{virialtheorem}
{
	short=virial theorem,
	long=\textit{The virial theorem describes the ratio of different types of energy in a system, being in dynamical equilibrium},
	extra=\\\TBDgl,
	class=glossary
}

\DeclareAcronym{darkmatter}
{
	short=Dark Matter,
	long=\textit{When \textsc{Fritz Zwicky} investigated the motions of the galaxies in the Coma cluster, he found that there is not enough matter to gravitationally bind the galaxies to the cluster. He concluded, that there is a type of matter that we cannot see. So he called it Dark Matter},
	extra=\\\TBDgl,
	class=glossary
}

\DeclareAcronym{cosmicweb}
{
	short=cosmic web,
	long=\textit{The large scale structure of the universe, consisting...},
	extra=\\\TBDgl,
	class=glossary
}

\DeclareAcronym{weakforce}
{
	short=weak force,
	long=\textit{The weak force...},
	extra=\\\TBDgl,
	class=glossary
}

\DeclareAcronym{axion}
{
	short=axions,
	long=\textit{The axions...},
	extra=\\\TBDgl,
	class=glossary
}

\DeclareAcronym{sterileneutrinos}
{
	short=sterile neutrinos,
	long=\textit{The sterile neutrinos...},
	extra=\\\TBDgl,
	class=glossary
}

\DeclareAcronym{selfinteractingDM}
{
	short=self-interacting DM,
	long=\textit{The self-interacting DM...},
	extra=\\\TBDgl,
	class=glossary
}

\DeclareAcronym{SMcosmo}
{
	short=Standard Model of Cosmology,
	long=\textit{The Standard Model of Cosmology is described in section \ref{sec:SMcosmo}},
	class=glossary
}

\DeclareAcronym{CMcosmo}
{
	short=Concordance Model of Cosmology,
	long=\textit{see \glo{SMcosmo}},
	class=glossary
}

\DeclareAcronym{debrogliewavelength}
{
	short=de Broglie wavelength,
	long=\textit{The de Broglie wavelength...},
	extra=\\\TBDgl,
	class=glossary
}

\DeclareAcronym{missingsatelitesproblem}
{
	short=missing-satellites problem,
	long=\textit{The missing-satellites problem...},
	extra=\\\TBDgl,
	class=glossary
}

\DeclareAcronym{cuspcoreproblem}
{
	short=cusp-core problem,
	long=\textit{The cusp-core problem...},
	extra=\\\TBDgl,
	class=glossary
}

\DeclareAcronym{toobigtofailproblem}
{
	short=too-big-to-fail problem,
	long=\textit{The too-big-to-fail problem...},
	extra=\\\TBDgl,
	class=glossary
}

\DeclareAcronym{missingmagneticmonopoles}
{
	short=missing magnetic monopoles,
	long=\textit{The missing magnetic monopoles problem...},
	extra=\\\TBDgl,
	class=glossary
}

\DeclareAcronym{strongcpproblem}
{
	short=strong CP problem,
	long=\textit{The strong CP problem...},
	extra=\\\TBDgl,
	class=glossary
}

\DeclareAcronym{thomasfermiregime}
{
	short=Thomas-Fermi regime,
	long=\textit{Thomas-Fermi regime...},
	extra=\\\TBDgl,
	class=glossary
}

\DeclareAcronym{kleingordonequation}
{
	short=Klein-Gordon equation,
	long=\textit{Klein-Gordon equation...},
	extra=\\\TBDgl,
	class=glossary
}

\DeclareAcronym{bulletcluster}
{
	short=Bullet Cluster,
	long=\textit{The Bullet Cluster...},
	extra=\\\TBDgl,
	class=glossary
}

\DeclareAcronym{QCDaxion}
{
	short=QCD axion,
	long=\textit{Axion particle proposed to solve strong CP problem in QCD},
	extra=\\\TBDgl,
	class=glossary
}
\makeatletter
\newcommand*{\hyperlinkcite}[1]{\hyper@link{cite}{cite.#1}}
\makeatother
\newcommand{\tb}[1]{\textbf{#1}}

\newcommand{\q}[1]{``{#1}''}
\newcommand{\qf}[1]{``{#1}''}
\newcommand{\qn}[1]{#1}

\begin{document}

\title{A proposal to improve the accuracy of cosmological observables}
\subtitle{and address the Hubble tension problem}
\date{\today}
\author{Horst Foidl\inst{1,2}\thanks{horst.foidl@outlook.com}
	\and
	Tanja Rindler-Daller\inst{1,2,3}\thanks{tanja.rindler-daller@univie.ac.at}
}

\institute{Institut f\"ur Astrophysik, Universit\"atssternwarte Wien,
	Fakult\"at f\"ur Geowissenschaften, Geographie und Astronomie,\\
	Universit\"at Wien, T\"urkenschanzstr.17, A-1180 Vienna, Austria
	\and
	Vienna International School of Earth and Space Sciences, Universit\"at Wien, Josef-Holaubek-Platz 2,
	A-1090 Vienna, Austria
	\and
	Wolfgang Pauli Institut, Oskar-Morgenstern-Platz 1, A-1090 Vienna, Austria
}

\date{Received December 14, 2023; accepted March 25, 2024}
%
\abstract
{Cosmological observational programs often compare their data not only with $\Lambda$ cold dark matter ($\Lambda$CDM), but also with extensions applying dynamical models of dark energy (DE), whose time-dependent equation of state (EoS) parameters $w$ differ from that of a cosmological constant. We found a degeneracy in the customary computational procedure for the expansion history of cosmological models once dynamical models of DE models were applied. This degeneracy, given the Planck-based Hubble constant $H_0$, provides an infinite number of cosmological models reproducing the Planck-measured cosmic microwave background (CMB) spectrum, including the one with a cosmological constant. Moreover, this degeneracy biases the comparison of $\Lambda$CDM with dynamical DE extensions.
}
{We present a complementary computational approach, that breaks this degeneracy in the computation of the expansion history of models with a dynamical DE component: the \qf{fixed early densities (EDs)} approach evolves cosmological models from the early Universe to the present, in contrast to the customary \qf{\qn{fixed $H_0$}} approach, which evolves cosmological models in reverse order. Although there are no equations to determine these EDs from first principles, we find they are accurately approximated by the $\Lambda$CDM model.
}
{We implemented a refined procedure, applying both approaches, in an amended version of the code CLASS, where we focused on representative dynamical DE models using the Chevallier-Polarski-Linder (CPL) parametrization, studying cases with monotonically increasing and decreasing $w$ over cosmic time.
}
{Our results reveal that a dynamical DE model with a decreasing $w$ of the form $w(a)=-0.9 + 0.1(1-a)$ could provide a resolution to the Hubble tension problem. Moreover, we find that combining the \qn{fixed EDs} approach and the customary \qn{fixed $H_0$} approach, while requesting to yield consistent results and being in agreement with observations across cosmic time, can serve as a kind of consistency check for cosmological models with a dynamical model of DE. Finally, we argue that implementing our proposed consistency check for cosmological models within current Markov chain Monte Carlo (MCMC) methods will increase the accuracy of inferred cosmological parameters significantly, in particular for extensions to $\Lambda$CDM. 
}
{Using our complementary computational scheme, we find characteristic signatures in the late expansion histories of cosmological models, allowing a phenomenological discrimination of DE candidates and a possible resolution to the Hubble tension, by ongoing and future observational programs.
}
\maketitle
\nolinenumbers
%
\section{Introduction}\label{sec:introaccuracy}

The current cosmological standard model of $\Lambda$ cold dark matter ($\Lambda$CDM) is based on many different observations of the Universe on large scales, especially the cosmic microwave background (CMB) (for example, readers can refer to the balloon observations of millimetric extragalactic radiation and geophysics (BOOMERanG) experiment by \citet{Bernardis2000} and \citet{MacTavish2006}, as well as observations with an increasing accuracy by the space missions Wilkinson Microwave Anisotropy Probe (WMAP) (\citet{Hinshaw2013}) and Planck (\citet{Collaboration2018})) -- , the large-scale structure of the cosmic web (for example, readers can refer to the Dark Energy Survey (DES);\citet{Abbott2022a}, the (extended) Baryon Oscillation Spectroscopic Survey (eBOSS and BOSS);\citet{Alam2021}, Large Synoptic Survey Telescope \citet{Collaboration2009}) -- , and measurements of the distance ladder using stellar standard candles (see e.g., \citet{Perlmutter1999,Schmidt1998,Riess1998,Perlmutter2003}).
However, the model comes with two major conundra, namely the still unknown nature of cold dark matter (CDM) and the nature of the cosmological constant $\Lambda$ which, according to measurements, each contribute roughly 25\% and 70\%, respectively, to the present-day energy density of the Universe. In attempts to better understand and constrain these components, an accurate measurement of the expansion rate, or Hubble parameter, over cosmic time is desired. Over recent years, discrepancies have been solidified between the measurement of this Hubble parameter at low redshifts $z$, the (local) Hubble constant $H_0$, and the value of the latter if high-$z$ data, notably encoded in the CMB, are extrapolated to the present. This issue goes under the header of the Hubble tension (problem), and seems to be in conflict with basic assumptions of the $\Lambda$CDM model. In particular, the question arises as to whether $\Lambda$ should be replaced by a dark energy (DE) component, whose energy density and equation of state (EoS) can vary with time, in order to fit cosmological observables at various scales. Since the nature of $\Lambda$ has defied a resolution as yet, despite theoretical efforts over decades, the Hubble tension adds even more urgency to investigate alternatives in the form of various DE models, which can be tested this way, using ever more precise measurements.

A recent review of the Hubble tension problem can be found in \citet{Hu2023}. There are two major approaches to test models of DE as possible answers to the Hubble tension problem. The first approach uses a parametrization ansatz of the EoS parameter of the DE component, such as the Chevallier-Polarski-Linder (CPL) parametrization (discussed shortly, see Eq.~\eqref{eq:EQCPLH0}), and it fits this model to observational data, determining the parameters of the dynamical model of DE. Alternatively, the EoS parameter of the DE component can be inferred directly from data. Recent studies (e.g., \citet{Dainotti2022},\citet{Dainotti2021,Dainotti2022a,Bargiacchi2023,Bargiacchi2023a,OColgain2021,Krishnan2021a}) indicate that the evolution of the expansion rate $H$ at low redshift is in conflict with the assumption of a cosmological constant preferring a dynamical model of DE. These results may indicate that the origin of the Hubble tension is likely to be in cosmology and not in local measurement issues (e.g., \citet{Krishnan2021a}), as observations with an increasing accuracy of $H_0$ in the local Universe seem to confirm the Hubble tension (e.g., \citet{Dainotti2023b,RidaKhalife2023}).

Many cosmological observation campaigns analyze their data, not only in light of testing $\Lambda$CDM, but also to examine the possible signature of DE candidates with a time-varying EoS. Examples include DES; readers can refer to the Year 3 (DES-Y3) results in \citet{Abbott2022a}, or the CMB measurements by \citet{Collaboration2018}. 
These and other campaigns often employ a dynamical model of DE, with a possible time-dependent EoS parameter $w = p/\rho$, where $p$ and $\rho$ constitute a pressure and energy density, in some averaged sense or fluid approach.
A useful representation of choice has been the so-called CPL parametrization by \citet{Chevallier2001} and \citet{Linder2003} for the EoS parameter,
%
\begin{equation} \label{eq:EQCPLH0}%
	w(a) = w_0 + \left(1 - a \right) w_a,
\end{equation}
which is defined for $a \in [0,1]$, where $a$ is the scale factor\footnote{In practice, cosmological codes, including the one we use, in the default configuration compute observables not earlier than at $a=10^{-14}$.} of the Friedmann-Lema\^{i}tre-Robertson-Walker (FLRW) model of the isotropic and homogenous background universe, while $w_0$ and $w_a$ are parameters that depend upon the specific DE model. However, in using this form in the comparisons with observations, they are merely free parameters. A cosmological constant $\Lambda$ is recovered as a special case if $w_0=-1$ and $w_a=0$, such that $w_{\Lambda}=-1$ throughout the cosmic evolution. 

The CPL parametrization offers a variation of $w$ such that DE models can be studied, which evolve linearly in $a$ from $w \simeq w_0+w_a$ in the early Universe to $w_0$ at the present at $a=1$. Now, campaigns such as \citet{Abbott2022a} and \citet{Collaboration2018} compute from their data the probability distribution of models in the $(w_0-w_a)$-plane (parameter subspace). They find that the probability of a cosmological constant is within the $1 \sigma$ area, \qf{fitted} by a multidimensional Monte Carlo integration to their observational data, using the Markov chain Monte Carlo (MCMC) method (see e.g.,  \citet{Carlin2008}). While a DE component cannot be definitely ruled out, a cosmological constant is in accordance with both the data in \citet{Abbott2022a} and \citet{Collaboration2018}. Furthermore, it is common practice to include data or constraints derived from previous literature in the analysis of the observed data (e.g., DES-Y3 includes results from CMB observations and low-z data from baryon acoustic oscillations (BAO)).

However, while statistically in agreement with $\Lambda$CDM, it appears that various cosmological observations may prefer different CPL models, in the sense that the (mean value of the) parameter $w_a$ is found to be either positive or negative, as we will discuss. It remains to be seen, whether these slight statistical preferences for one or the other eventually converge to a common result, or whether differences pertain pointing to physical explanations, such as a changing EoS of DE. The work in this paper relates to this question, as follows.

The process of inferring parameters of cosmological models from observations depends upon the way they are extracted from measurements. In case the cosmological observable can be measured directly, such as $H_0$ in the local universe (e.g., \citet{Riess2022}), there is no or only little dependence on a specific cosmological model. High-redshift data (e.g., CMB temperature spectra by \citet{Collaboration2018}) is analyzed by specifying a cosmological model (e.g., $\Lambda$CDM) and by \q{fitting} the model to the data.
Inherent to this procedure is the computation of a cosmological model ($\Lambda$CDM or any of its extensions) and the {extrapolation} of the model parameters to the present (see e.g., \citet{Alam2021}). These present-day values then enter the customary computation procedure of FLRW models. 
This customary computational procedure starts the backward-in-time integration of the equations determining the expansion history at the present, applying a present-day value of the Hubble parameter $H_0$, that is using a specific value of $H_0$. We call it the \qf{\qn{fixed $H_0$}} approach in the forthcoming. Thus, the initial densities in the early Universe are determined by the considered model and the provided value of $H_0$.
We present a complementary approach, that is based on the assumptions that the initial densities of the cosmic components were determined by processes in the very early Universe, evolving even prior to the point in time, where the computation of cosmological models customarily starts. We call this approach the \qf{fixed early densities (EDs)} approach, that is particularly relevant for cosmic components having a time-dependent EoS parameter, as we exemplify below. 
 
Indeed, there are no equations to determine the initial energy densities in the early Universe from first principles. But we use the fact that these early densities can be approximated to high accuracy by the $\Lambda$CDM concordance model, given that in the radiation-dominated and then CDM-dominated epochs, any DE component would be very much subdominant, similar to $\Lambda$.
More precisely, the two approaches can be characterized as follows. The customary \qn{fixed $H_0$} approach starts the computation at the present with the provided value of $H_0$ and integrates the equations backward-in-time to the early Universe. Thus, the densities in the early Universe (and thereby $H(z)$) vary with the choice of the model and its parameters. On the other hand, the \qn{fixed EDs} approach starts the computation in the early Universe with the initial densities determined by the $\Lambda$CDM concordance model. Thus, the densities at the present (and thereby $H_0$), that are determined by a forward-in-time integration of the equations, vary with the choice of the model and its parameters. The important difference between both approaches is that in the \qn{fixed EDs} approach $H_0$ can be checked (and is checked) against observations in the local Universe. Whereas in the \qn{fixed $H_0$} approach, $H(z)$ in the early Universe is not checked against observations of the early Universe. Moreover, as exemplified in Sect.~\ref{sec:compmodelsH0}, the subdominance of DE in the early Universe makes it not a promising tool to perform those checks. Thus, the \qn{fixed $H_0$} approach is not able to provide an answer to the Hubble tension problem, regardless of the model tested. Instead, by construction, it rules out a cosmological origin of the Hubble tension problem and relegates it to issues in the local measurements of $H_0$.

We find that the customary \qn{fixed $H_0$} approach is affected by a degeneracy in the parametrization of dynamical models of DE (e.g., the CPL parametrization). Owing to this degeneracy, for a given value of $H_0$, one can find a (infinite) number of combinations of the CPL parameters $w_0$ and $w_a$, such that the computed CMB temperature spectrum is in agreement with the measured CMB temperature spectrum and the expansion history agrees with that of $\Lambda$CDM, when using conformance parameters. So, care must be taken, when doing a MCMC analysis, not to inadvertently restrict the sampling of cosmological model parameters, in particular $H_0$, and thereby bias the results.

Therefore, we present a refined computation procedure for the expansion history applying the \qn{fixed EDs} approach, that breaks this degeneracy.
In the process we also see that the refined computational procedure lends itself to the introduction of a novel consistency check of cosmological models. This consistency check requires consistent results between {both approaches}, while demanding agreement with observational data across the entire cosmic expansion history. We implement a proof-of-concept for a proposed modification of the standard MCMC method. We find that only models including a DE component with a decreasing EoS parameter can provide an answer to the Hubble tension problem. In this paper, we do not investigate the physical motivation of such alternatives to $\Lambda$, but focus on the computational approach. 
We argue that an implementation of the mentioned consistency check within MCMC methods should be a way to reduce dramatically the available parameter space of models, leading to a significant increase in the determined accuracy of model parameters from such an extended MCMC, in turn. We describe the steps that are required to pursue this extension.

This paper is organized as follows. 
In Sect.~\ref{sec:basicEqsH0}, we recapitulate the basic equations involved in the evolution of the background universe. In Sect.~\ref{sec:timedependingEoS}, 
we illustrate and discuss in qualitative terms the consequences that arise in the computation of the expansion history for cosmological models having components with time-dependent EoS parameters. We introduce the \qn{fixed EDs} approach and compare it to the customary \qn{fixed $H_0$} approach, with respect to a constant as well as a time-dependent EoS parameter $w$.
In Sect.~\ref{sec:compmodelsH0}, we present quantitative results for representative cosmological models with a dynamical model of DE, upon implementing our refined computational procedure in an amended version of the Cosmic Linear Anisotropy Solving System (CLASS). This way, we can accurately calculate the expansion histories and spectra, and discuss our results in comparison with those of the standard approach.
We consider two different CPL-based model universes, one inspired by the mean values of the DES-Y3 results with a negative $w_a$ parameter, and an alternative CPL model with a positive $w_a$ parameter. 
In light of our findings, we include in  Sect.~\ref{sec:observations} a discussion of the consequences for the Hubble tension problem, and we suggest a multi-step enhancement to be implemented in the standard MCMC routines, in order to increase the determined accuracy of the parameters of cosmological models with a dynamical model of DE. A summary and conclusion is presented in Sec~\ref{sec:discussionconclusionsH0}.
The results of additional calculations and consistency checks can be found in two appendices.

\section{Basic equations}\label{sec:basicEqsH0}
We start with a recapitulation of the basic equations in the computation of the background evolution of FLRW models, whose observables depend upon cosmic time $t$ or scale factor $a$, but not on spatial coordinates. 
If we assume that all cosmic components of a given FLRW model can be described as \q{cosmic fluids,} we can assign in each case (\q{i}) an EoS that relates the respective energy densities $\rho_i$ and pressures $p_i$ via the corresponding EoS parameter $w_i$,
%
\begin{equation} \label{eq:EQeosnbgH0}%
	p_i(t) = w_i(t) \rho_i(t).
\end{equation}
In general, the EoS parameters $w_i$ can themselves change with cosmic time $t$ or scale factor $a$, respectively.
However, in the current concordance model of $\Lambda$CDM all
its cosmic components are characterized by a constant, that is time-independent\footnote{In certain cases, for instance in studies of phase transitions in the early Universe, when the reduction of relativistic degrees of freedom in the wake of the Universe's expansion is considered in detail, the assumption of constant EoS parameters is relaxed.} $w_i$. 
Applying the first law of thermodynamics to the adiabatic expansion of a comoving volume in the Universe yields the energy conservation equation
%
\begin{equation} \label{eq:EQeconsnbgH0}%
	\frac{\partial \rho_i}{\partial t} + 3 H \left(\rho_i + p_i\right) = 0.
\end{equation}
Applying Eq.~\eqref{eq:EQeosnbgH0} and assuming that all EoS parameters $w_i$ are constant, and that there is no transformation between different components,
one arrives at 
%
\begin{equation} \label{eq:EQdensevolgenH0}%
	\rho_{i}(a) = \Omega_{i,0} \rho_{\text{crit},0} \: a^{-3(1+w_i)}.
\end{equation}
The factor in front is the (normalized) critical density defined below; the subscript \q{$0$} customarily refers to present-day values.
Thus, under the above assumptions, the evolution of the various $\rho_i$ as a function of the scale factor $a$ follows a simple law (the dependence on $t$ is indirect via $a=a(t)$). In the flat $\Lambda$CDM model, we have constant EoS parameters as follows:
$w_r=1/3$, $w_b=0=w_{cdm}$ and $w_{\Lambda}=-1$, giving the energy densities for radiation $\syidx{SYMed}_{r} \propto a^{-4}$, for baryons $\syidx{SYMed}_{b} \propto a^{-3}$, for CDM $\syidx{SYMed}_{\text{CDM}} \propto a^{-3}$, and for the cosmological constant $\rho_{\Lambda} \propto a^{0} = const$.

These energy densities enter in the standard procedure of integrating the Friedmann equation 
%
\begin{equation} \label{eq:EQfriedmannLCDMnbgLCDM}%
	\syidx{SYMH}^{2}(t) = \frac{8 \pi \syidx{SYMG}}{3 \syidx{SYMc}^{2}} \left[ \syidx{SYMed}_{r}(t) + \syidx{SYMed}_{b}(t) + \syidx{SYMed}_{\text{CDM}}(t) + 		\syidx{SYMed}_{k}(t) +		\syidx{SYMed}_{\syidx{SYMDE}}(t) \right],
\end{equation}
%
from the early Universe to the present. We include the curvature term in the formula for completeness, but
$\rho_k$ is set to zero in $\Lambda$CDM, and we will also set $\rho_k=0$ in this paper. $H(t)$ is the so-called Hubble parameter (\qn{Hubble function} would be the more correct description. We will use the term \qn{expansion rate} for it interchangeably), defined as
%
\begin{equation} \label{eq:EQH}%
	H(t) = \frac{\dot{a}}{a},
\end{equation}
where the dot denotes the derivative with respect to cosmic time $t$. The \glo{criticaldensity} is given by
%
\begin{equation} \label{eq:EQcritdenstH0}%
	\rho_{\text{crit},t} = \frac{3 H^2(t) c^{2}}{8 \pi G}.
\end{equation}
The customary density parameters
%
\begin{equation} \label{eq:iCDMEQdensparH0}%
	\Omega_{t,i} = \frac{\rho_{i}(t)}{\rho_{\text{crit},t}},
\end{equation}
are nothing but the background energy densities, relative to the critical density, whose present-day value is given by
%
\begin{equation} \label{eq:EQcritdensH0}%
	\rho_{\text{crit},0} = \frac{3 H_{0}^{2} c^{2}}{8 \pi G},
\end{equation}
with the present-day Hubble constant $H(0):=H_0$.
\begin{figure}  [!b]
	\includegraphics[width=1.0\columnwidth]{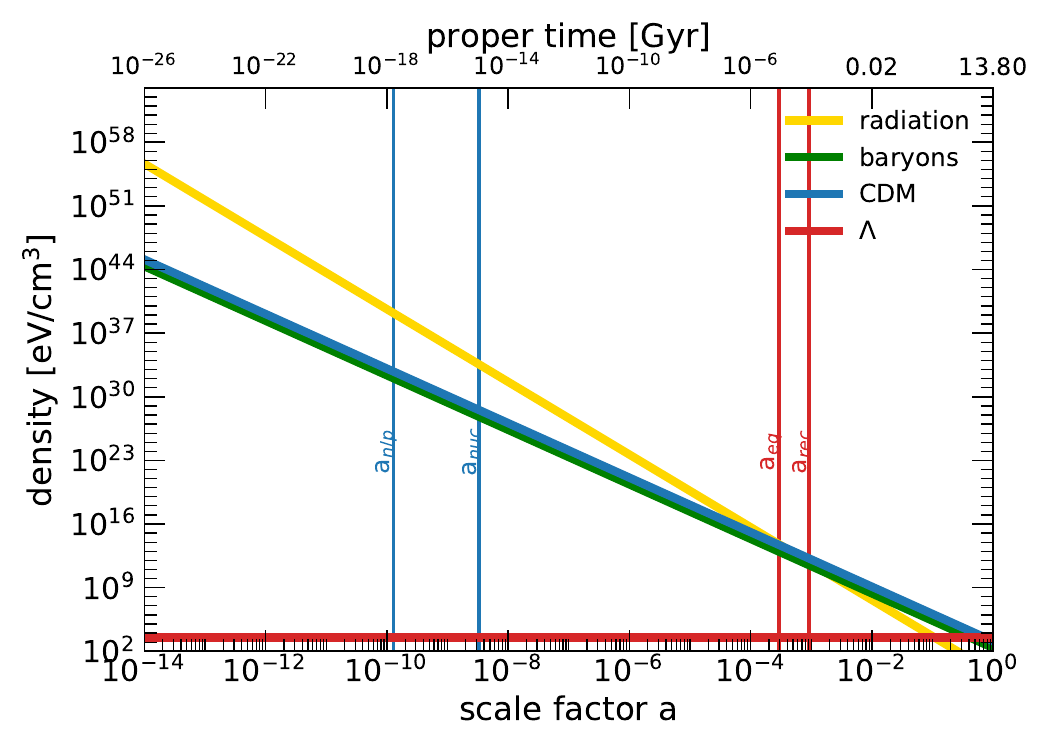}
	\includegraphics[width=1.0\columnwidth]{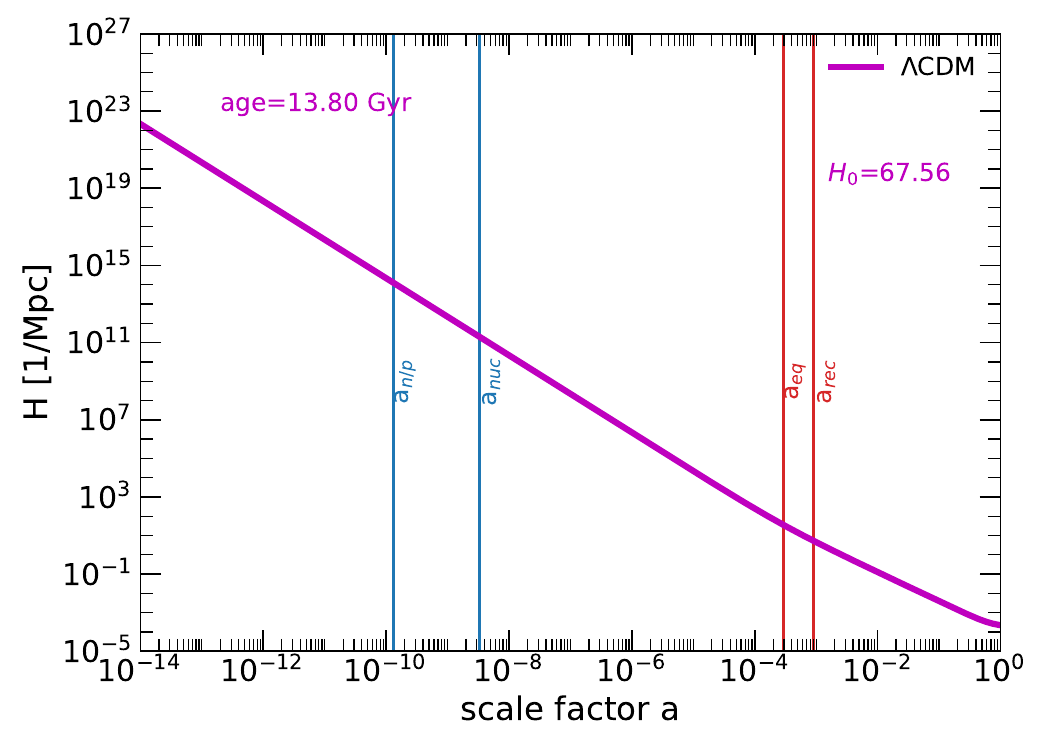}		
	\caption[Evolution of densities and expansion rate in the LCDM concordance model]
	{\tb{Evolution of energy densities and expansion rate in the $\Lambda$CDM concordance model.} Top panel: evolution of the energy densities, determined by Eq.~\eqref{eq:EQdensevolgenH0}. Bottom panel: evolution of the expansion rate $H$ in natural units, determined by the total of the energy densities given by Eq.~\eqref{eq:EQfriedmannLCDMnbgLCDM}. The vertical lines in blue bracket the epoch of big bang nucleosynthesis, between neutron-proton freeze-out at $a_{\text{n/p}} \sim 1.3 \cdot 10 ^{-10}$ and nuclei production $a_\text{nuc} \sim 3.3 \cdot 10^{-9}$, while the vertical lines in red (at larger scale factor) indicate the time of matter-radiation equality $a_{\text{eq}}$, followed by recombination $a_{\text{rec}}$, in terms of the scale factor $a$.
	}
	\label{fig:LCDM}
\end{figure}
The Friedmann equation in \eqref{eq:EQfriedmannLCDMnbgLCDM} can be alternatively written as an algebraic closure condition, and at the present time it reads
%
\begin{equation} \label{eq:EQclosureLCDMfull}%
	1 = \Omega_{r,0} + \Omega_{b,0} + \Omega_{\text{CDM},0}  + \Omega_{k,0} + \Omega_{\Lambda,0}
\end{equation}
(including the curvature term in the formula for completeness).
So, we can interpret Eq.~\eqref{eq:EQdensevolgenH0} as the result of a backward-in-time integration (because $\rho_{\text{crit},0}$ depends on $H_0$) of the energy conservation equation \eqref{eq:EQeconsnbgH0} for the individual components of a given model, as a function of scale factor $a$. This backward-in-time integration is performed, because we have no equation which determines the initial conditions (ICs) in the early Universe from first principles. 
However, below we will exemplify that CMB measurements, which provide us with information of the conditions at place at a redshift of $z \sim 1090$, along with extrapolating this high-precision data to present-day values using the $\Lambda$CDM concordance parameters, can approximate the initial densities to a high accuracy.
So, Eq.~\eqref{eq:EQdensevolgenH0} determines the energy densities in the early Universe, while customarily performing a forward-in-time integration of $H$ when solving the Friedmann equation ~\eqref{eq:EQfriedmannLCDMnbgLCDM}. 

The observation of the CMB by the Planck mission \cite{Collaboration2018} confirm previous measurements and more accurately determine the set of parameters of the $\Lambda$CDM concordance model. As an illustration,
Figure \ref{fig:LCDM} depicts the evolution of the energy densities of the cosmic components, using the present-day values
$\Omega_{r,0} = 9.16714 \times$10$^{-5}$,
$\Omega_{b,0} = 0.0482754$,
$\Omega_{\text{CDM},0} = 0.263771$,
$\Omega_{\Lambda,0} = 0.687762$,
and setting $\Omega_{k,0} = 0$.
The present-day Hubble constant $H_0 = 67.56$~km/s/Mpc determines the present-day critical density $\rho_{\text{crit},0}$, that is used in Eq.~\eqref{eq:EQdensevolgenH0}, in order to evolve the densities backward-in-time to the early Universe. The integration of Eq.~\eqref{eq:EQfriedmannLCDMnbgLCDM} determines the expansion history and yields an age of $13.8$~Gyr for the Universe in the $\Lambda$CDM model.

\section{Cosmic components with a time-dependent EoS parameter: The case for dark energy}\label{sec:timedependingEoS}
While the current standard model $\Lambda$CDM is capable of modeling the expansion history of the Universe to good accuracy, certain discrepancies to data have been reported, such as the current value of the Hubble parameter, discussed below. More importantly, the very nature of $\Lambda$ remains an open question, and the community has considered extensions to $\Lambda$CDM, which include the possibility of a DE component, in lieu of $\Lambda$, which comprises a time-dependent EoS parameter, the details of which depend on the underlying DE model.
As a result, cosmological models are often fit to observational data, which include a DE component having a variable EoS parameter, while the concordance $\Lambda$CDM model with its constant $w_{\Lambda}=-1$ is included as well. Prominent examples of observational campaigns which compare their data, among others, to individual DE candidates and a cosmological constant, include the CMB observations by \citet{Collaboration2018} and big galaxy surveys such as the DES-Y3 results (\citet{Abbott2022a}). Both of them use the popular and useful CPL parametrization of Equ.~\eqref{eq:EQCPLH0}. Similarly to the general solution known for a constant EoS parameter
%
\begin{equation} \label{eq:EQrhoLCDM}%
	\rho \propto a^{-3 (1 + w_{const})},
\end{equation}
the CPL parametrization of DE provides an analytical solution to the integration of the energy conservation equation \eqref{eq:EQeconsnbgH0}, 
which reads as
%
\begin{equation} \label{eq:EQrhoCPL}%
	\rho \propto a^{-3 (1+ w_0 + w_a)} e^{-3 w_a (1-a)}.
\end{equation}
This offers a very straightforward way to incorporate CPL-based models of DE into existing cosmological codes. 
\begin{figure}  [!htb]
	\fbox{\includegraphics[width=1.0\columnwidth]{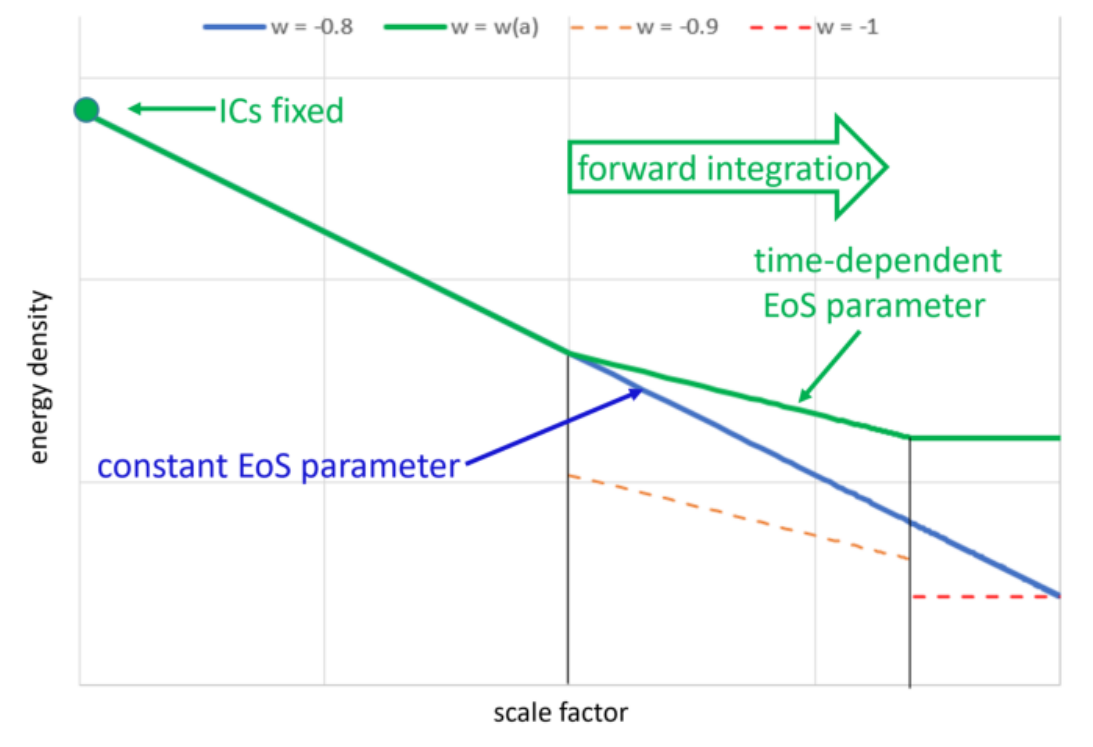}}
	\caption[Comparing the evolution of densities for constant and nonconstant EoS]
	{\tb{Illustrative comparison between the evolution of energy density for a component with a constant or nonconstant EoS (arbitrary scaling).} The \qn{blue} component has a constant EoS parameter $w=-0.8$. The \qn{green} component has a nonconstant EoS parameter $w$, which starts out with $w=-0.8$ (equal to the blue component) and evolves in two steps (indicated by the black vertical solid lines) to $w=-0.9$ (the slope is indicated by the dashed orange line) and finally to $w=-1$ (the slope is indicated by the dashed red line). The final density at the present time is higher for the \qn{green} component. }
	\label{fig:intForwardiCDM}
\end{figure}

Before we present the details, we want to explain the consequences that arise from \qn{fixed EDs} vs \qn{fixed $H_0$} in a more illustrative way in this section first.   
In order to discuss the evolution of the energy density and its contributing effect onto the expansion history of an exemplary component with a time-dependent EoS parameter, we compare it to the evolution of a component with a constant EoS parameter, sketched in Fig.~\ref{fig:intForwardiCDM}.

The initial density of the exemplary cosmic component is determined by processes in the very early Universe. Assuming that the EoS parameter remains constant, its evolution is shown by the blue solid line.
Alternatively, we consider a case where an initially constant EoS parameter evolves to a different EoS parameter for this component (i.e., a time-dependent EoS), depicted by the green solid line. Obviously, for both cases the evolution is initiated at identical ICs. For the alternative with a time-dependent EoS parameter, we choose an EoS parameter that drops at two points in time (respective scale factors), first from $w=-0.8$ to $w=-0.9$ and finally to $w=-1$, the EoS parameter of a cosmological constant. With each drop the slope of the energy density evolution flattens (determined by the {forward-integration} of the energy conservation equation \eqref{eq:EQeconsnbgH0}). As a result, the final energy density at the present time is higher, compared to the energy density resulting from the assumption of a constant EoS parameter. 
Moreover, we recognize that a lower (higher) EoS parameter will decrease (increase) the deceleration of the expansion rate $H$, resulting in an increase (decrease) of $H_0$, compared to a component described by a constant EoS parameter with identical ICs in the early Universe. We note that this simple illustration already indicates that a DE component with a decreasing EoS parameter could possibly provide an answer to the Hubble tension problem (in \citet{Wagner2022}, a different ansatz is presented, the result of which is comparable to our approach of forward-in-time integration, for it also provides a $H_0$ closer to local measurements. They synchronize the cosmic time of early- and late-universe fits of cosmological models, indicating an increased expansion rate in the local Universe).

\begin{figure}  [!htb]
	\fbox{\includegraphics[width=1.0\columnwidth]{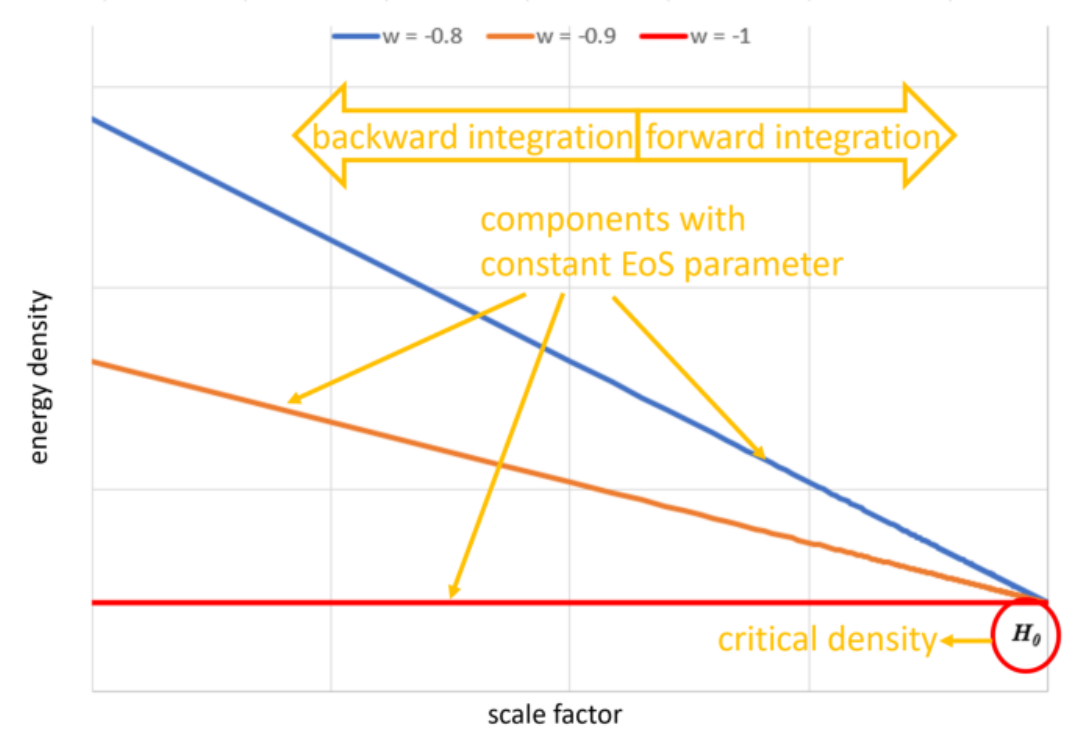}}
	\caption[The evolution of densities with constant EoS]
	{\tb{Illustrative evolution of energy densities of components, each with constant EoS (arbitrary scaling).} $w=-0.8$ (blue), $w=-0.9$ (orange), and $w=-1$ (red). 
	}
	\label{fig:intLCDM}
\end{figure}

In contrast, we sketch in Fig.~\ref{fig:intLCDM} examples of components which each have constant EoS parameter; where the case $w=-1$ describes the $\Lambda$-component in the $\Lambda$CDM model.
No matter what the value for $w$ is (blue, orange and red lines), the fact that $w=const$ implies that the order of integration of the energy conservation equation \eqref{eq:EQeconsnbgH0} does not impact the final results, as the gradients of the densities never change. The \qn{fixed EDs} approach vs \qn{fixed $H_0$} approach reveals no differences (without the need for a suitable choice of ICs for the backward integration). We can choose forward-in-time or backward-in-time integration, whatever may be suitable to our problem. 
In particular, we have this freedom in $\Lambda$CDM, where present-day density parameters $\Omega_{i,0}$ and expansion rate $H_0$ can be used in a backward-in-time integration of the energy conservation equation \eqref{eq:EQeconsnbgH0} for the energy densities of the individual components via the simplified Eq.~\eqref{eq:EQdensevolgenH0}. In effect, this procedure then determines the ICs in the early Universe, that represent the starting point for the evolution of the background densities, when the forward-in-time integration of the Friedmann equation in \eqref{eq:EQfriedmannLCDMnbgLCDM} is performed. This whole procedure is viable and leads to identical results for both approaches, {as long as} all components of a given cosmological model have a constant EoS parameter.  

However, once we introduce a component with nonconstant EoS parameter, such as a DE component in lieu of $\Lambda$, we need to be careful. 
Assuming initial densities to be determined in the early Universe and the fact that the Universe evolved from the early Universe to present, a component with a nonconstant EoS parameter, sketched by the green solid line in Fig.~\ref{fig:intForwardiCDM}, requires a forward-integration of Eq.~\eqref{eq:EQeconsnbgH0}. So  we cannot use the customary backward-integration, without a suitable choice of $H_0$, 
see also Figs.~\ref{fig:intICwrong} and \ref{fig:intICcorrect} below. Instead, integrating the energy conservation equation \eqref{eq:EQeconsnbgH0}, starting from the initial densities in the early Universe up to the present, obtains the correct evolution of the densities, which enter the Friedmann equation \eqref{eq:EQfriedmannLCDMnbgLCDM}. Related to this careful computation of the backward- and forward-in-time evolution is the choice of the correct starting point(s) of the integrations, depending on the corresponding cosmic time related to the measurement of $H_0$, as we will elaborate shortly. 

Although the Friedmann equation determines the expansion history as the total of the evolution of the energy densities over time (or scale factor) and eventually their initial values, it does of course not explain these initial values. 
The latter were determined by processes in the (very) early Universe and they impact the subsequent evolution of the background universe ever since, including the values at the present time. Although we cannot compute the EDs from first principles, in practice we can determine (or approximate) them by using observations which provide us with physical information at desirably high redshifts. These are foremost the observations of the CMB\footnote{In the future, the possible observation of primordial gravitational waves could provide us with data at even higher $z$.}, for example, by \citet{Collaboration2018}, providing us with information on the density parameters for the constituents of the Universe at the time when the CMB has formed, that is the time of recombination at $z \sim 1090$. Using these data extrapolated to the present, along with the choice of picking $\Lambda$CDM as our concordance model, we end up with a present-day value of the expansion rate of $H_0 \simeq 67$ km/s/Mpc. We may call it the \qf{concordance value for $H_0$}.  
By applying these parameters of the $\Lambda$CDM concordance model, we can compute the corresponding initial densities, these are the \qf{concordance EDs,} now by the customary backward-in-time integration, that is fixing $H_0$, as all constituents of the model have a constant EoS parameter. In the subsequent section, we will justify that these \qn{concordance EDs} are indeed a very accurate way to determine the energy densities at place in the early Universe, making the \qn{fixed EDs} approach feasible.

Furthermore, the evaluation of $H_0$ in indirect observations (see e.g., \citet{DiValentino2021}) requires two steps. First, measurements of observables refer to a specific cosmic time, for example, the time when the CMB was emitted. Then, the implication of the measurement is extrapolated to redshift zero (see e.g., \citet{Alam2021}), which is performed by assuming a specific cosmological model, for example, $\Lambda$CDM for the Planck-based $H_0$. 
Extrapolated quantities, such as the value of $H_0$, {are bound to} this specific cosmological model, used in the extrapolation. In turn, this \qn{concordance value for $H_0$} implies respective \qn{concordance EDs,} the values for the initial energy densities, {and vice versa}. 

However, it is important to stress that this extrapolation step is not required for {direct} measurements of $H_0$, for example, in the local Universe. Thus, these measurements are not sensitive to the choice of a specific cosmological model.

\begin{figure}  [!htbp]
	\fbox{\includegraphics[width=1.0\columnwidth]{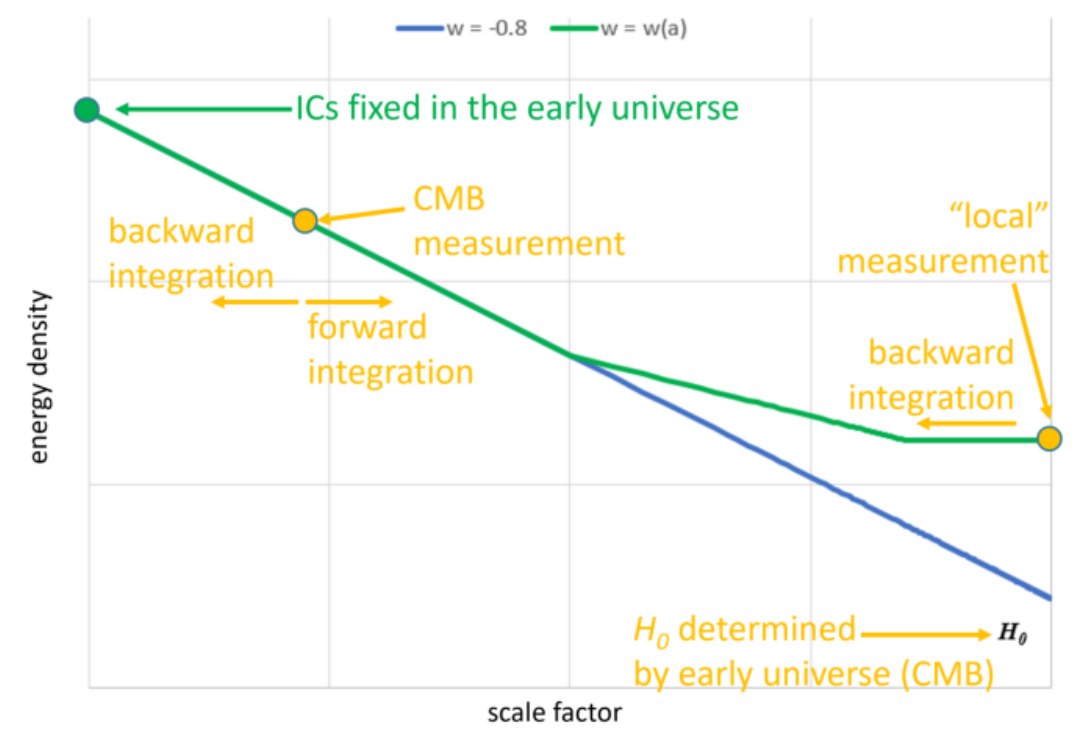}}
	\caption[Integration in general]
	{\tb{Illustration of integration orders in determining energy densities of cosmic components.} The actual ICs are fixed by processes in the very early Universe, but we have only access to observables at specific later times (respective redshifts or scale factors). The values of the latter determine the value of $H$, and by extrapolation to the present the value of $H_0$. If $H_0$ determined by the CMB is used, it is this starting point (referred to as \q{CMB measurement}) which must be used in the integration of Eq.~\eqref{eq:EQeconsnbgH0}: values for the energy densities before this point in time have to be calculated using backward integration, while values for the energy densities after this point in time have to be calculated using forward integration. The \qn{green} component has a time-dependent $w$, while the \qn{blue} component has a constant $w=-0.8$.
	}
	\label{fig:intForwardBackward}
\end{figure}

Now, we set up a general rule, illustrated in Fig.~\ref{fig:intForwardBackward}, how to compute the expansion history for components with time-dependent EoS parameter via integration of the energy conservation equation \eqref{eq:EQeconsnbgH0}, in order to solve the Friedmann equation \eqref{eq:EQfriedmannLCDMnbgLCDM}, such that the calculations are in agreement with the observed CMB spectra and lead to consistent results for the expansion history.
When we apply the CMB-based determination of $H_0$, it is the time of recombination $a_\text{rec}$ (referred to as \q{CMB measurement} in the illustration), which shall be the starting point of the integration of Eq.~\eqref{eq:EQeconsnbgH0}: backward integration to get the energy densities prior to $a_\text{rec}$, all the way back to the very early Universe, and forward integration to get the energy densities at all later times past $a_\text{rec}$. 
On the other hand, if we choose a value of $H_0$ determined by observations of the local Universe, when we may assume $z \approx 0$, then the starting point for the integration is the present, and the integration of the energy densities is performed backward-in-time.

To elaborate more on these rules, which we will adopt in the next section, we present two additional illustrations.
The first one depicted in Fig.~\ref{fig:intICwrong} displays a situation where the recombination time should be our reference point. Hence, we use the CMB-based value of $H_0$. 
\begin{figure}  [!b]
	\fbox{\includegraphics[width=1.0\columnwidth]{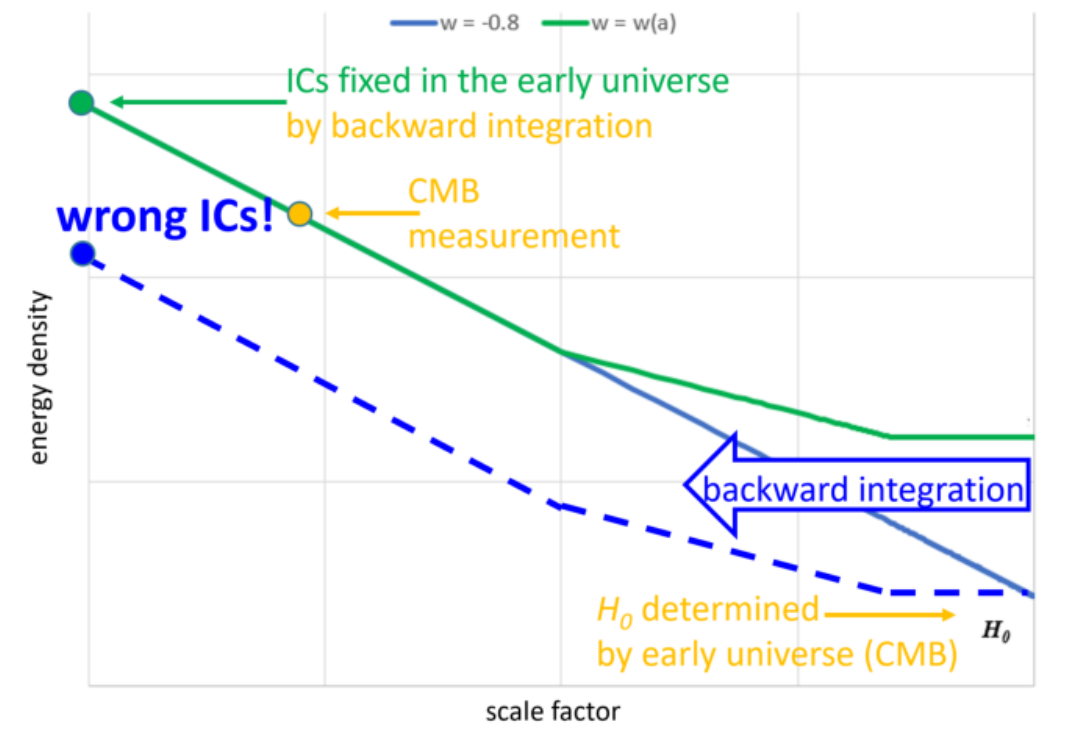}}
	\caption[Integration with wrong ICs]
	{\tb{Integration with wrong ICs.} This illustration complements the one depicted in Fig.~\ref{fig:intForwardBackward}. The customary backward integration (blue dashed) of the \qn{green} component with time-dependent EoS parameter leads to incorrect ICs, when we start from the value of $H_0$ determined from observations of the early Universe.
	}
	\label{fig:intICwrong}
\end{figure}
However, when we, as is customarily, fix $H_0$ and apply a standard backward integration, using the present-day value of $H_0$ extrapolated from CMB measurements, we \qf{arrive} at \q{wrong} ICs, and not the ones which gave actually rise to the CMB that we measure.   
As a result, even a subsequent correctly performed forward integration of Eq.~\eqref{eq:EQeconsnbgH0} will result in an \q{incorrect} evolution of the expansion history, not \q{fitting} to the CMB measurement\footnote{In fact, the argument of \q{fitting} to the CMB measurements depends on the entire set of the model parameters and is discussed in detail in the next section.}.
The second illustration is shown in Fig.~\ref{fig:intICcorrect}, and depicts a situation, where the present time is the reference point, using a value of $H_0$ determined by observations of the local Universe. (These measurement values of $H_0$ typically tend to be higher than those determined by the observation of the CMB, known as the Hubble tension problem, which we will discuss in the next sections).
\begin{figure}  [!htb]
	\fbox{\includegraphics[width=1.0\columnwidth]{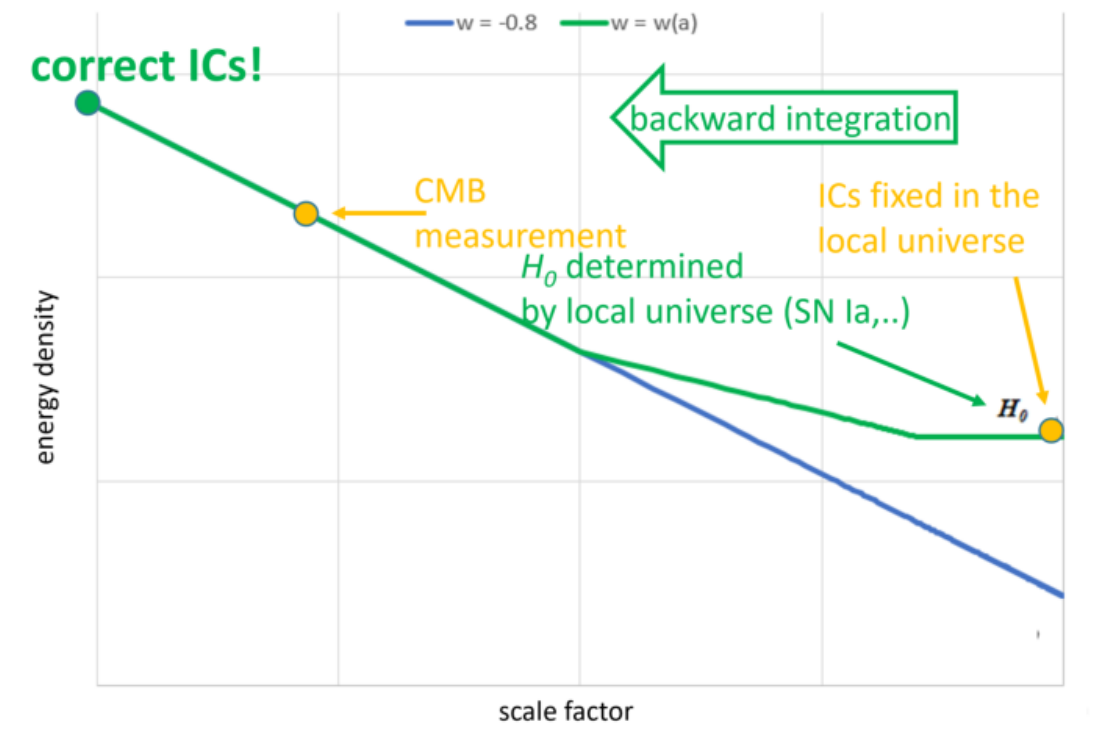}}
	\caption[Integration with correct ICs]
	{\tb{Integration with correct ICs.} This illustration complements the one depicted in Fig.~\ref{fig:intForwardBackward}. The customary backward integration of the \qn{green} component with time-dependent EoS parameter \q{ends up} at the correct ICs, when we start from the value of $H_0$ determined from observations of the local Universe.
	}
	\label{fig:intICcorrect}
\end{figure}
The reference point for the integration is now the present and we again fix $H_0$. According to the rules, we now apply an integration backward-in-time ending up at the \q{correct} ICs, which determined the observed CMB temperature spectrum. A subsequent forward integration will then result in a correct evolution of the expansion history, as expected. We note that this result is in accordance with Fig.~\ref{fig:intForwardiCDM}, where we fixed the early densities. A DE component with time-dependent EoS parameter can yield consistent results in both the \qn{fixed EDs} and the \qn{fixed $H_0$} computations of the expansion history, when the above mentioned rules, illustrated in Fig.~\ref{fig:intForwardBackward}, are met.

In the next section, we go beyond this qualitative picture and present detailed results for representative model universes having a DE component with a time-dependent EoS parameter, also known as a dynamical model of DE. The refinements to the computation procedure to include the \qn{fixed EDs} approach, which we apply in the next section, can be summarized as follows.

If the \qf{early} CMB-based extrapolation to $H_0$ is used:
	\noindent computation of the EDs with the $\Lambda$CDM concordance parameters, because the reference point for the integration is the early Universe, and $\Lambda$CDM binds the \qn{concordance EDs} to the \qn{concordance value for $H_0$}. This is followed by the forward integration, starting at the early universe with the computed EDs,
	using the parameters of the investigated model universe.

If the \qf{local} value of $H_0$ is used:
	\noindent backward integration, starting at $z=0$ with the given value of $H_0$,
	using the parameters of the investigated model universe.

\section{Computation of exemplary cosmological models with dynamical dark energy}\label{sec:compmodelsH0}
%
In this section, we make more precise our procedure, where we suggest to add a novel parametrization to cosmological codes, in order to perform the computations of the expansion histories of cosmological models, including components with time-dependent EoS. The novel parameter flags the provided value for $H_0$ as inferred from measurements in the early Universe or in the local Universe, respectively, where \qn{early} values of $H_0$, according to our approach, demand \qn{fixed EDs} and forward integration of Eq.~\eqref{eq:EQeconsnbgH0}, whereas \qn{local} values demand \qn{fixed $H_0$} and backward integration.

As already mentioned, \qn{early} values of $H_0$ require a two-step procedure. In the first step, we compute the EDs, that determined the observations of the configured value of $H_0$ (e.g., the Planck-based value). In doing so, we must apply the same model parameters (e.g., the $\Lambda$CDM concordance values) that were used to extrapolate the observational data to the present-day $H_0$, in order to \qn{arrive} at the identical early densities, that evolved to those quantities that gave rise to the measurement values. The \qn{concordance EDs} are bound to the \qn{concordance value for $H_0$} via the $\Lambda$CDM concordance model, as we exemplified in the preceding section. The computation of the EDs corresponds to the illustration of Fig.~\ref{fig:intForwardBackward}, following the blue path from the present to the ICs in the early Universe. The subsequent integration follows the green path, where we have to forward-integrate the energy densities via the energy conservation equation \eqref{eq:EQeconsnbgH0} for each considered component, starting at the computed EDs, instead of using Eq.~\eqref{eq:EQdensevolgenH0}, by applying the parameters of the model {under consideration}, for example, a model with a dynamical form of DE.
Obviously, local measurement values of $H_0$ need no extrapolation of observational data to the present-day value of said $H_0$. Hence, there are no special requirements on the computation of the EDs, but we have to perform a backward integration of Eq.~\eqref{eq:EQeconsnbgH0}, applying the parameters of the model {under consideration}. This corresponds to Fig.~\ref{fig:intICcorrect}, where the computation backward-in-time and the subsequent forward-in-time integration follow the green path.
In either case, the evolution of the background universe can be rightfully considered as an initial value problem (IVP), determined by the initial energy densities in the early Universe, which give rise to the configured value of $H_0$ as determined by  high-$z$ or low-$z$ observations, respectively. 

While we have emphasized the issue of consistent backward and forward integrations with respect to the adopted measurement value of $H_0$ in models with time-dependent EoS, we must also stress another point. As a matter of fact, the data analysis of cosmological observations also often employ variations in the density parameters $\Omega_{i,0}$, for example, variations in the matter content $\Omega_{m,0}$, which are different from the values of the concordance $\Lambda$CDM model. This sampling of parameters is also related to the adopted routines, typically MCMC methods.  
Therefore, in highlighting our proposed procedure, we actually need to consider two different sources of variations in the parameters of a considered model universe: 1) at least one of the components has a time-dependent EoS parameter, as such a component is often used in the data analysis of cosmological observations, typically a DE component, and 2) the density parameters $\Omega_{i,0}$ of components may be different from that model which was used in the extrapolation of $H_0$ (i.e., different from the density parameters of the concordance model), as typically encountered in the MCMC data analysis of cosmological observations. 

In our paper, we explore the impact of both sources of variations in the computations of model universes.
To this end, we use the Boltzmann code CLASS\footnote{The code CLASS is publicly available at \href{https://lesgourg.github.io/class_public/class.html}{https://lesgourg.github.io/class\_public/class.html}}, which was designed to provide a flexible coding environment for implementing cosmological models, to calculate their background evolution and linear structure formation. The modular concept of CLASS and the coding conventions make it possible to enhance the existing code, without the risk of compromising existing functionality. Furthermore, it offers the  configuration of a CPL-based DE component.
The underlying code concepts can be found in \citet{Lesgourgues2011}. 
This code uses the \q{Planck 2018} cosmological parameters from \citet{Collaboration2018} as the default parameter set. Additionally, the configuration provides different sets of precision configuration files, to reflect varying requirements on precision needed in the results and available computation time. The precision configuration offering the highest accuracy is proofed to be in conformance with the Planck results within a $0.01\%$ level.  

We developed an amended version of CLASS, that includes our suggested enhancements in the procedure of computing the EDs and the integration of the energy conservation equation. The novel parametrization of an early-based (CMB) versus locally determined value of $H_0$ makes sure that the corresponding switch is carried out in the computation of the EDs of the DE component and the subsequent integration of Eq.~\eqref{eq:EQeconsnbgH0}. If the parameter is flagged as \qn{early,} the concordance $\Lambda$CDM parameters, that is the Planck-based value of $H_0$ and the $\Lambda$CDM concordance parameters from \citet{Collaboration2018} are used  (mentioned in the description of Fig.~\ref{fig:LCDM}). If the parameter is set to \qn{local,} then the calculation proceeds with the assumption that a locally determined value for $H_0$ is being used.
We used the CPL-based DE model in CLASS for the computation of the exemplary models presented in this section. 

We first mention our results, concerning the impact of a variation of the (non-radiation) density parameters on the computation of model universes, in light of \qn{early} and \qn{local} values of the adopted value of $H_0$. The bottom line is that we find no significant impact by our refined computational procedure, even for exotic models with very different values of $\Omega_{i,0}$, compared to $\Lambda$CDM. Our computational scheme gives results not different from those applying the standard procedure. A marked difference only occurs, if the density parameter of radiation is changed.
We relegate these results and their explanation into Appendix \ref{app:wConst}. This, in fact, justifies the use of the \qn{concordance EDs} as the early densities in our computations. This can also be seen in the figures we present in this section: regardless of huge variations in CPL model parameters, there are no deviations from the expansion history of $\Lambda$CDM at the time of recombination.

Now, our investigation of models having components with nonconstant EoS parameter, corresponding to the illustrations of the previous section, does reveal differences between our scheme versus the standard customary approach, when it comes to the late stages of the expansion history. We cover the relevant range of a possible MCMC sampling of $H_0$ by performing two computations with a CMB-based value of $H_0$ and a locally based value of $H_0$, respectively. 
We exemplify these differences by computing two different model universes with dynamical DE, applying the CPL parametrization \eqref{eq:EQCPLH0}. In the previous section, we recognized that an increasing or decreasing evolution of the EoS parameter with cosmic time determines a decrease or an increase of $H_0$ compared to the evolution given by a constant $w$. For this reason we use two models. One model with an increasing EoS parameter and the second one with a decreasing EoS parameter.

The first model is inspired by the DES-Y3 results, having an increasing EoS parameter, that is $w_a < 0$. In order to examine our suggested procedure, we are interested in probing the scheme for models which differ from the concordance $\Lambda$CDM model, if ever so slightly. 
We considered first a comparison of the results of DES-Y3 with $\Lambda$CDM, as follows. We stress that DES-Y3 is in accordance with $\Lambda$CDM to within 1-$\sigma$. But they investigate also the CPL model of DE (see Eq.~\eqref{eq:EQCPLH0}) and find values of $w_0=-0.95^{+0.08}_{-0.08}$ and $w_a=-0.4^{+0.4}_{-0.3}$. We used the mean values of these parameters to describe a cosmology with DE and call it the \qn{DES-Y3 CPL model} for simplicity. For these CPL mean values, the EoS of DE evolves from an initial value of $w_{DE}= -1.35$ to a final value of $w_{DE}=-0.95$, that is a model with increasing EoS parameter.

For the second model we used a decreasing EoS parameter, that is $w_a > 0$.
While the DES-Y3 and Planck results prefer a nonpositive value of the CPL parameter $w_a$ clearly below zero (for DES-Y3 see \citet{Abbott2022a}, their Figure 4, the light blue contours and the red contours, where they include data from \citet{Collaboration2018} Figure 30; only the combination with low-z data, seen in their Figure 5 green contours, shifts the result to the \q{less negative} value 
$w_a = -0.4$), other observations in the more local Universe show a preference for a positive value of $w_a$, that suggests a decreasing $w$ for DE. For example, BAOs measured in the SDSS Data Release 9 spectroscopic galaxy survey by \citet{Anderson2012}, in addition with measurements of luminosity distances from a large sample of SNe Ia by \citet{Guy2010}, \citet{Conley2011}, and \citet{Sullivan2011} yield $w_0 = -1.17^{+0.13}_{-0.12}$ and $w_a = +0.35^{+0.50}_{-0.49}$ (\citet{Hinshaw2013}). Such as with DES-Y3, the results are statistically in agreement with $\Lambda$CDM; in the case here a cosmological constant is within the $95$\% confidence region in the $w_0$-$w_a$ plane. 
So, given the possibility of discrepant preferences for the CPL parameters of DE from different cosmological observations, we also present the analysis of our computational scheme for models with positive value for $w_a$, that is a decreasing EoS parameter.
As a case in point, we used the following CPL parameters, $w_0=-0.9$ and $w_a=+0.1$, for which the EoS parameter evolves from $w_{DE} = -0.8$ to a present-day value of $w_{DE}=-0.9$. This choice is substantiated by inspection of Figure 5 of the DES-Y3 results in \citet{Abbott2022a}, where the green contours in that Figure show low-$z$ data alone. For these data, the probability of a cosmological constant is right at the edge of the $1\sigma$-area. Moreover, the center of this area is such that $w_0$ lies above $-1$ and $w_a$ is clearly larger than zero, thus also preferring DE over a cosmological constant.

\begin{figure*}  [!htbp]
	\includegraphics[width=1.0\columnwidth]{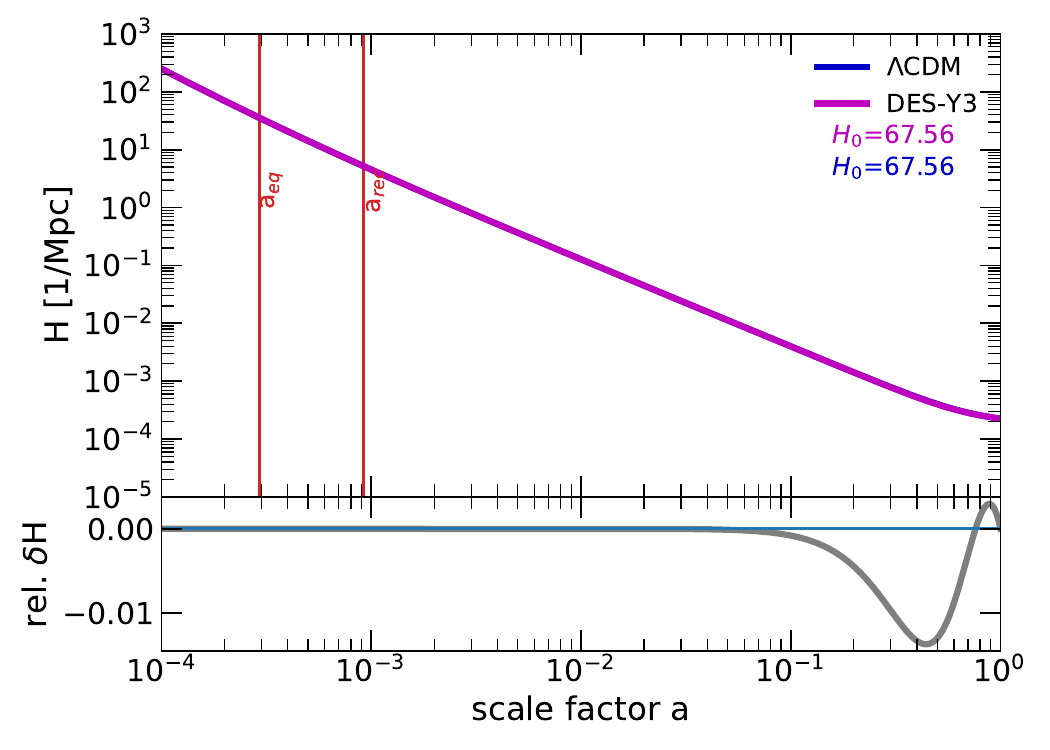}
	\includegraphics[width=1.0\columnwidth]{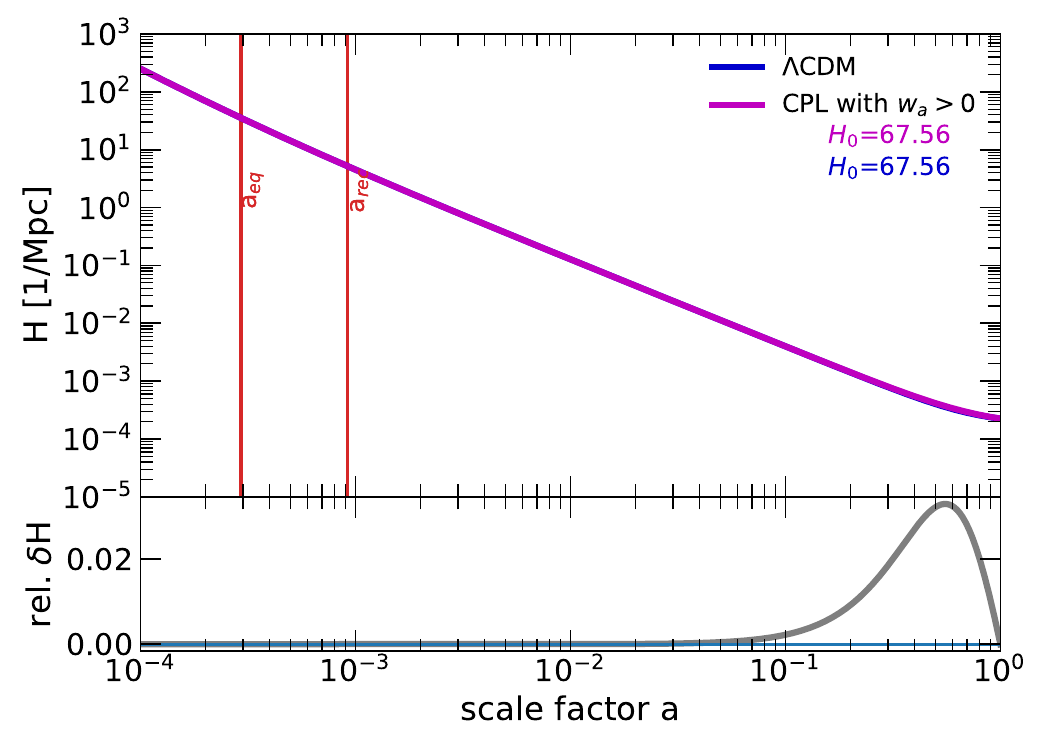} 	
	\includegraphics[width=1.0\columnwidth]{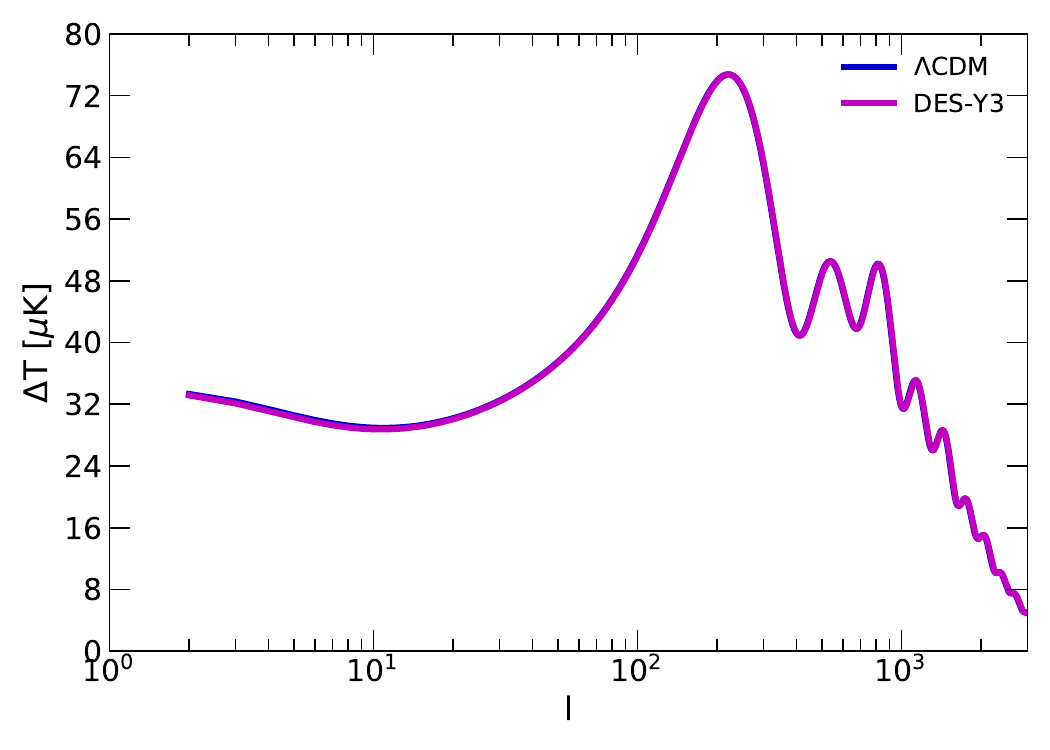}
	\includegraphics[width=1.0\columnwidth]{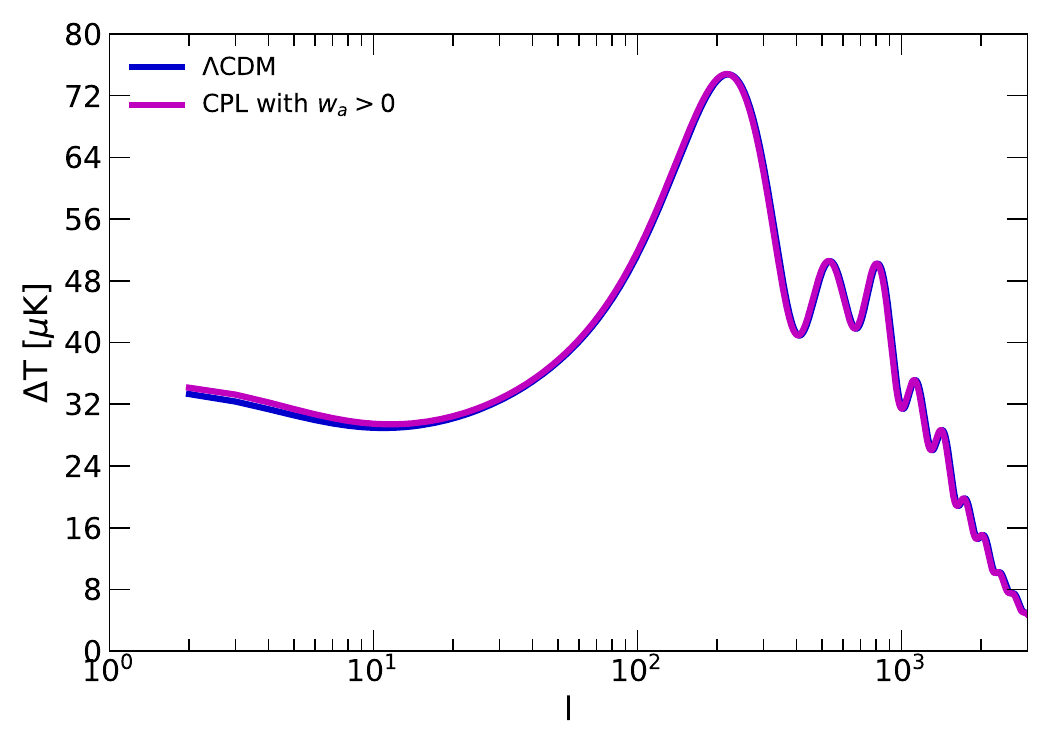} 		
	\includegraphics[width=1.0\columnwidth]{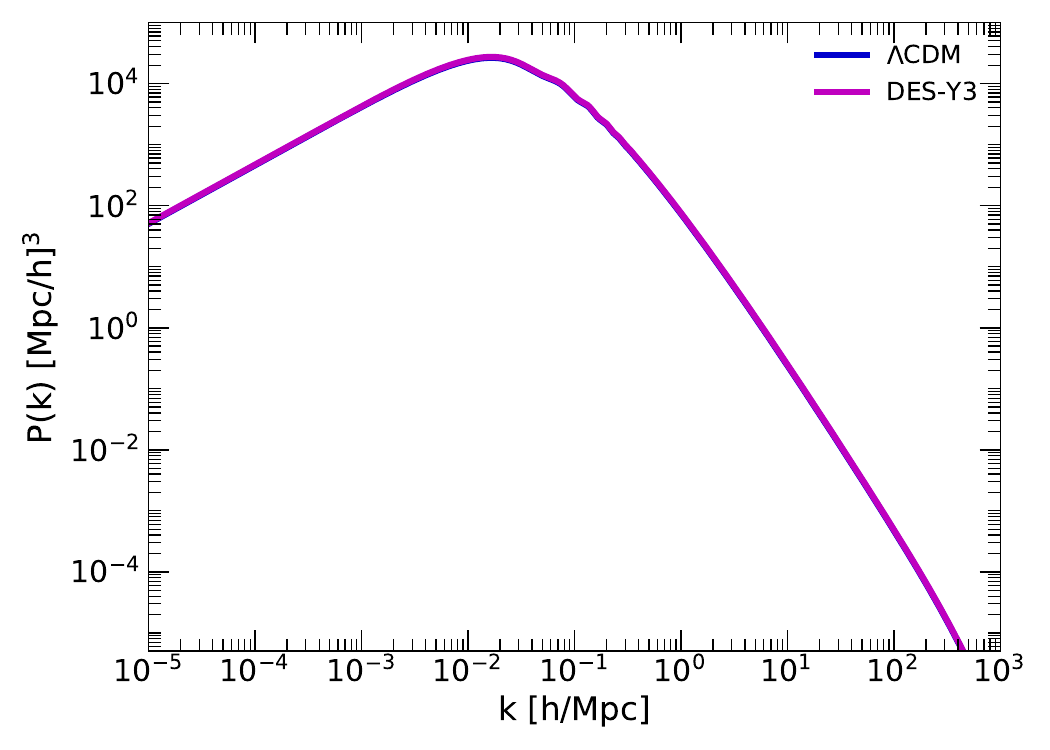}
	\includegraphics[width=1.0\columnwidth]{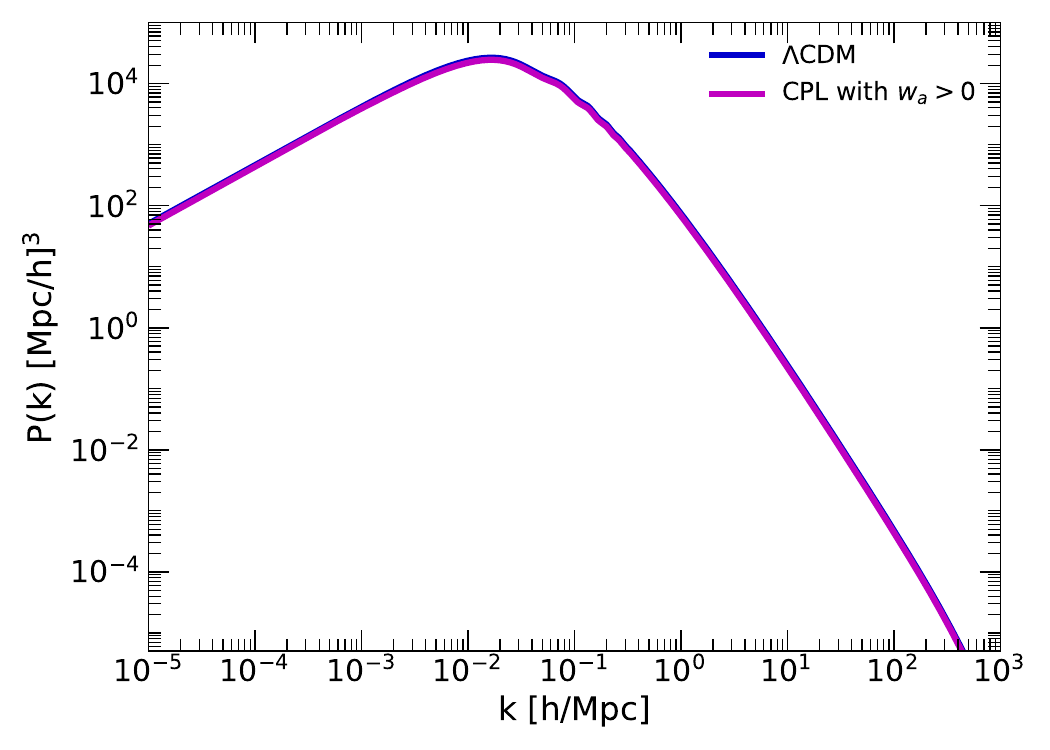}	
	\caption[CPL models computed with the standard procedure]
	{\tb{CPL models (magenta) and \boldmath$\Lambda$CDM (blue) both computed with the standard procedure.} 
		The left-hand side displays the results of the \qn{DES-Y3 CPL model} with $w_0=-0.95$ and $w_a=-0.4$. The right-hand side displays the results for the CPL model with $w_0=-0.9$ and $w_a=+0.1$. 
		We use the original version of CLASS and the value of $H_0$, determined by CMB observations from \citet{Collaboration2018}.
		The top panels display the expansion histories, where the vertical lines in red indicate the time of matter-radiation equality at scale factor $a_\text{eq}$, followed by recombination $a_\text{rec}$ (both determined by the cosmological model). The bottom insets show the relative deviation of the respective expansion histories (gray curve), which only matters at late times (respective high scale factors). The middle and bottom panels show the CMB temperature spectra (temperature anisotropies) as a function of mode number $l$, and the matter power spectra as a function of wave number $k$, respectively.
	}
	\label{fig:proc_customary}
\end{figure*}
\subsection{Standard procedure}\label{sec:proc_customary}

We started with the computation of both models using the standard computational procedure, that is fixed $H_0$, shown in Fig.~\ref{fig:proc_customary}; the left-hand side displays the results for the \qn{DES-Y3 CPL model}; the right-hand side those for the CPL model with $w_a > 0$. Both models are compared to the $\Lambda$CDM concordance model.

\subsubsection{Standard procedure (increasing $w$)}

The expansion history of the \qn{DES-Y3 CPL model,} compared to $\Lambda$CDM, is shown in Fig.~\ref{fig:proc_customary} top left-hand panel. We used the original CLASS code with the default value of $H_0=67.56\pm0.42$ km/sec/Mpc, obtained by \citet{Collaboration2018} (see their Table 2, column \q{TT,TE,EE+lowE+lensing}, where CLASS additionally uses a best fit to WMAP data).
As expected from the \qn{fixed $H_0$} approach, we see an almost perfect agreement with the expansion history of the \glo{LCDMmodel}, and only at late stages, beginning at $a \sim 10^{-1}$, a decrease in the expansion rate of the CPL model of $\sim 1$\% compared to $\Lambda$CDM is seen. It subsequently \qn{overshoots} the expansion rate of $\Lambda$CDM, and finally is forced back (!) to the configured value of $H_0$ (see the gray line at the bottom of the panel), as we have fixed  $H_0$ and the densities in the early universe are then given by the model. 
Also, the calculation of the CMB temperature spectrum (middle panel) and matter power spectrum (bottom panel) in the left-hand side column displays a perfect agreement with $\Lambda$CDM. 

Considering the illustration of Fig.~\ref{fig:intICwrong}, where the customary computational procedure (by fixed $H_0$) \qn{arrives} at the wrong early densities, one may be surprised to see such a perfect agreement with $\Lambda$CDM. 
Nevertheless, the explanation for the perfect agreement, in expansion history as well as in the spectra, is straightforward and can be illustrated partly with the top panel of Fig.~\ref{fig:LCDM}, which displays the evolution of the energy densities in $\Lambda$CDM. We recognize the well-known fact that $\Lambda$ remains a very subdominant component well past $a_\text{rec}$. Of course, the same is true for the DE component of the \qn{DES-Y3 CPL model}: although its density (not shown) is not a constant line, it is also very much subdominant compared to the other (early) densities. 
The spectra are in agreement with $\Lambda$CDM, also thanks to the expansion history, being \qf{forced} to the configured value of $H_0$. 
This is supported by the findings of \citet{Vagnozzi2023}, stating that modification of physics beyond $\Lambda$CDM in the early Universe is not sufficient to resolve the Hubble tension.

\subsubsection{Standard procedure (decreasing $w$)}

Similarly to the \qn{DES-Y3 CPL model}, we show in the right-hand side column of Fig.~\ref{fig:proc_customary} the results for the CPL model with a positive $w_a$, using the Planck-based value of $H_0$ in the standard procedure. We again compare these results from the public version of CLASS with $\Lambda$CDM.
The top panel displays the expansion history. We recognize again an almost perfect agreement with $\Lambda$CDM. However, beginning with $a \sim 10^{-1}$, the expansion rate in the CPL model increases to $\sim 3$~\% above the value of $\Lambda$CDM at $a \sim 6 \times 10^{-1}$, but subsequently is forced back (!) to the configured value of $H_0$.
Also, we show the CMB temperature spectrum and the matter power spectrum computed with the standard procedure, in the middle and bottom panel, respectively. For reasons already mentioned for the \qn{DES-Y3 CPL model}, we see an almost perfect agreement with $\Lambda$CDM. 

\subsubsection{Conclusions for the standard procedure}\label{sec:results_standard}

Given the subdominance of a DE component in the early Universe, the deviation in the total sum of densities compared to the one with $\Lambda$ is small, resulting in negligible deviations in the early expansion history. On the other hand, in the late stages of expansion, the customary computational procedure is insensitive to the evolution of $w$ (see e.g., \citet{OColgain2021,Staicova2022,Vagnozzi2023}), as the impact of $w_a$ converges to zero and $H_0$ is \qn{forced} to the configured value. This explains why we do not see a decrease in $H_0$, in contrast to the expectations from Fig.~\ref{fig:intForwardiCDM} for an increasing EoS parameter. Hence, we see that the computational procedure applied to the CPL-parametrized DE component is insensitive to the parameter $w_a$.

We performed an additional test, and computed a model with negative $w_a$, using (DES-Y3)-only data; see the light blue contours in Figure 4 of \citet{Abbott2022a}. We used the (mean) values $w_0 = -0.6$ and $w_a = -1.8$, corresponding approximately to the central point of the 1$\sigma$-region. These values are clearly different from $w_0 = -1$ and $w_a = 0$ of a cosmological constant. Yet, we received results in almost perfect agreement with $\Lambda$CDM for the expansion history, the CMB temperature spectrum and the matter power spectrum. We draw the conclusion, that the standard computational procedure, not only is insensitive to the parameter $w_a$, but there is a severe $w_0$-$w_a$-degeneracy. In fact, given the Planck value of $H_0$, one can find probably an infinite number of combinations of $w_0$ and $w_a$, including the one of a cosmological constant, that perfectly match the CMB temperature spectrum, where the density parameters for radiation, CDM and baryons remain unmodified. We also recognize that DES-Y3 use the density parameters obtained from the base cosmology, while using $w_0$ and $w_a$ in the list of priors, when fitting the CPL model.

In fact, we computed three models with  DE components that clearly are different from $\Lambda$CDM, and all of them are in perfect agreement with the CMB temperature spectrum. Consequently, we may conclude, that the mere agreement with the $\Lambda$CDM concordance model, due to the $w_0$-$w_a$-degeneracy, does not by itself confirm the cosmological constant $\Lambda$.
Hence, the Planck value of the Hubble constant $H_0$ cannot be considered as a unique value. But it is simply the value determined by the assumption of a cosmological constant, that is $w_0=-1$ and $w_a=0$ (and the flatness of space). This degeneracy in the customary computation procedure, exemplified here for the CPL parametrization, exists for any kind of dynamical model of DE.

Moreover, the inability of discriminating CPL models in terms of the parameters $w_a$ and $w_0$ affects the results when fitting CPL models to observational data. This eventually provides an explanation for the aforementioned unspecific results of data analyses, for values of $w_a$ above and below zero; see also \citet{Dainotti2023}, \citet{Bargiacchi2022}, and \citet{Dainotti2023a}, which report the same nonpositive values for $w_a$, as seen in DES-Y3 (\citet{Abbott2022a}) and \citet{Collaboration2018}.

\subsection{Amended procedure}\label{sec:proc_amended}

We proceeded with the same model comparison, CPL model versus $\Lambda$CDM, but this time applying our complimentary computational procedure by employing the choice of using \qf{early-based} and \qf{locally based} values for $H_0$, parameterized according to our amended version of the CLASS code. In order to break the $w_0$-$w_a$-degeneracy of the standard procedure, an additional assumption is needed. We will exemplify below that the expansion rate $H$ in the early Universe is a viable quantity to break the degeneracy, as it is approximated to high accuracy by the $\Lambda$CDM concordance model, which is immune to the $w_0$-$w_a$-degeneracy.
We performed this comparison for both CPL models; first for the DES-Y3 inspired model with increasing $w$ and then for the model with decreasing $w$.

\subsubsection{Amended procedure (increasing $w$)}\label{sec:DES_results}
\begin{figure*}  [!htb]
	\includegraphics[width=1.0\columnwidth]{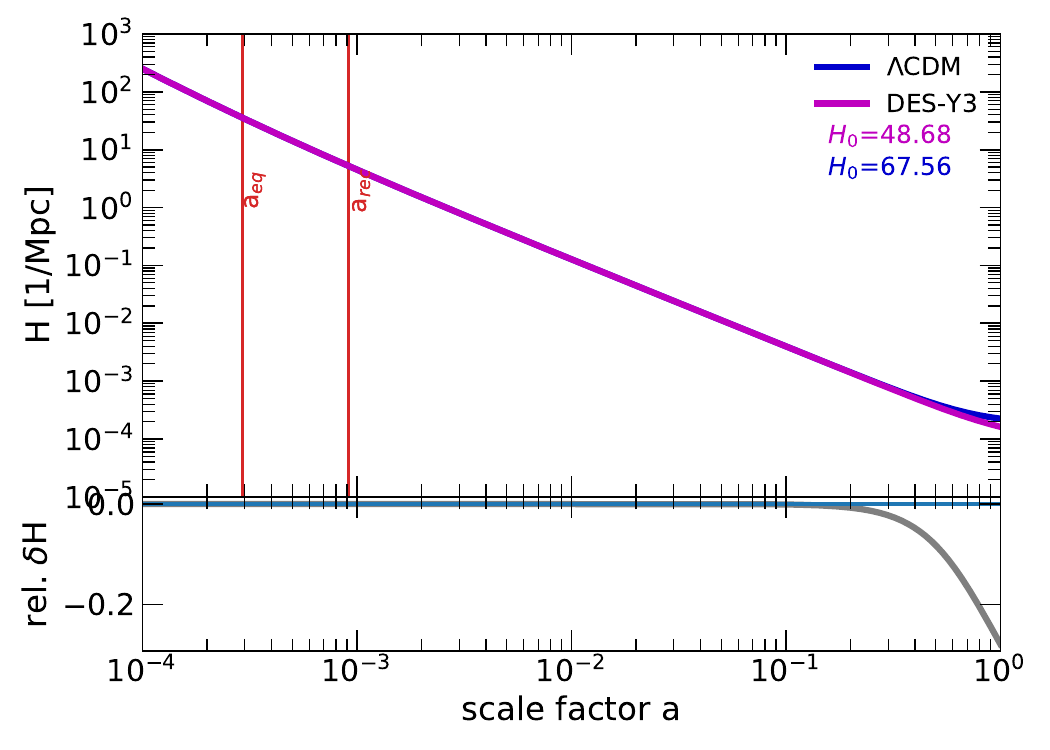}
	\includegraphics[width=1.0\columnwidth]{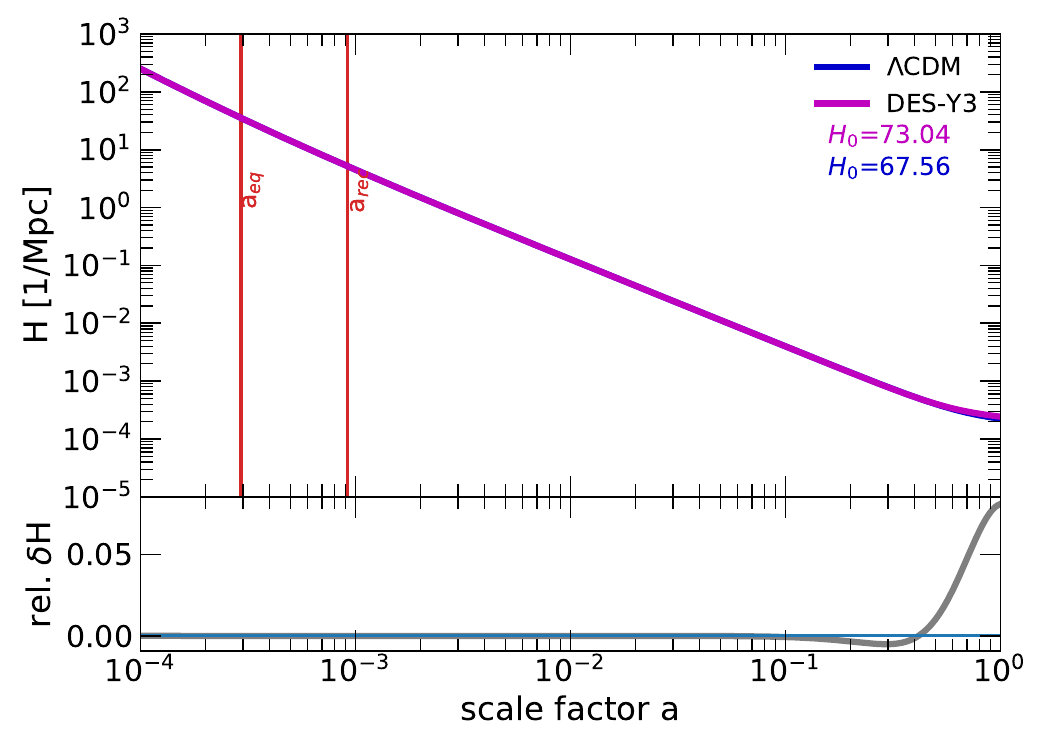}	
	\includegraphics[width=1.0\columnwidth]{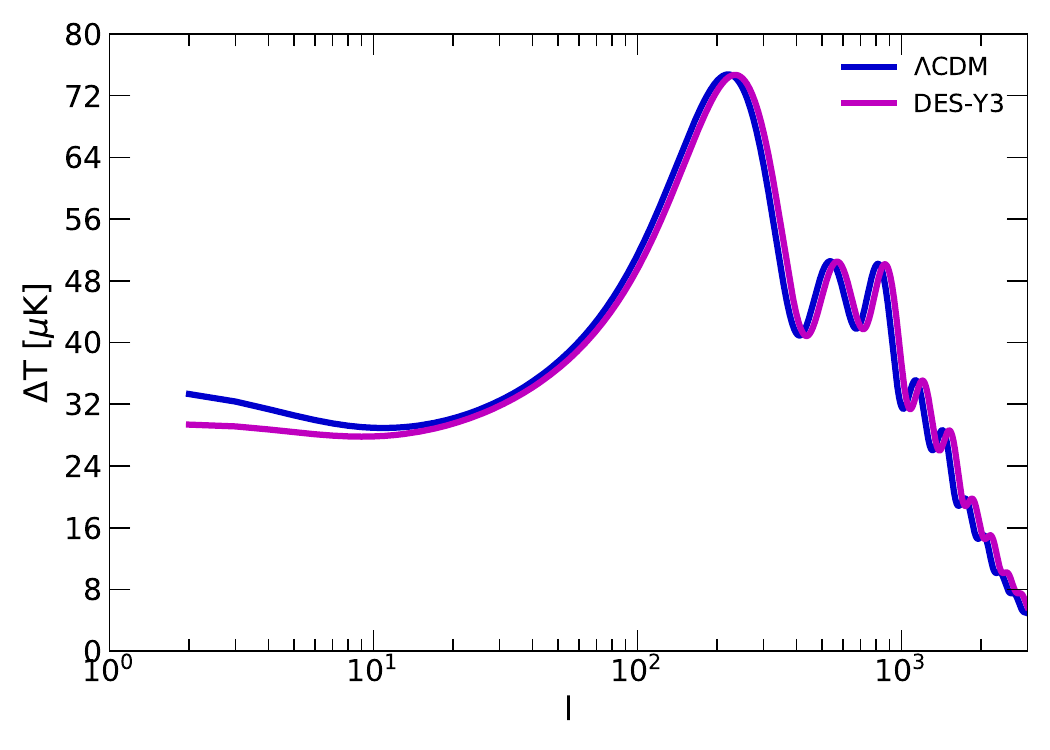}
	\includegraphics[width=1.0\columnwidth]{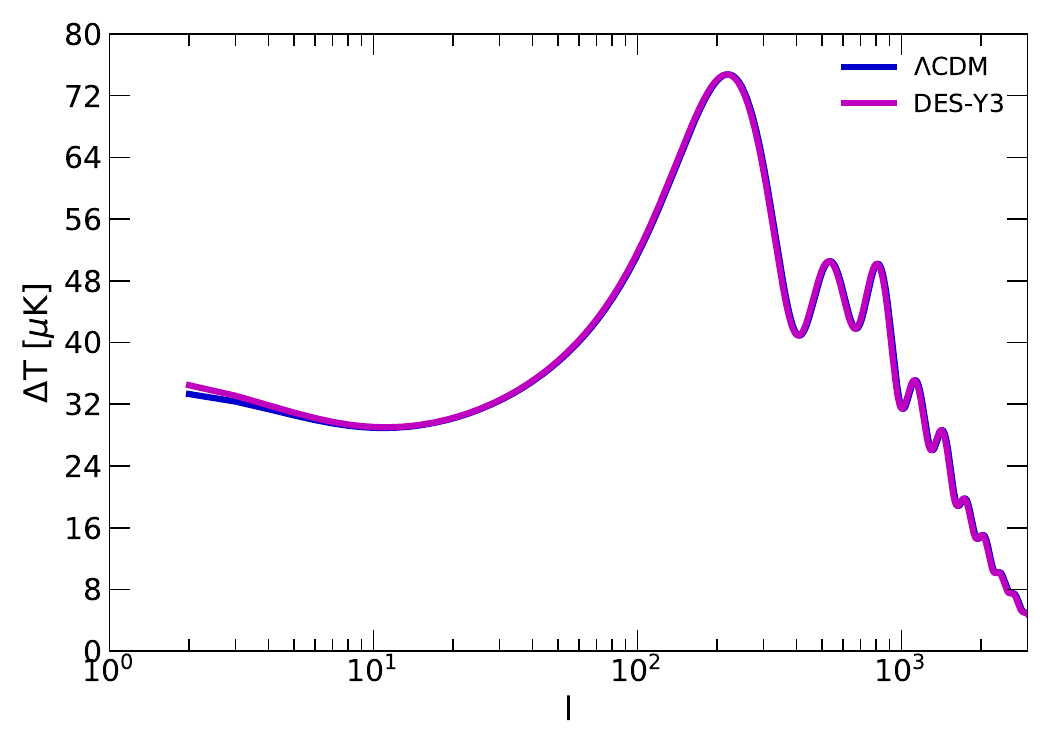}	\includegraphics[width=1.0\columnwidth]{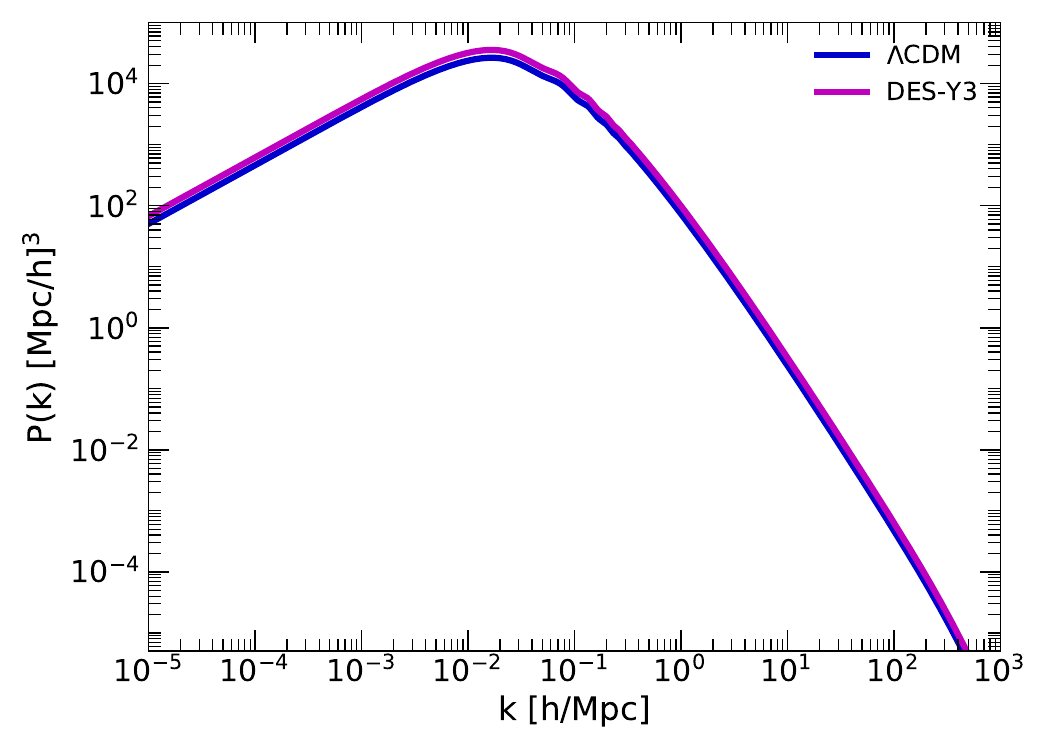} 
	\includegraphics[width=1.0\columnwidth]{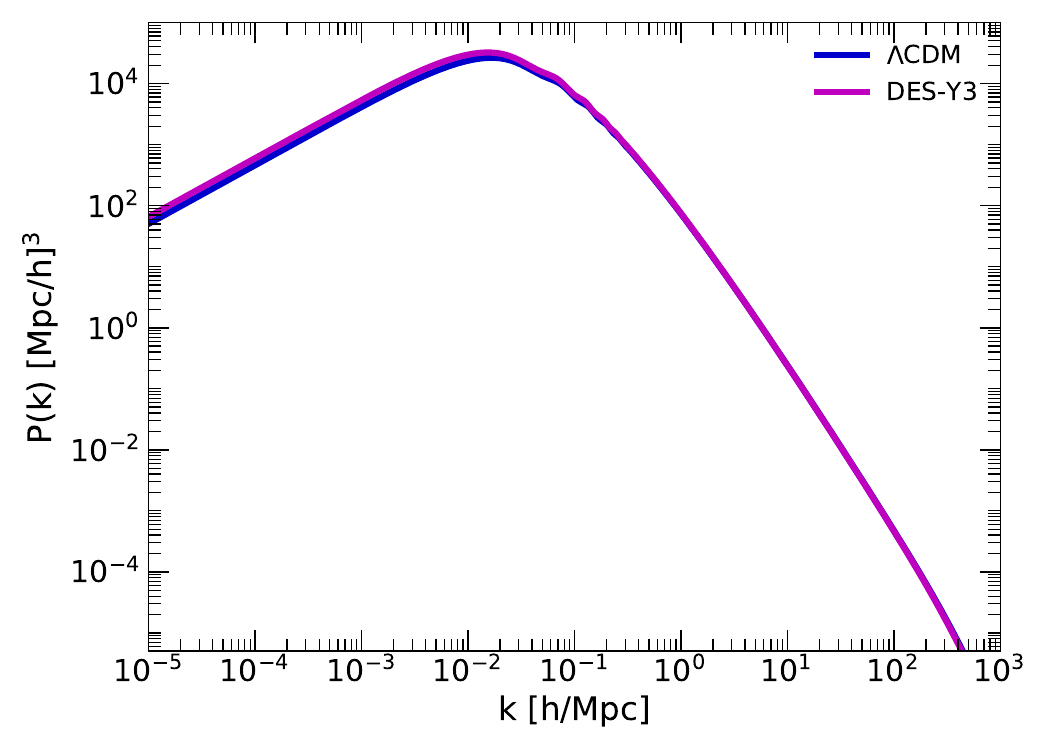}			
	\caption[DES-Y3 CPL model computed with the enhanced procedure]
	{\tb{DE component with increasing $w$: the \qn{DES-Y3 CPL model} with $w_0=-0.95$ and $w_a=-0.4$ (magenta) compared to $\Lambda$CDM (blue), both computed with our scheme.} 
		Panels on the left-hand column were computed using the \qn{early}   parametrization of $H_0=67.56$ km/sec/Mpc using \citet{Collaboration2018}, while panels on the right-hand column employ the \qn{local} parametrization of $H_0=73.04$ km/sec/Mpc from \citet{Riess2022}. Top panels: expansion histories; for the vertical lines in red see Fig.~\ref{fig:proc_customary}. We note the deviations in the expansion histories from $\Lambda$CDM at late times (respective high scale factors) and the different run of these deviations (see gray curves in the bottom insets of the top panels). Middle panels: CMB temperature spectra. Bottom panels: matter power spectra. The results based on the CMB-determined $H_0$, shown in the left-hand column, display a clear deviation from $\Lambda$CDM, whereas the results based on the locally determined $H_0$ are in better agreement with $\Lambda$CDM.
	}
	\label{fig:DES_integration}
\end{figure*}
The results for the \qn{DES-Y3 CPL model} are shown in Fig.~\ref{fig:DES_integration}.
We make a side-by-side comparison of the plots, using the \qn{early} CMB-based $H_0=67.56\pm 0.42$ km/sec/Mpc from \citet{Collaboration2018}, that is the \qn{fixed EDs} approach (left-hand column), and the locally based, that is the \qn{fixed $H_0$} approach (right-hand column) using $H_0=73.04\pm1.2$ km/sec/Mpc from \citet{Riess2022}.
Top panels show the expansion histories, and the middle and bottom panels show the CMB temperature and matter power spectra, respectively.
Let us first focus on the top panels. The top-left panel depicts the results for the CMB-based value of $H_0$. According to the nonpositive parameter $w_a$ of the CPL model for DE, we clearly see a significantly decreased value of $H_0 = 48.68$ km/sec/Mpc, compared to the configured value $H_0=67.56$. Of course, that value for $H_0$ is not in agreement with observations, but in light of the expectation from Fig.~\ref{fig:intForwardiCDM}, we can understand what happened: the increasing EoS parameter results in a decrease of $H_0$ toward the present time. 
Again, around $a_{\text{rec}}$, there is no difference between the DES-Y3 CPL model and $\Lambda$CDM, for both $\Lambda$ and DE are then subdominant.
The top-right panel shows results for the locally based value of $H_0$.
The relative evolution of $H$ in the DES-Y3 CPL model displays a slight s-shape (gray curve in the inset). This calculation corresponds to the situation illustrated in Fig.~\ref{fig:intICcorrect}, where the backward integration \qn{arrives} at the correct EDs.

The spectra in the left-hand column also display clear deviations from $\Lambda$CDM, due to the deviation of the expansion history in the late Universe. While the peak structure in both cases is barely affected (as the density parameters of radiation, CDM and baryons are identical to $\Lambda$CDM), there is an overall shift, horizontally for the CMB spectrum and vertically for the matter power spectrum, due to the reduced growth of structures in the late stages of the evolution, for the latter. The horizontal shift to the right-hand side in the temperature spectrum is connected to the deviation in the expansion rate, as follows. The multipole moments shown in the x-axis, are in Fourier space. Hence, it is not a shift, but stretching (compressing) of the spectrum from the right-hand side to the left-hand side, with increasing (decreasing) expansion rate.
More pronounced differences in the spectra occur at large spatial scales (corresponding to low $l$ and low $k$, respectively).  

Comparing now to the right-hand column, which uses the local value of $H_0$, we see much better agreement with $\Lambda$CDM for all observables, although some minor deviations in the spectra remain at large spatial scales (probably of no discriminating significance given the error bars of CMB measurements at low $l$). As mentioned earlier, this outcome corresponds to Fig.~\ref{fig:intICcorrect}, where the backward-in-time integration \qn{arrives} at the correct densities in the early Universe. 

To reiterate, the simulation run for the early-based (high $z$) value of $H$ employing our scheme does not \qn{force} the expansion rate to the configured \qn{concordance value} of $H_0=67.56$~km/sec/Mpc, as we have fixed the early densities while $H_0$ is determined by the (CPL) model. Therefore, we see \q{exacerbated results} on the left-hand column, with $H_0=48.68$~km/sec/Mpc, compared to the standard procedure depicted in Fig.~\ref{fig:proc_customary}. But in light of the explanations of Fig.~\ref{fig:intForwardiCDM}, we conclude that the results of the amended procedure are correct, that is more realistic in terms of the physical expectations. 
On the other hand, the results displayed in the right-hand column conform much better to $\Lambda$CDM, because our starting point in the backward integration uses the locally based $H_0$ (at $z\approx 0$), which is numerically close to the \qn{concordance value for $H_0$} to begin with.
Yet, the two simulation runs using different $H_0$, depending on whether early-based or locally based measurement values have been used to parameterize the integration runs, do not display consistent results for the two approaches of \qn{fixed EDs} vs fixed $H_0$; we see clear differences between the left and right columns in Fig.~\ref{fig:DES_integration}, signaling that backward and forward evolution of the expansion history does not yield consistent results. Therefore, this model is not able to provide an answer to the Hubble tension problem.

\subsubsection{Amended procedure (decreasing $w$)}\label{sec:iCDM_results}

\begin{figure*}  [!htb]
	\includegraphics[width=1.0\columnwidth]{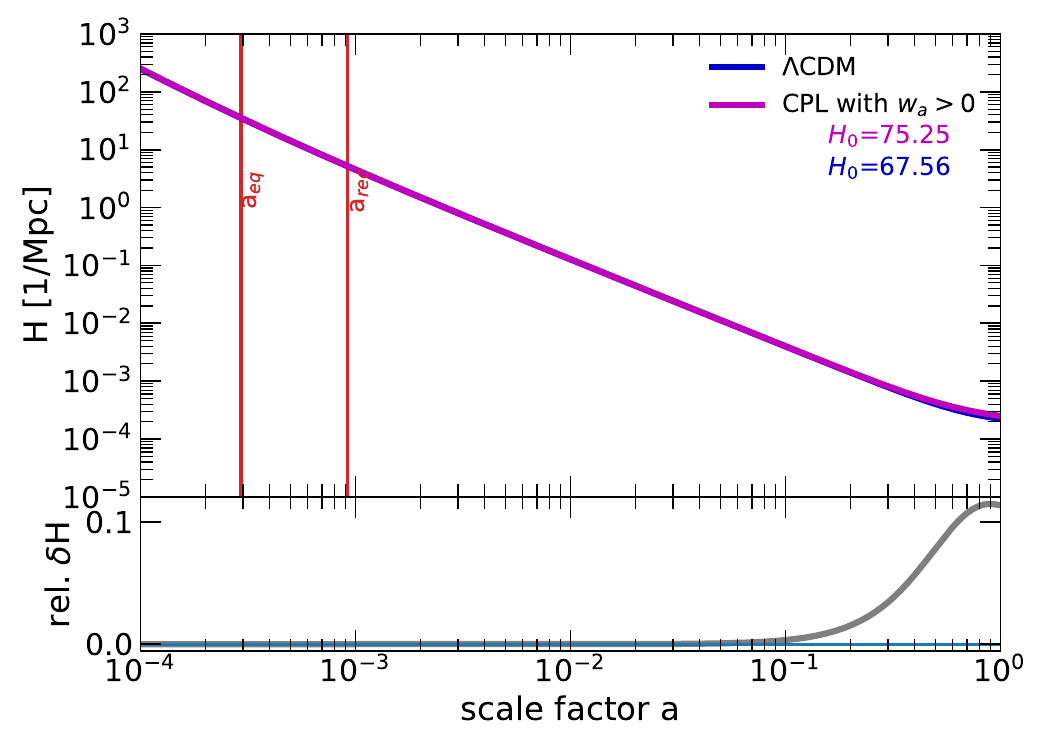}
	\includegraphics[width=1.0\columnwidth]{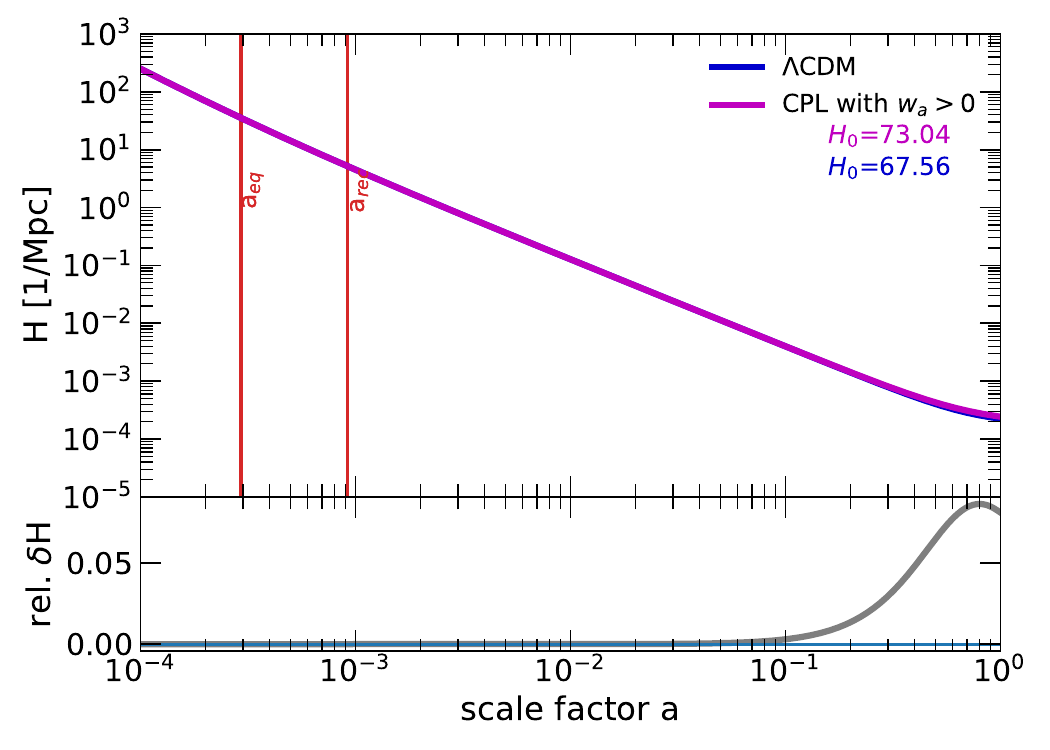}
	\includegraphics[width=1.0\columnwidth]{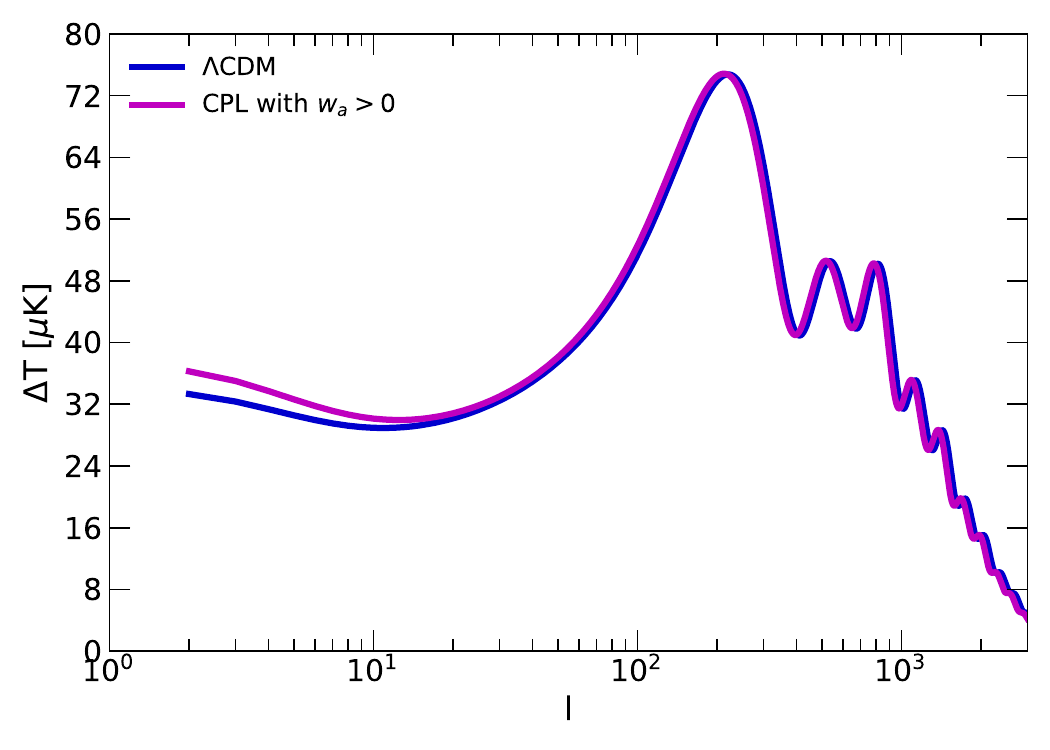}
	\includegraphics[width=1.0\columnwidth]{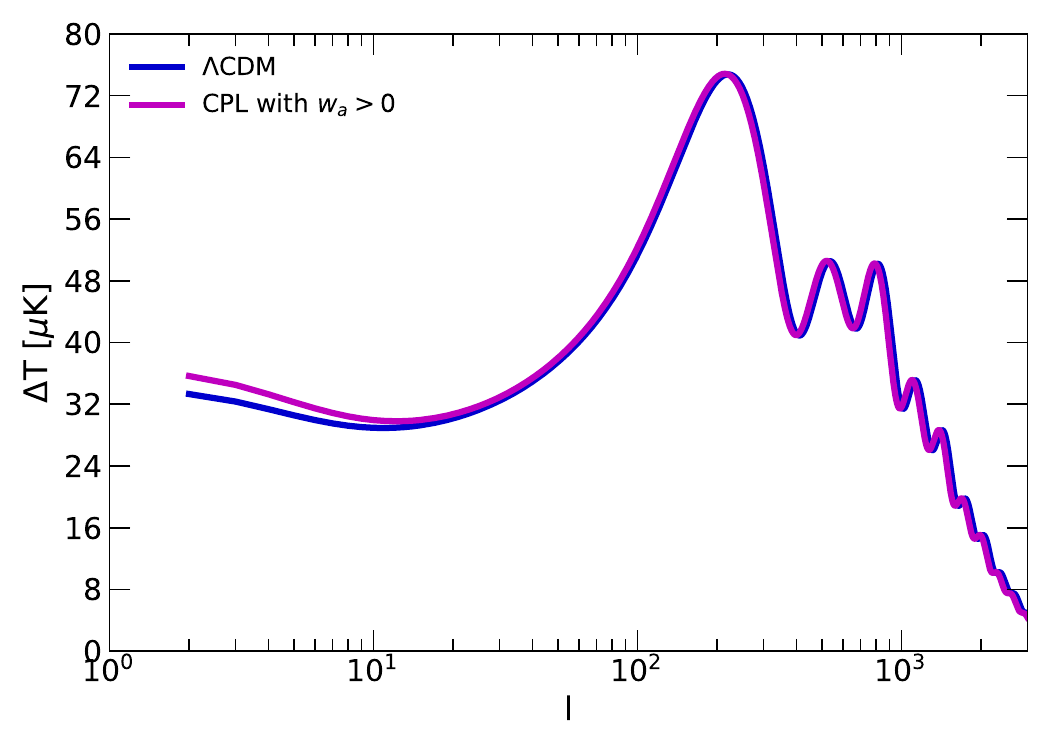}
	\includegraphics[width=1.0\columnwidth]{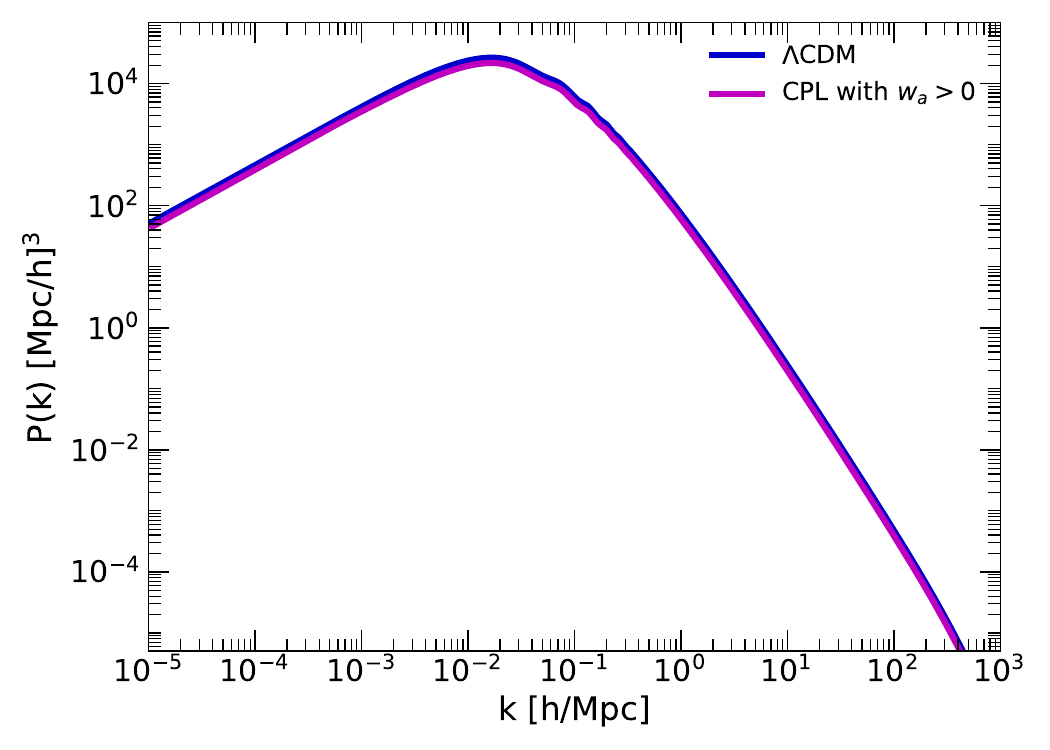}
	\includegraphics[width=1.0\columnwidth]{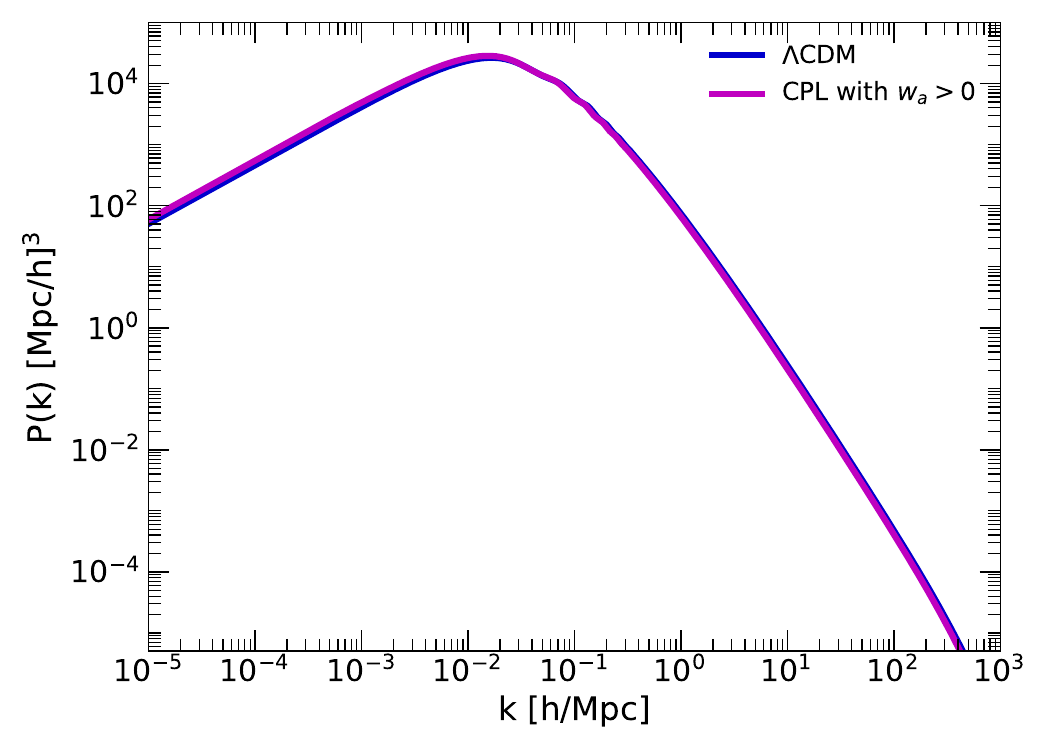}			
	\caption[Expansion history for a DE component with decreasing EoS parameter]
	{\tb{DE component with decreasing \boldmath$w$: a CPL model with $w_0=-0.9$ and $w_a=+0.1$ (magenta) compared to $\Lambda$CDM (blue), both computed with our scheme.} Panels on the left-hand column were computed using the \qn{early} parametrization of $H_0=67.56$ km/sec/Mpc from \citet{Collaboration2018}, while panels on the right-hand column employ the \qn{local} parametrization of $H_0=73.04$ km/sec/Mpc from \citet{Riess2022}. Top row: expansion histories; for the vertical lines in red see Fig.~\ref{fig:proc_customary}. Middle panels: CMB temperature spectra. Bottom panels: matter power spectra. In both columns, we see good overall agreement with $\Lambda$CDM, except for the fact that the chosen CPL model may provide a resolution to the Hubble tension problem. A difference remains in the present-day Hubble constant between left-hand and right-hand columns, see main text.
	}
	\label{fig:iCDM_integration}
\end{figure*}

Now, Figure~\ref{fig:iCDM_integration} proceeds with the comparison of the other considered CPL-DE model with $w_a > 0$ vs $\Lambda$CDM, computed by the amended procedure.
We make a side-by-side comparison of the plots, using the \qn{early} CMB-based $H_0=67.56\pm 0.42$ km/sec/Mpc from \citet{Collaboration2018}, that is the \qn{fixed EDs} approach, (left-hand column) and the locally based, that is the \qn{fixed $H_0$} approach (right-hand column) using $H_0=73.04\pm1.2$ km/sec/Mpc from \citet{Riess2022}.
In the top panels, we see the expansion histories. Again, the expansion rate displays no difference to $\Lambda$CDM around and well past $a_{\text{rec}}$. 
For the early-based value of $H_0$, we see beginning at $a \sim 10^{-1}$ a rise of $H$ resulting in $H_0=75.25$~km/s/Mpc, in contrast to the \qn{concordance value} $H_0=67.56$~km/s/Mpc.
This result is in agreement with the expectation from Fig.~\ref{fig:intForwardiCDM} that a decreasing EoS parameter $w$ leads to a higher Hubble constant at the present, when fixing the early densities. 
On the other hand, the right-hand panel uses the \qn{local} value of $H_0=73.04$ km/sec/Mpc from \citet{Riess2022}. 
As expected and similarly to the right-hand column of the previous Fig.~\ref{fig:DES_integration}, the resulting value of $H_0$ is identical to the configured value, as this is the starting point for the backward integration, when fixing $H_0$  (see Fig.~\ref{fig:intICcorrect}), to determine the evolution of the energy densities. 
The comparison of the CMB temperature spectra and the matter power spectra obtained from the amended computational procedure, using \qn{early} vs \qn{local} values of $H_0$, with $\Lambda$CDM are also depicted in Fig.~\ref{fig:iCDM_integration}. For both spectra, we see good agreement with $\Lambda$CDM, except for some deviations at large spatial scales.

Again, the amended code gives us results for the expansion histories, as well as for the spectra, with respect to the different (\qn{local} vs \qn{early}) parametrizations of $H_0$, which are in accordance with the expectations described in Sect.~\ref{sec:timedependingEoS}. In contrast to the model with an increasing EoS parameter, we see here, for the model with a decreasing EoS parameter, consistent results for each case, \qn{fixed EDs} or fixed $H_0$, along with being in accordance with observations across cosmic time. We might thus conclude that this model points toward a resolution to the Hubble tension problem. Of course, more precise statements require further investigation, for example, provided via an extended MCMC analysis as discussed in Sect.~\ref{sec:observations} and \ref{sec:discussionconclusionsH0}.

\subsubsection{Conclusions for the amended procedure}\label{sec:results_amended}

Summarizing the results of the two CPL-based DE models in this section, Fig.~\ref{fig:DES_integration} vs Fig.~\ref{fig:iCDM_integration}, we can see that the latter case -- a CPL-DE model with $w_a > 0$, that is a decreasing EoS parameter -- gives much better agreement with both, the concordance $\Lambda$CDM model and the present-day locally determined Hubble constant from \citet{Riess2022}. In other words with each, either \qn{fixed $H_0$} or fixed EDs, respectively, via the backward and forward integrations of this CPL model of Fig.~\ref{fig:iCDM_integration} yield results which are (much more) consistent with observations, compared to the model in Fig.~\ref{fig:DES_integration}. In principle, we could attribute this finding to the mere chance of having picked favorable CPL parameters (were it not for other reasons which motivated our choice in the first place; a publication on a physics-based model of DE is currently in progress). Indeed, the CPL parametrization is purely a convenient way of representing DE models with a time-dependent $w$, and it is not obvious that any choice of its free parameters will fit perfectly to observations (see e.g., \citet{OColgain2021} and \citet{Dinda2022}). However, it so happens that a choice of $w_0=-0.9$ and $w_a=+0.1$, upon fixing EDs, yields a value of $H_0 = 75.25$ km/s/Mpc, which is quite close to the value of \citet{Riess2022} (which is the starting point of the backward integration in each model that we calculated).       

We can draw two major conclusions from this section. First, our results suggest
that a DE component with a decreasing EoS parameter $w$ may provide a resolution to the Hubble tension problem, because the results of both backward-in-time (fixed $H_0$) and forward-in-time integrations (fixed EDs), each with their respective flagging of $H_0$ (early or local), are in good agreement with the local measurements of $H_0$ and the expansion rate at the time of recombination.
Second, we argue that the requirement of finding consistent results in both cases should be used as a restrictive criterion of cosmological models. In fact, the procedure can help to find model parameters that are free of the Hubble tension problem, in a very natural way, in that they also provide a present-day Hubble constant in accordance with local measurements. We note that, in order to obtain consistent results, the data analysis (MCMC) procedure must ensure that a sufficient range of $H_0$ is sampled for the fixed $H_0$ approach. In the next section, we discuss how this approach should be implemented via a three-step procedure in a kind of extended MCMC method, that is such that eventually only models without any difference between forward and backward, (i.e., \qn{fixed EDs} and fixed $H_0$) evolution will be sampled, while being in agreement with local measurements. We expect that such an approach will considerably increase the determined accuracy of the parameters of cosmological models; see also e.g. \citet{Dainotti2022}, \citet{Dainotti2023}, \citet{Dainotti2021}, \citet{Dainotti2022a}, \citet{Bargiacchi2023}, \citet{Bargiacchi2023a}, \citet{OColgain2021}, \citet{Staicova2022}, \citet{Krishnan2021a}, and \citet{Lenart2023} whose results indicate that measurements (up to $z \sim 7.5$ using quasars) might favor a dynamical model of DE.

More precisely, with regard to the computation of cosmological models and in light of the conclusions for the standard procedure of Sect.~\ref{sec:results_standard}, we demand consistent results in the computation of a cosmological model, obtained from early densities approximated by concordance $\Lambda$CDM values (\qn{fixed EDs} approach) with those obtained via $H_0$ as measured in the local Universe (the customary \qn{fixed $H_0$} approach), which breaks the $w_0$-$w_a$-degeneracy and fulfills the general requirement to give consistent results in the forward and backward evolution of the model and being in agreement with observations.
Consequently, in addition to the above CPL-based DE models, we performed this consistency check for $\Lambda$CDM, where we used our amended code with a CPL parametrization, that is setting $w_0=-1$ and $w_a=0$, using the \qn{early} vs \qn{local} parametrization of $H_0$, respectively. Unsurprisingly, we recognize a Hubble tension problem in the concordance $\Lambda$CDM model, see Appendix~\ref{app:checkmodels}.

\section{Toward an improvement of the accuracy of cosmological observables}\label{sec:observations}
In the previous sections, we discuss our complementary approach of \qn{fixed EDs} vs the customary approach of \qn{fixed $H_0$} in the computations of cosmological models with a DE component having a time-dependent EoS, and we present a refined procedure for the self-consistent computation of the expansion history and linear spectra. In this section, we advance one step further and we argue that under the assumption of a cosmological origin of the Hubble tension, the very requirement of recovering consistent results between backward-in-time (i.e., fixed $H_0$) and forward-in-time (i.e., fixed EDs) evolution of cosmological models should be used, in order to check the consistency of a cosmological model (see also Appendix~\ref{app:checkmodels} and  Sect.~\ref{sec:results_amended}). 
This consistency check, in turn, can be used to extend the standard MCMC procedure used to analyze cosmological data. As a result of this requirement, we expect the available parameter space to be significantly reduced, leading to an increase in the accuracy of the derived parameters of cosmological models, in particular for extensions to $\Lambda$CDM.

The Hubble tension problem basically refers to the discrepancy in the measurement results for $H_0$ between observations of the CMB -- whose analysis is based on {extremely well-understood theories} -- and observations in the local Universe -- based on {extremely well-checked measurements}. Generally, the latter provide higher values of $H_0$ than the CMB-based extrapolated values for $H_0$ (see e.g., \citet{Hu2023}).
At the same time, from a theory perspective, we require consistent results for model universes evolved from the early Universe to the present {and} vice-versa from the present to the early Universe, while being in agreement with observational data across the entire expansion history.
In Appendix~\ref{app:checkmodels}, we highlight the case of $\Lambda$CDM as an example to check the consistency of cosmological models and their parameters, respectively. This procedure is built on the assumption that the evolution of the background universe is a well-posed IVP, describing the evolution from the early Universe to the present (\qn{fixed EDs} approach). However, the same shall be true in the converse direction (since we consider the Universe as a closed and conservative system), that is using ICs provided by measurements of the local (present) Universe, the equations of motion can be evolved backward-in-time to the early Universe (customary \qn{fixed $H_0$} approach), for example, back to the time of recombination, yielding identical results to the forward evolution. 

The question arises as to how the consistent treatment of both approaches can be combined in the analysis of cosmological data. In general, cosmological observations, that apply the MCMC method, explore the parameter space of (typically) FLRW models, and determine the probability distribution of each of the model parameters, by \q{matching} the properties of computed model universes to observational data. For example, the data analysis of DES-Y3 match the computed matter power spectra with observed BAO data, in order to determine the set of model parameters with the highest likelihood. The CMB observation by the Planck mission uses a similar procedure: computed temperature spectra are matched with the measured CMB spectrum, yielding the set of model parameters with the highest likelihood, the well-known concordance parameters of the \glo{LCDMmodel}.

Now, we already emphasized that many observational campaigns (such as DES-Y3, or the Planck mission) include in their analysis (i.e., in their comparison with data) candidates for DE with a time-dependent EoS parameter via a CPL parametrization (see Eq.~\eqref{eq:EQCPLH0}), along with the cosmological constant of $\Lambda$CDM. Since these candidate models are affected by some issues identified in the previous section (see Sect.~\ref{sec:proc_customary}), when only the customary \qn{fixed $H_0$} approach is applied, we suggest that the aforementioned consistency check, based on the \qn{fixed EDs} approach, is being applied, in order to avoid these issues. As a result, we expect that the viable parameter space of models shrinks dramatically, as for example, dynamical models of DE with an increasing EoS parameter will probably be ruled out, such that a significant improvement in the accuracy of the determined parameters of cosmological models can be achieved.
The standard MCMC computational method should be extended to include a three-step procedure, where each step applies the refined computational procedure introduced above, as follows. 

First, we remind that in the standard MCMC method each of the model parameters is assigned a presumed range of values (known as the priors), that determine the entire parameter space of a given model to be \qn{fitted} to data. The parameter space (i.e., a grid of parameter sets) is sampled iteratively, advancing from one parameter set to the next. In each step, the model is computed (comparable with our model computations in the preceding section) and matched to observational data. The quality of this match is expressed by computing the likelihood, for example, by a $\chi$-square test. The likelihood controls the strategy of the sampler, which also assures not to step out of the defined parameter space but to remain within it. In a final step, the probability distribution (known as the posterior distribution) is computed, applying Bayesian statistics. We propose to extend the computational step for the sampled models to a three-step procedure, instead of the standard two-step procedure.


In the first step, the cosmological model under consideration (by the sampler) is computed using the value of $H_0$, inferred from measurements or sampled as a member of the priors, respectively, now flagged as \qn{local}, performing a backward integration of the densities, starting from the \qn{local} ICs, as demanded by the \qn{fixed $H_0$} approach (this corresponds to Fig.~\ref{fig:intICcorrect}). Keeping the $w_0$-$w_a$-degeneracy in mind, one must not restrict the prior of $H_0$ by for example, including probability densities from previous literature or previous computations with restricted parameter space, etc. (see e.g., DES-Y3 results for the $w_0$-$w_a$ model).

The novel second step introduces the \qn{fixed EDs} approach. In this step, the sampled cosmological model is computed according to the \qn{fixed EDs} approach by applying the \qn{concordance value for $H_0$} (flagged as \qn{early}) and the concordance model parameters. We note that these model parameters are not those determined by the MCMC method. They do not enter the lists of priors and posteriors. They only determine the initial densities, as demanded by the \qn{fixed EDs} approach (this corresponds to Fig.~\ref{fig:intForwardBackward}). 
Starting at these computed EDs, the subsequent forward integration applies the sampled parameter set, entering the computation of the posteriors.

The final step also extends the customary procedure of the MCMC analysis. While customarily, the sampled model parameters are \qn{fitted} to the measured data (e.g., CMB temperature spectrum or the matter power spectrum, respectively), we now consider all information available for the sampled model: the expansion histories, CMB temperature and matter power spectra from the computations. Both results, from \qn{fixed $H_0$} (step one) and \qn{fixed EDs} (step two), are checked for consistency. Only those parameter sets, that display consistent results for all three quantities, taking the accuracy of the measurements into account, are considered in the subsequent computation of the likelihood and the probability distribution, respectively. 
Since this additional consistency check will lead to a reduction of the allowed parameter space (probably ruling out dynamical models of DE with an increasing EoS parameter, see also next paragraph), it will increase significantly the accuracy of the final parameters, resulting from the MCMC analysis.


Encouraged by the results depicted in Fig.~\ref{fig:iCDM_integration} for a DE component with decreasing EoS parameter, that is $w_a > 0$ in the specific CPL parametrization, we see the Hubble tension problem to be explained phenomenologically in a very natural way by applying the \qn{fixed EDs} approach.
We expect, based on the proof-of-concept implementations in the preceding section, that such an analysis will find $w_0 > -1$ and $w_a > 0$, which is in agreement with low-$z$ data depicted by the green contours in Figure 5 of DES-Y3 \citet{Abbott2022a}. Also, Y1 results presented in Figure 11.13 of \citet{Collaboration2009} depicts a probability distribution in the $w_0-w_a$-plane compatible with our expectation.
In contrast, DES-Y3 results (\citet{Abbott2022a}) and the CMB measurements by \citet{Collaboration2018}, when comparing a CPL-based model of DE to a cosmological constant, yield a high probability for $w_a < 0$  (see also \citet{Dinda2022, Vagnozzi2023,OColgain2021, Staicova2022}). 
Our calculation of an exemplary model in Sect.~\ref{sec:DES_results} with $w_a < 0$, where we picked their central values of the CPL parameters calling it \qn{DES-Y3 CPL model} for mere convenience, does not pass the recommended consistency check (see Fig.~\ref{fig:DES_integration}). 

In the description of step one of the extended MCMC routine, we argued that one must be careful when including information from previous measurements into the Markov chain, in order not to anticipate a value for $H_0$. Of course, using results from previous observations is an important means to improve the accuracy of one's own data analysis. But instead of including previous probability distributions, we did this by including the measured CMB temperature spectrum in the computation of the likelihood in step three by comparing it to the computed temperature spectrum of the sampled model. To this end, we implemented a proof-of-concept version of the extended MCMC procedure and fitted our model with $w_0 = -0.9$ $w_a = +0.1$ to the CMB temperature spectrum of the concordance model using our amended version of CLASS. The resulting parameters $w_0 = -0.875$ and $w_a = 0.075$ removed the minimal horizontal shift of the spectrum, seen in Fig.~\ref{fig:iCDM_integration} and gives a perfect agreement with the measured spectrum (with the concordance $\Lambda$CDM model, respectively).

Finally, we stress again that the cosmological model, which is used in deriving extrapolations of observables, such as $H_0$, is bound to these values. In other words, the EDs giving rise to the CMB measurements, the extrapolated $H_0$ and the corresponding cosmological model (parameters) are inextricably linked together, where none of them can be used without being impacted by the other. This fact also explains results of {indirect} measurements, which (apparently!) seem not to apply CMB-based measurements, and yet yield values of $H_0$ close to the Planck \qn{concordance value}, as seen for example, in Figure 1 by \cite{DiValentino2021}.
Such a case can be seen in \citet{Alam2021}, their Table~5 with $H_0=67.35\pm 0.97$~km/s/Mpc, based on Big Bang nucleosynthesis (BBN) and BAO data. They apply the MCMC method, while using $\Omega_{r,0}$ from the concordance model (thus, excluding radiation from the MCMC-sampled parameter space), {and assuming a cosmological constant}. Hence, the computations of the sampled cosmological models as part of the MCMC method \qn{arrive} at a value of $H_0$ close to that of the concordance model (see also Appendix \ref{app:wConst}). In this sense, the use of CMB data is implicit in these mentioned cases, due to the relation between the concordance \glo{LCDMmodel}, $H_0$ and the corresponding initial densities, that have been implicitly applied in their model assumptions, namely $\Omega_{r,0}$ of the concordance model {and} a cosmological constant.
As this could be seen as a bad example of the characteristics connected to the customary \qn{fixed $H_0$} approach, it confirms at the same time our approach to approximate the initial densities in the early Universe as determined by the $\Lambda$CDM concordance model.

\section{Summary and conclusion}\label{sec:discussionconclusionsH0}

A cosmological model should fulfill certain criteria. First, computations of the model forward-in-time and vice versa backward-in-time must yield identical results. Consistent results in forward and backward evolution of a model basically checks the consistency of the computation and not of the model. Apart from some exotic models of cosmic components, this is not an issue in the computation of the expansion history. The second condition is even more important: the results have to agree with observations across cosmic time. In this sense, the \glo{LCDMmodel} is not consistent, for it does not agree with local measurements of the Hubble constant (Hubble tension problem). On the other hand, in the early Universe, $\Lambda$CDM is in conformance with measurements of the CMB. Recent studies analyzing the evolution of the expansion rate $H$ at low redshifts indicate that these data are not in accordance with a cosmological constant and that the Hubble tension might be of cosmological origin, due to a DE component with a time-dependent EoS parameter, rather than connected to issues in the local measurements (e.g., \citet{Krishnan2021a,Dainotti2023b,RidaKhalife2023}).

In Sect.~\ref{sec:timedependingEoS} we discuss in a qualitative way if the customary computation procedure for the expansion history of a cosmological model can be applied to a DE component with a time-dependent EoS parameter. We find that special considerations must be taken into account, compared to a component with a constant EoS parameter. We conceived a refined procedure to compute the evolution of the energy densities, in that we implemented a case distinction between the provided $H_0$ from measurements based either on the \qn{early Universe} (CMB-based), or the \qn{local Universe} (e.g., based on standard candles). Most important, we find that the \qn{concordance EDs,} computed by the customary backward-in-time integration starting at the \qn{concordance value for $H_0$}, provides an accurate approximation of the initial densities in the early Universe. These are used as the starting point for the subsequent forward-in-time integration of the energy conservation equation \eqref{eq:EQeconsnbgH0}. Hence, for early-based values of $H_0$, the required forward-in-time integration starts at the computed early densities, whereas for locally based values of $H_0$ the integration is performed backward-in-time.

In Sect.~\ref{sec:compmodelsH0} we implement this computation procedure into the code CLASS and perform computations of the expansion histories, CMB spectra and matter power spectra of representative cosmological models.
Hereby, we find that the customary computation procedure, \qn{fixed $H_0$,} for the expansion history of a cosmological model is affected by a $w_0$-$w_a$-degeneracy of the CPL parametrization applied to the DE component, that is due to the subdominance of DE in the early Universe while becoming dominant only in the late Universe. We show in Sect.~\ref{sec:proc_customary} that this degeneracy results in an agreement with $\Lambda$CDM over a wide range of the parameters $w_0$ and $w_a$. In fact, this degeneracy applies not only to the CPL parametrization, but to all dynamical models of DE which offer the same degrees of freedom in the evolution of $w(z)$. For Planck-based value of $H_0$, one can find a number of combinations of $w_0$ and $w_a$ that result in an agreement with $\Lambda$CDM. This biases the fitting of models to data, as well (\citet{Dinda2022, Vagnozzi2023,OColgain2021, Staicova2022}). It also explains the nonpositive values for $w_a$ reported in the DES results \citet{Abbott2022a} and  \citet{Collaboration2018}; see DES-Y3 in \citet{Abbott2022a} and the light blue contours and the red contours in their Figure 4, where they include data from \citet{Collaboration2018} Figure 30. Only the combination with low-$z$ data, seen in their Figure 5 green contours, shifts the result to the \q{less negative} value of $w_a = -0.4$, compatible with a cosmological constant. 

In Sect.~\ref{sec:proc_amended}, we propose and analyze a complementary computational approach -- the \qn{fixed EDs} approach -- for the computation of the expansion histories for cosmological models, that breaks the $w_0$-$w_a$-degeneracy, affecting dynamical models of DE with a time-dependent EoS parameter. This approach is built on the assumption that the evolution of the background universe is a well-posed IVP, describing the evolution from the early Universe to the present, given by initial densities in the early Universe, determined by processes at play prior to the cosmic time, where the computation of the cosmological models is initiated. 
Although we cannot determine the initial densities at place in the early Universe from first principles, we can use the fact that the initial densities are approximated by the $\Lambda$CDM concordance model, applying the customary \qn{fixed $H_0$} approach, with high accuracy, as $\Lambda$CDM with its constant $w$ is not affected by the $w_0$-$w_a$-degeneracy.
As a result, the \qn{fixed EDs} approach can eventually provide a possible resolution to the Hubble tension problem by a DE component with a decreasing EoS parameter, {under the assumption} that the Hubble tension is of cosmological origin and not based on issues in the \qn{local} measurements.
We find that the model with decreasing $w(a)= -0.9+0.1(1-a)$ comes close not only to a consistent evolution, when backward-in-time (fixed $H_0$) and forward-in-time (fixed EDs) calculations are compared to each other, {but it also predicts} a present-day value of the Hubble parameter close to that inferred by observations of the local Universe. Thus, such a DE model may provide a resolution to the Hubble tension problem (see also Fig.~\ref{fig:intForwardiCDM} for illustration). In contrast, the model with increasing $w(a)$ does not display a consistent evolution, when backward-in-time (fixed $H_0$) and forward-in-time (fixed EDs) calculations are compared to each other.

In short, the two approaches can be characterized as follows. The customary \qn{fixed $H_0$} approach starts the computation at the present with a given value of $H_0$ and integrates the equations backward-in-time toward the early Universe. Thus, the densities in the early Universe (and thereby $H(z)$) vary with the choice of the model and its parameters. Instead, the \qn{fixed EDs} approach starts the computation in the early Universe with initial densities determined by the $\Lambda$CDM concordance model, being identical for all considered models. We found these densities to approximate the \q{real} densities to high accuracy, making this approach feasible. Thus, the densities at the present (and thereby $H_0$), determined by a forward-in-time integration of the equations, vary with the choice of the model and its parameters. The important difference between both approaches is that in the \qn{fixed EDs} approach the resulting $H_0$ can be checked (and is checked) against observations in the local Universe. Whereas, in the customary \qn{fixed $H_0$} approach $H(z)$ in the early Universe is not checked against observations in the early Universe. In addition, as exemplified in Sect.~\ref{sec:compmodelsH0}, due to the subdominance of DE in the early Universe, it is not promising to perform those checks. Moreover, the customary \qn{fixed $H_0$} approach is, by construction, not able to provide an answer to the Hubble tension problem, regardless of the model tested. Hence, it rules out a cosmological origin of the Hubble tension problem by construction, relegating it to issues in the local measurements.

We like to stress that our \qn{fixed EDs} approach is no criticism of the correctness of $H_0$ as determined by \citet{Collaboration2018}. On the contrary, the assumption of a cosmological constant was the only reasonable choice at the time of measurement, and moreover, is immune to the consequences of a dynamical model of DE. We simply treat it what it is: the information of  conditions at place at a redshift of $z \sim 1090$ extrapolated to present-day values, {under the assumption} of a cosmological constant\footnote{In Sect.~\ref{sec:proc_customary} we present three models with a DE component, being clearly different from a cosmological constant, that all reproduce the CMB spectrum of $\Lambda$CDM.}. Only the existence of these results makes the \qn{fixed EDs} approach feasible. In other words, the finding of the \qn{concordance EDs} is the great achievement of the CMB measurement by the Planck mission, with the establishment of the $\Lambda$CDM concordance parameters.

Subsequently to the finding of Sect.~\ref{sec:proc_amended}, we argue that employing both, \qn{fixed EDs} approach and \qn{fixed $H_0$} approach, provides a consistency check for cosmological models. In Sect.~\ref{sec:observations} we propose that this novel consistency check be included in the standard MCMC method, that is applied in many observational programs to compare dynamical models of DE to a cosmological constant and test extensions to $\Lambda$CDM. It boils down to the requirement to sample only those models which exhibit consistent results for \qn{fixed EDs} and \qn{fixed $H_0$} calculations of the cosmic evolution, in the final computation of the probability distribution. Including this recommended consistency check would extend the standard MCMC method, to include a three-step procedure starting with the customary calculation of the backward-evolution assuming locally based $H_0$, followed by a calculation of the forward-evolution starting at \qn{concordance EDs}, and a final consistency check where only consistent parameter sets are considered in the computation of the probability distribution. This final step would then narrow down the parameter space dramatically, increasing significantly the accuracy of the determined model parameters.

We present our conclusions as follows. The results depicted in Figs.~\ref{fig:DES_integration}, \ref{fig:iCDM_integration} and \ref{fig:LCDM_consistency} illustrate well the known fact that CMB observations alone are not very suitable to discriminate individual models of DE: around the recombination time, the expansion history displays simply too little deviations from the evolution of the expansion history based on a cosmological constant, despite huge variations in the DE model parameters. However, by the same token, this nonsensitivity provides us with robust information of the initial energy densities predominant in the early Universe and thus makes the \qn{fixed EDs} approach feasible. At the other end of the evolution, differences in the late stages of the expansion history provide a tool to discriminate individual DE candidates. As seen by the gray solid curves in the bottom insets of the top panels in the Figs.~\ref{fig:DES_integration}, \ref{fig:iCDM_integration} and \ref{fig:LCDM_consistency}, the deviation in the expansion history from that of a cosmological constant gives a characteristic signature for individual DE candidates. This is a phenomenological approach, but it can put constraints on the evolution of the EoS of DE candidates. On the grounds of these constraints, new models for DE could be developed, in turn, in the future, either by determining the parametrization of dynamical models of DE, or by reconstructing the evolution of DE's EoS parameter directly from data.

Of course, we also have to keep in mind that the CPL parametrization is merely a convenient representation, and one could consider a different evolution of the EoS parameter, altogether (see e.g., \citet{Dinda2022} and \citet{OColgain2021}. Also, some observations or findings may give indications in which direction to go. For instance, hydro-dynamical and geometrical models of the cosmic web presented by \citet{Icke1984}, \citet{Icke1987} and \citet{Icke2001} indicate that its evolution can be modeled by an effective EoS parameter evolving linearly in redshift, rather than in scale factor. 
The Hubble tension problem has also triggered more proposals with respect to possible DE candidates, for example, models of early dark energy (EDE), for a review see \citet{Poulin2023}. One of the drawbacks of these attempts is that mitigating the Hubble tension problem increases the so-called $\sigma_8$ tension. In \citet{Reboucas2024}, this is addressed by assuming a modification of EDE with a piecewise-constant EoS parameter, providing a solution to both problems.

The change in the expansion history and growth factor due to a possible DE component belong to the most significant observables that allow its EoS to be constrained. Therefore, observations densely sampled in redshift $z$ with high accuracy of the determined Hubble parameter $H$, especially obtained for $z \le 2$, will be required from ongoing and future observational campaigns to constrain the evolution of the EoS of DE, specifically to detect potentially a decreasing EoS parameter. If statistically significant deviations from a cosmological constant would be found and backed with multiple observations, it would have consequences, not only for a potential resolution of the Hubble tension problem, but more fundamentally for $\Lambda$CDM and its status as the current cosmological standard model.
After all, the Hubble tension problem might result because of the assumption of a cosmological constant. As we show in this paper, an exemplary cosmological model having a DE component with a decreasing EoS parameter of $w_0=-0.9$ and $w_a=+0.1$ comes close in fulfilling the new consistency criterion, along with yielding good agreement with the locally determined Hubble constant. This is in contrast to other DE models. Also, the $\Lambda$CDM concordance model exhibits disagreements. In fact, demanding the proposed consistency check seems to rule out a cosmological constant (see Appendix~\ref{app:checkmodels}), {if} we accept the Hubble tension problem as real (see also e.g., \citet{Dainotti2022,Dainotti2023,Dainotti2021,Dainotti2022a,Bargiacchi2023,Bargiacchi2023a,OColgain2021,Krishnan2021a}, reporting that measurements might favor a dynamical model of DE, instead of a cosmological constant).
In light of our results, we thus argue that the Hubble tension problem might be resolved, once the assumption of a cosmological constant is abandoned in favor of a DE component with a decreasing EoS parameter of certain specificity.
Of course, in order to test such a DE model more carefully (also beyond the CPL parametrization), high-accuracy determinations of the Hubble parameter over a range in redshift bins are required, as just emphasized. In fact, many ongoing and future large-scale structure surveys are probing DE candidates from low-$z$ to mid-$z$ range by determining a prospective evolution in the EoS parameters. In the interests of convenience, we name some of the most important surveys (no ranking implied), including the

\noindent Dark Energy Spectroscopic Instrument (DESI) survey, see \citet{Seo2003}: measurement of bright galaxies up to a redshift of $z=0.4$, luminous red galaxies up to $z=1$, emission-line galaxies up to $z=1.6$ and quasars up to $z=3.5+$,

\noindent Dark Energy Survey (DES) (\cite{Abbott2022,Abbott2022a}) applies gravitational lensing for 319 million objects in the range $z=0.2$ to $1.05$, divided into six bins,

\noindent Hobby-Eberly Telescope Dark Energy Experiment (HETDEX) observes  1 million Lyman-$\alpha$ emitters at $1.9<z<3.5$, and 1 million OII-emitting galaxies at $z<0.5$ (\citet{Hill2012,Gebhardt2021}),

\noindent Euclid survey (see e.g., \citet{Rhodes2019}) uses weak lensing observations of galaxies up to $z \ge 2$, BAOs for $z \ge 0.7$ and local observations out to distances of 5 Mpc. The deep survey primarily targets galaxies at $z > 6$, where follow-up observations by the James Webb Space Telescope are envisaged,

\noindent Nancy Grace Roman Space Telescope's High Latitude Time Domain Survey: measurement of thousands of SNe Ia up to $z=2$, see for example, \cite{Hounsell2018}, and 

\noindent Vera C. Rubin Observatory, previously referred to as the Large Synoptic Survey Telescope (LSST): measurement of SNe Ia with mean $z \sim 0.45$ to max $z=0.8$, see for example, \citet{Ivezic2019}.


In addition, {direct} measurements in the \qn{local} Universe remain very important and promising probes, because they are not sensitive to the specific characteristics of presumed cosmological models, as described by for example,  \citet{Kenworthy2019}, \citet{Bernal2016} and  \citet{Riess2022}.

\begin{acknowledgements}
The authors are grateful for helpful discussions with Oliver Hahn, Bodo Ziegler, Glenn van de Ven and Dragan Huterer.
T.R.-D. acknowledges the support by the Austrian Science Fund FWF through the  FWF Single-Investigator Grant (FWF-Einzelprojekt) No. P36331-N, and the support  by the Wolfgang Pauli Institute in hosting this grant.
\end{acknowledgements}



\begin{appendix}
\section{The impact of variations in the density parameters ${\Omega_{i,0}}$}\label{app:wConst}

In our study of the impact of a DE component with time-dependent EoS onto the computation of the evolution of its energy density, we have used in this paper the same density parameters as for the $\Lambda$CDM concordance model. 

In this appendix, we check our amended code for cases with deviations of the density parameters $\Omega_{i,0}$, compared to the values of $\Lambda$CDM, as a testbed how our modification fares in such situations, and compare the results to those of the standard customary approach.
In order to provide hard tests, we considered exotic models, but for components with constant EoS parameter. 

\begin{figure}  [!htbp]
\includegraphics[width=1.0\columnwidth]{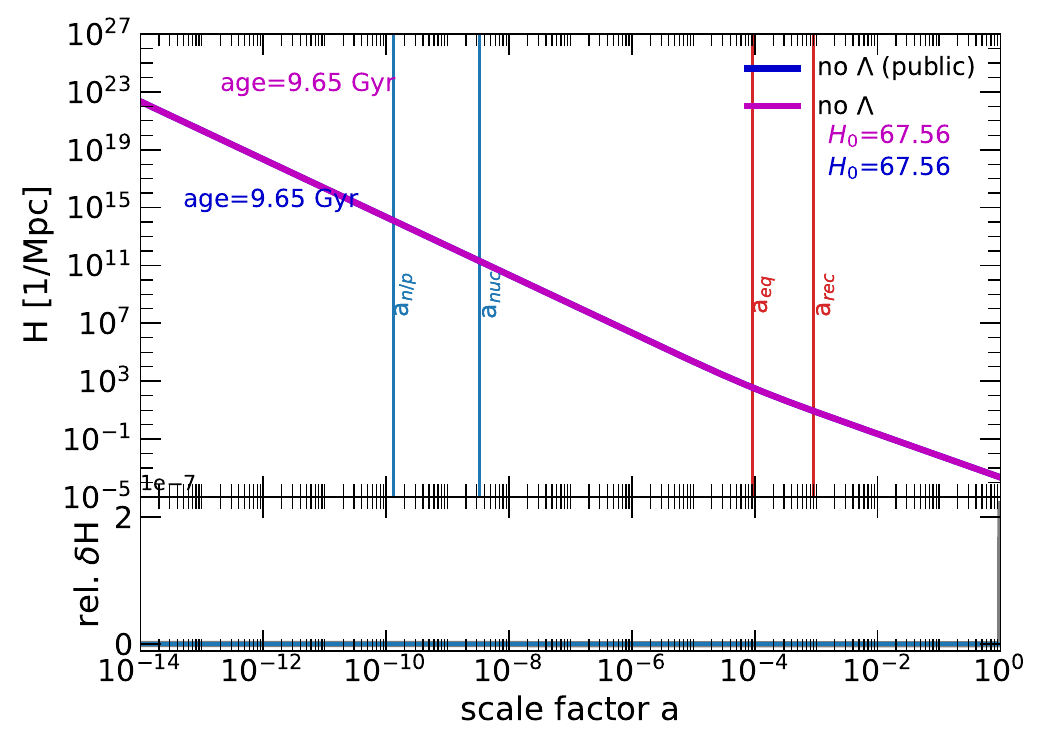} 	 	
\includegraphics[width=1.0\columnwidth]{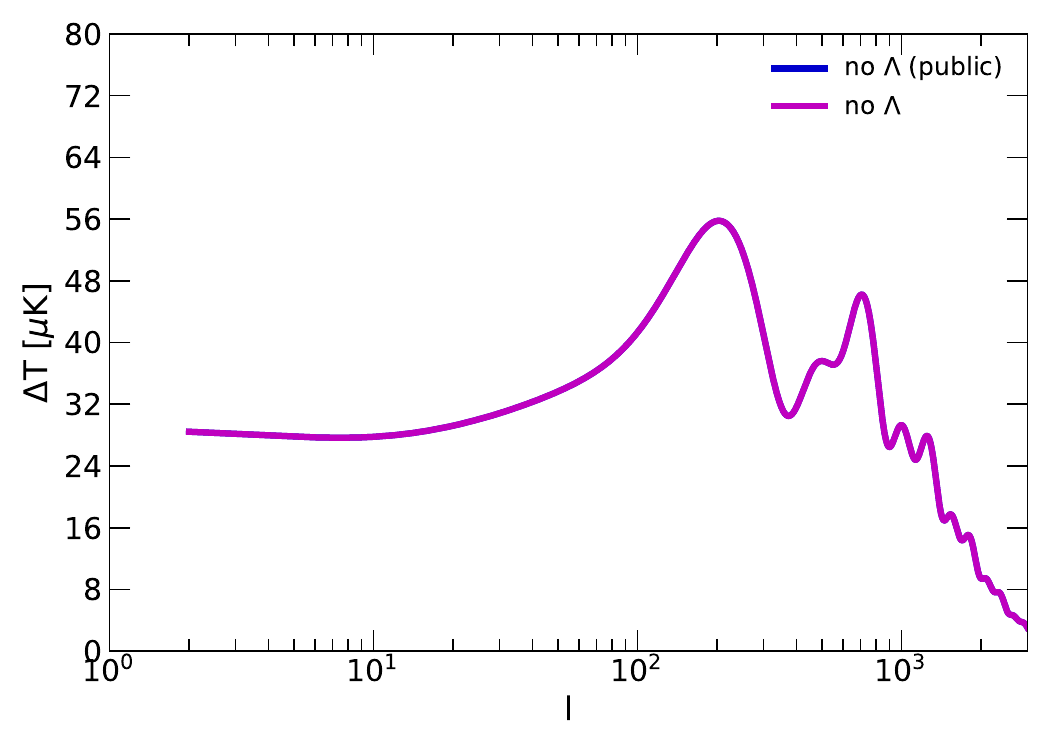}
\includegraphics[width=1.0\columnwidth]{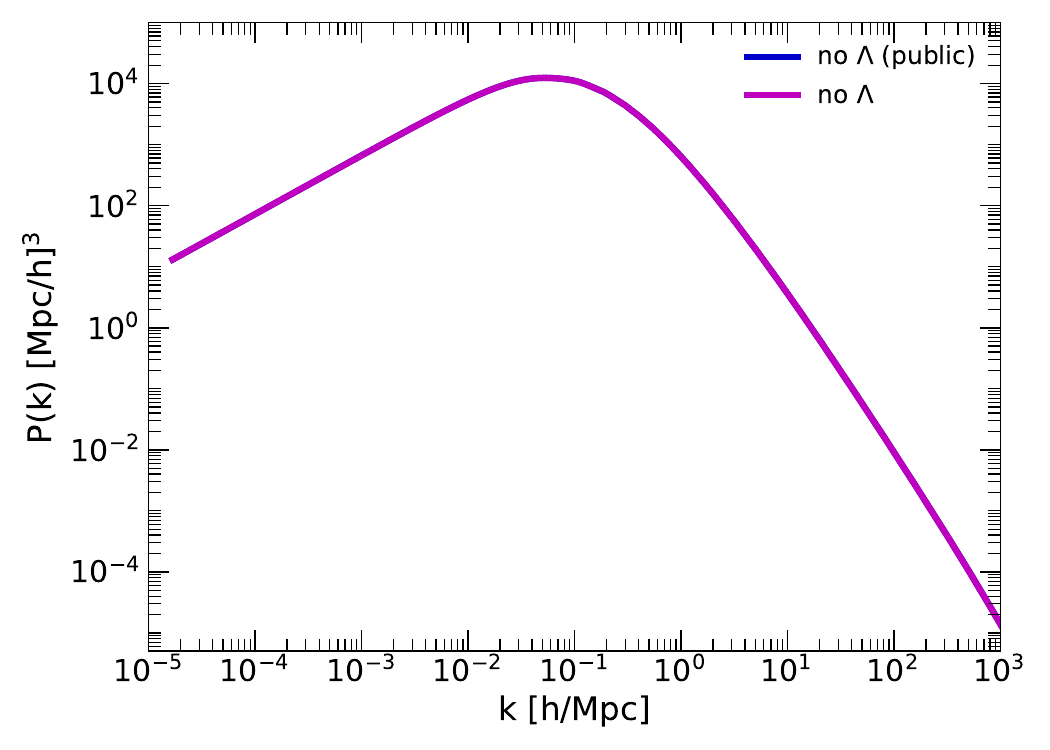}	
\caption[Impacting the initial densities (no $\Lambda$)]
{\tb{Model without \boldmath$\Lambda$.} The panels display from top to bottom the expansion history, the CMB temperature spectrum and the matter power spectrum of a model universe, which omits $\Lambda$ but increases the CDM component accordingly, such that the total of the energy densities is equal to the critical density (i.e., a case close to the Einstein-de-Sitter model). The blue curves depict the results obtained by the public version of CLASS. The magenta curves show the results from our amended version, which uses the ICs of the concordance model. The results of both computations agree very well (the magenta curves lie on top of the blue curves).
}
\label{fig:wConst-L}
\end{figure}

We first considered a model without a $\Lambda$-component, while obeying Eq.~\eqref{eq:EQclosureLCDMfull} with vanishing curvature term, and we increased the density parameter for the CDM accordingly. This case is close to the Einstein-de-Sitter model. The results can be seen in Figure ~\ref{fig:wConst-L}, that were calculated using our amended code as well as the original, public code of CLASS.
The expansion histories agree very well, resulting in identical values for $H_0$ for both code versions. The same agreement can be seen in the respective power spectra of the model universe. There is only a minor difference of $\sim 10^{-15}$ in the initial densities in the public version, compared to the amended version.
We also checked another exotic model, where we modified the density parameters of both baryons and CDM such to make them equal, while leaving the total matter content unchanged. (This model has the same expansion history than $\Lambda$CDM, but the spectra differ.)  For both code versions, we get the same results (not shown here). 

For both exotic models, the initial densities are not much different, because we changed the matter components and/or $\Omega_\Lambda$ that are subdominant in the early Universe.
Therefore, we get basically the same results for both codes.  

\begin{figure}  [!b]
\includegraphics[width=1.0\columnwidth]{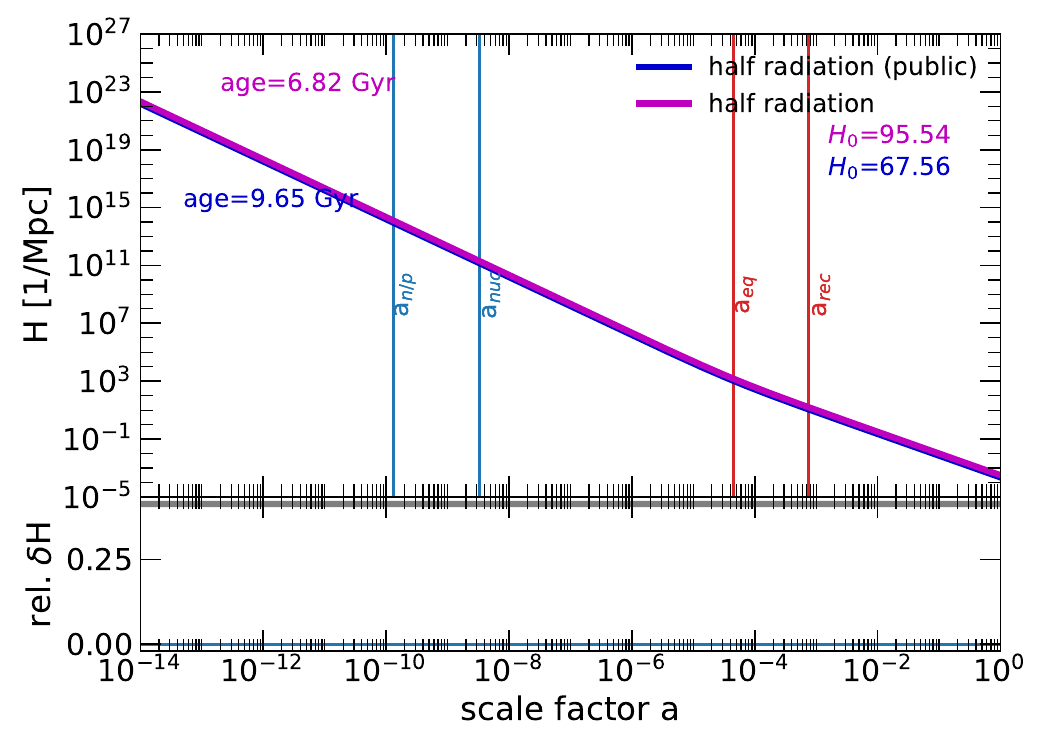} 	
\includegraphics[width=1.0\columnwidth]{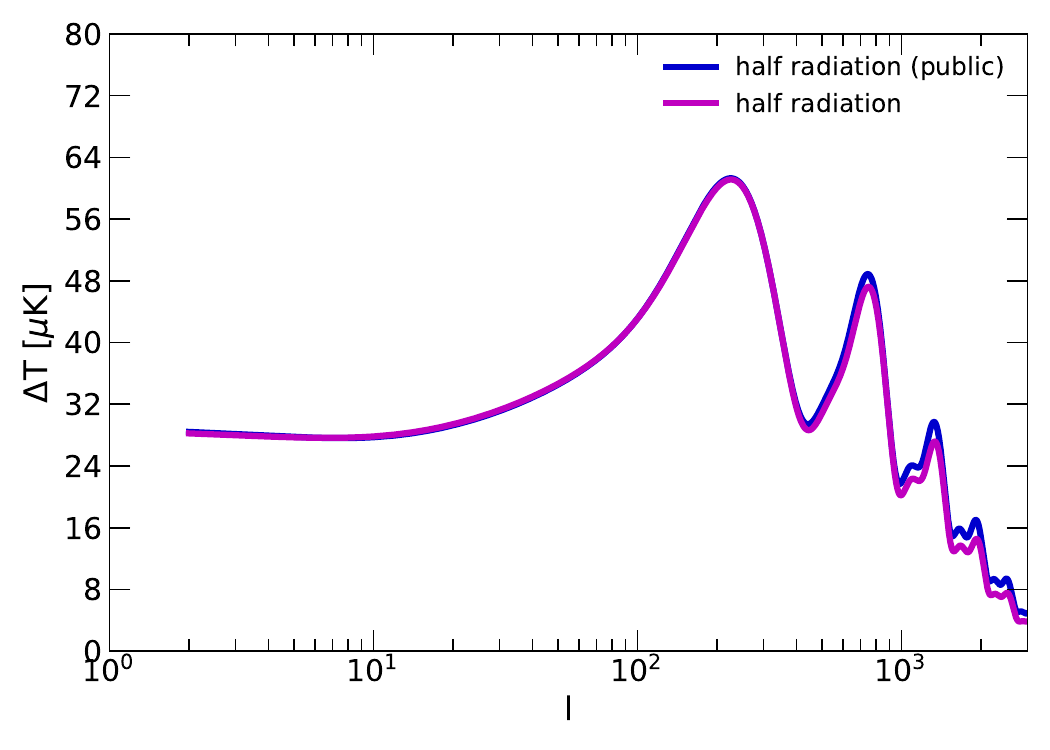}
\includegraphics[width=1.0\columnwidth]{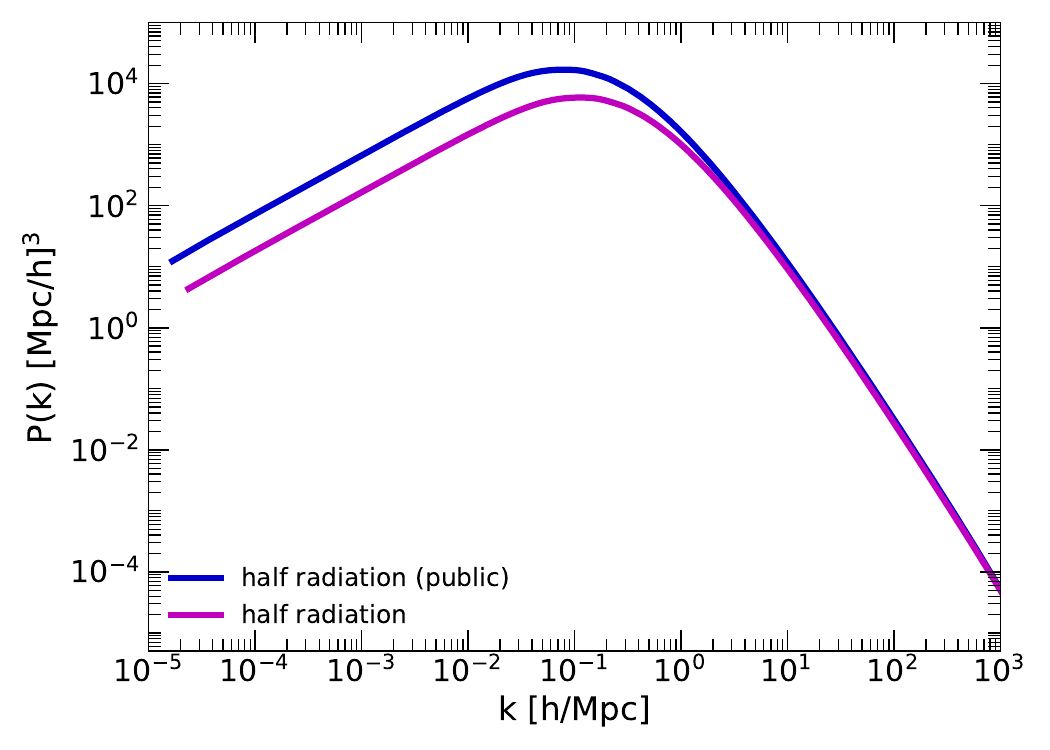}		
\caption[Impacting the initial densities (low radiation)]
{\tb{Model without \boldmath$\Lambda$ and less radiation.} The panels display from top to bottom the expansion history, the CMB temperature spectrum and the matter power spectrum of a model universe without $\Lambda$, whose density parameter for radiation is reduced to half its value of the concordance model, while its CDM density parameter has been accordingly increased, such that the total of the energy densities is equal to the critical density. The blue curves depict the results obtained by the public version of CLASS. The magenta curves show the results from the amended version, which uses the ICs of the concordance model. The results of both computations significantly deviate from each other.
}
\label{fig:wConst-R}
\end{figure}

However, once we picked a case where the ICs were greatly affected, we see a difference, which is illustrated by the following model displayed in Fig.~\ref{fig:wConst-R}. Here, we omitted $\Lambda$ (as in the case of Figure ~\ref{fig:wConst-L}), but additionally we reduced the density parameter for radiation by half, compared to its standard value, while increasing the density parameter for the CDM component accordingly. Since radiation is the dominant component in the early Universe, such an extreme change affects the ICs critically (top panel of Fig.~\ref{fig:wConst-R}).
Now, we can see a clear difference between the results from the public and the amended CLASS code in the expansion history, as well as in the spectra.
Also, matter-radiation equality $a_{\text{eq}}$ and recombination $a_{\text{rec}}$ are shifted to a later cosmic time, and the present-day expansion rate of $H_0 = 95.54$ km/sec/Mpc, is much higher compared to the configured value of $H_0 = 67.56$ km/sec/Mpc.

Of course, the adoption of the \qn{concordance EDs}, while modifying the density parameter for the radiation component is not in agreement with observations, for both versions of the code. Obviously, this is not a viable procedure in the computation of cosmological models, because it breaks the connection between the EDs and $H_0$ established by the concordance model, as already discussed. 
In fact, the importance of adopting (keeping) the standard value for the radiation density parameter, especially by excluding it from MCMC parameter space searches, has been known and is handled that way by observational campaigns which apply the standard scheme, see for example, \citet{Abbott2022a} and \citet{Alam2021}.
To summarize, provided the density parameter of radiation is the same as in the standard, concordance model, the variation of the remaining parameters leads to the same results for both computational procedures, for both the public CLASS code and our amended CLASS code.

\section{Consistency check for $\Lambda$CDM}\label{app:checkmodels}
\begin{figure*}  [!htb]
\centering
\includegraphics[width=1.0\columnwidth]{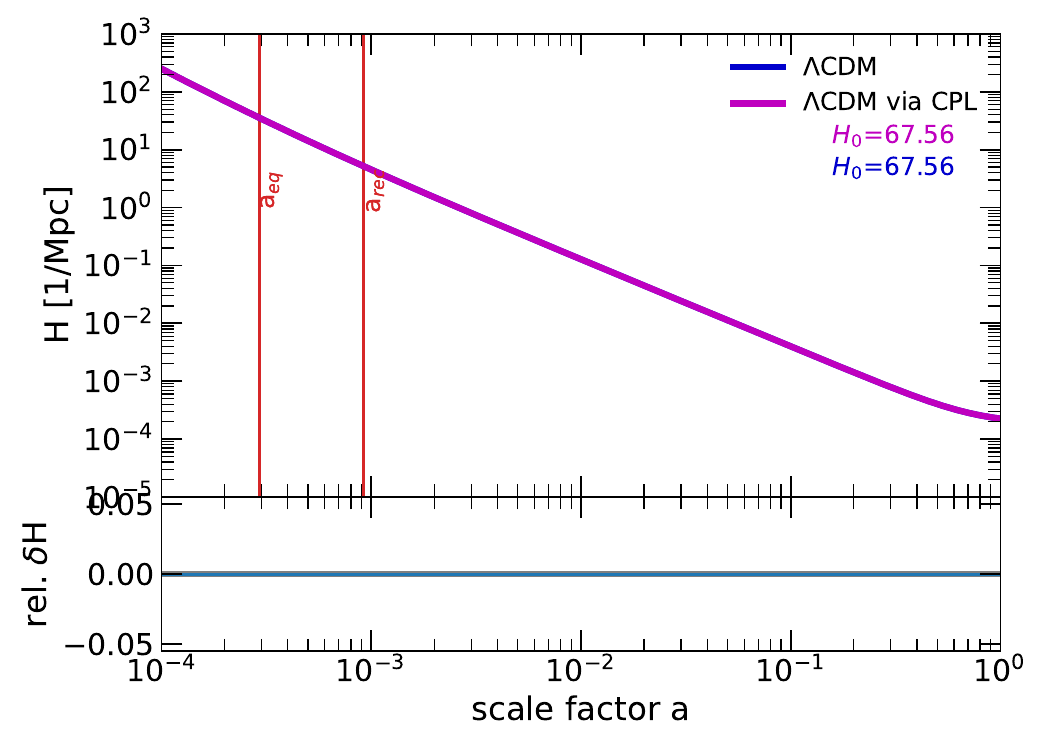}
\includegraphics[width=1.0\columnwidth]{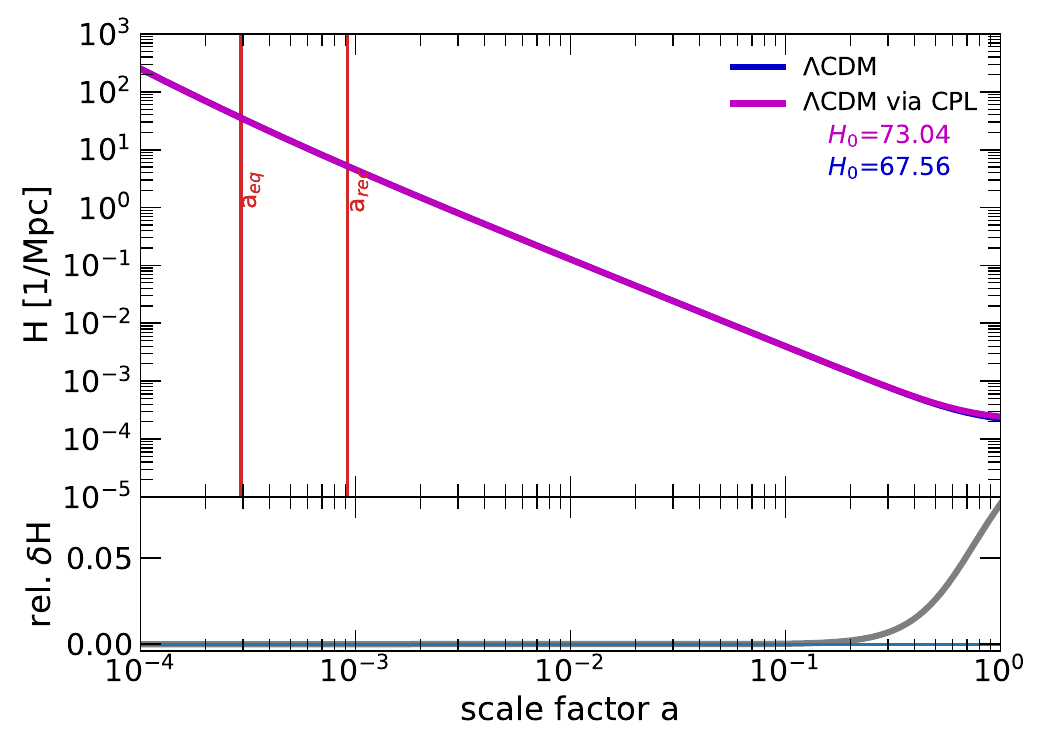}	
\includegraphics[width=1.0\columnwidth]{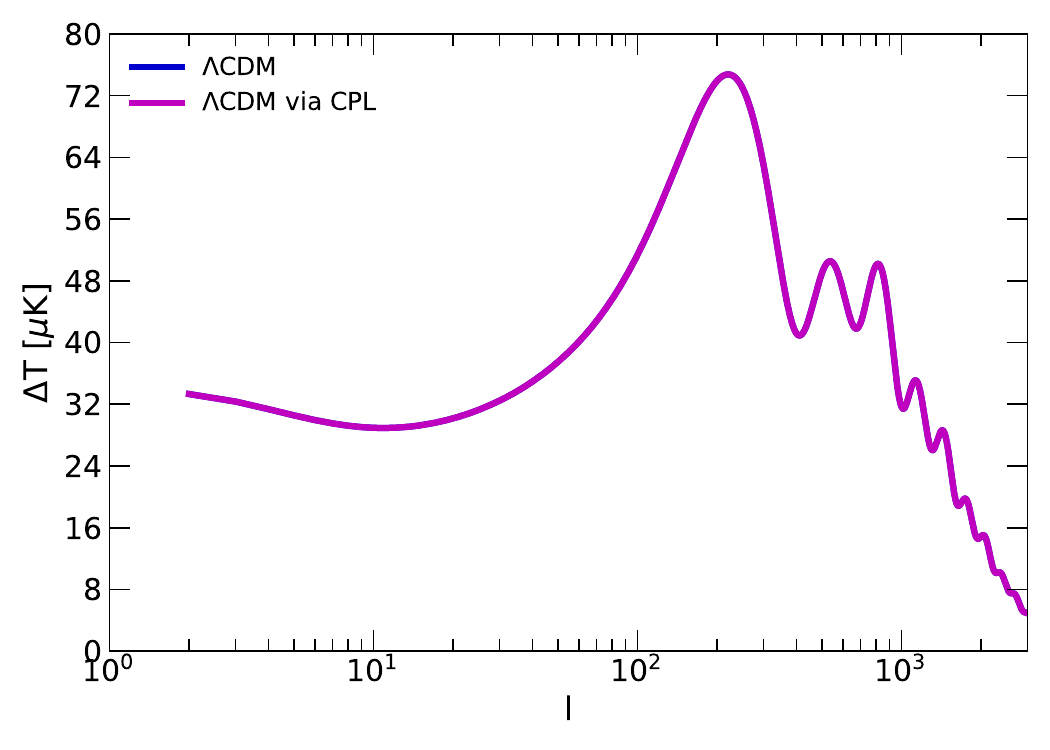}
\includegraphics[width=1.0\columnwidth]{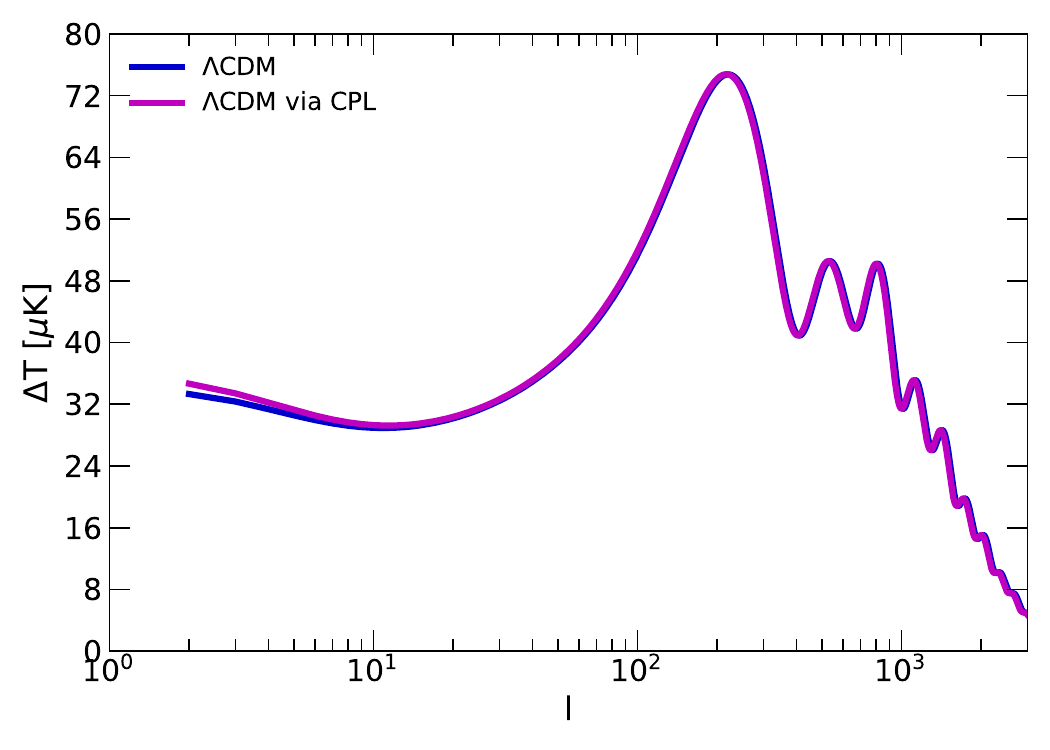}
\includegraphics[width=1.0\columnwidth]{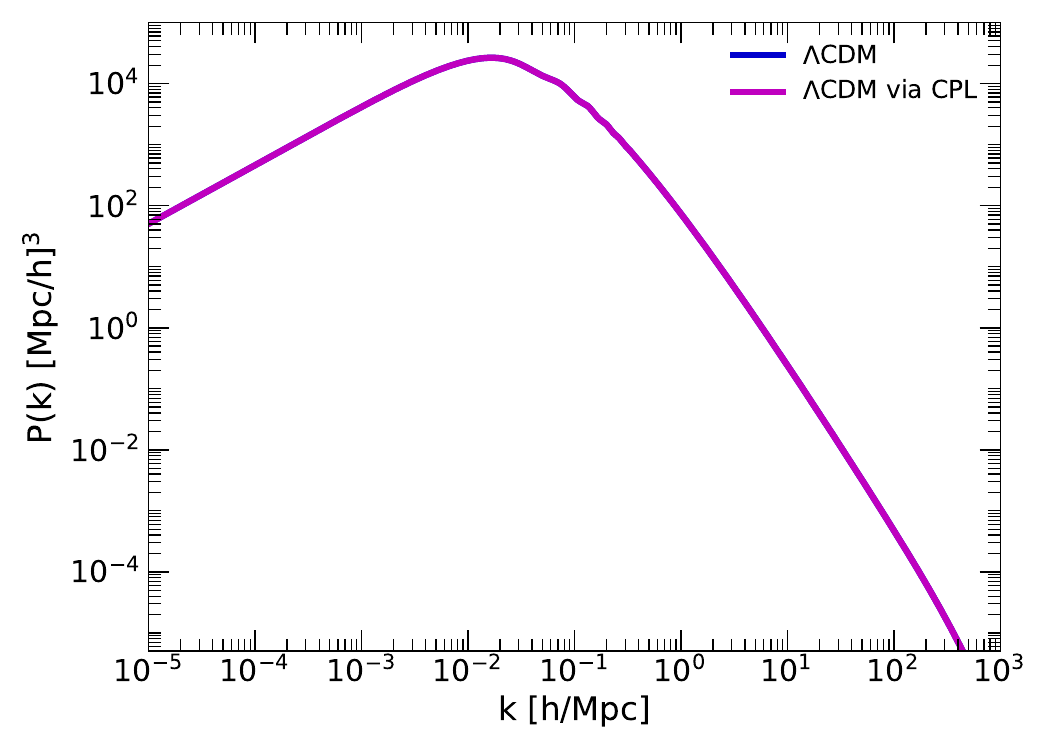} 
\includegraphics[width=1.0\columnwidth]{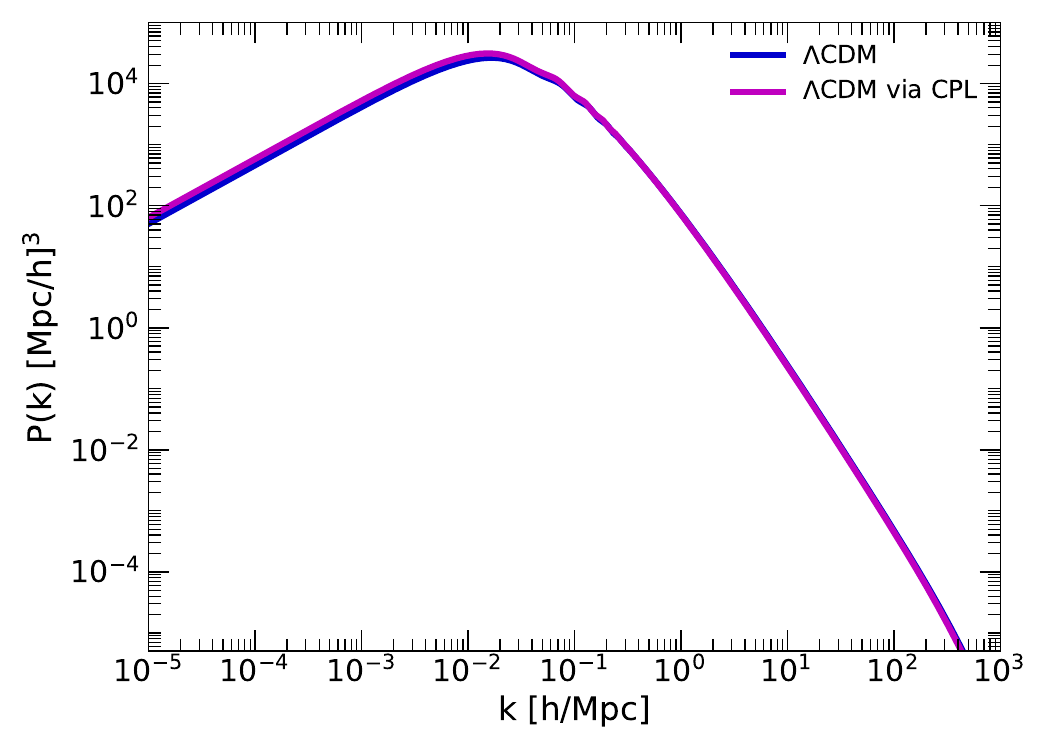}					
\caption[LCDM consistency]
{\tb{\boldmath$\Lambda$CDM consistency check.} We use our amended version of CLASS to compute observables of the \glo{LCDMmodel}, using the CPL parametrization with associated parameters $w_0=-1$ and $w_a=0$. Left-hand panels: calculation based on the parametrization which uses  the \qn{early} CMB-based value of $H_0=67.56$ km/sec/Mpc from \citet{Collaboration2018}. Since this is equivalent to the standard input in CLASS for $\Lambda$CDM, we see no difference in the results.   Right-hand panels: calculation based on the parametrization which uses the \qn{local} value of $H_0=73.04$ km/sec/Mpc from \citet{Riess2022}. The top row displays the expansion histories, while the middle and bottom rows display the CMB temperature spectra and matter power spectra, respectively. See main text for explanations.
}
\label{fig:LCDM_consistency}
\end{figure*}

We tested our amended CLASS code by computing observables for the $\Lambda$CDM model, and compared them with the results of the original, public version of CLASS. Our amended code used the CPL parametrization, by setting $w_0=-1$ and $w_a=0$ in order to recover a cosmological constant. We computed two versions, one based on the parametrization which used the \qn{early} CMB-based value of $H_0$ from \citet{Collaboration2018} (left-hand panels), and another one based on the parametrization which used the \qn{local} value of $H_0$ from \citet{Riess2022} (right-hand panels). 
The results and comparison for the expansion histories, as well as for the CMB temperature spectra and matter power spectra, can be found in Fig.~\ref{fig:LCDM_consistency}.

The left-hand panels show no difference in the results, which is expected, since we parameterize the model in a way that is equal to the standard CLASS input for the concordance $\Lambda$CDM model. On the other hand, the right-hand panels show a difference: the expansion history displays similar deviations from the \glo{LCDMmodel} to those shown in Fig.~\ref{fig:iCDM_integration}. But the maximum deviation happens at a scale factor of $a=1$, in contrast to Fig.~\ref{fig:iCDM_integration}, where the maximum deviation is at a scale factor of $a \sim 0.8$. There are also some minor deviations in the spectra in the right-hand panels, limited to large spatial scales.
Assuming that the Hubble tension problem is real, we can recognize that $\Lambda$CDM suffers this problem, for the backward and forward evolutions do not yield consistent results. In other words, we see differences in the expansion history (gray curve) from the concordance model when we adopt a locally determined $H_0$ in our model, see right-hand top panel. In this sense, $\Lambda$CDM does not pass the recommended consistency check for cosmological models and their parameters, respectively. Thus, there remains a Hubble tension problem in $\Lambda$CDM.

\end{appendix}

\end{document}